\documentclass[pra,reprint,amsmath,amssymb]{revtex4-1}

\usepackage{graphicx}
\usepackage{hyperref}

\begin{document}

\title{Thermodynamics of dilute Bose gases: Beyond mean-field theory for binary mixtures of Bose-Einstein condensates}

\author{Miki Ota and Stefano Giorgini}

\affiliation{INO-CNR BEC Center and Dipartimento di Fisica, Universit\`a di Trento, 38123 Trento, Italy}

\email{miki.ota@unitn.it}

\date{\today}

\begin{abstract}
We study the thermodynamic properties of binary Bose mixtures, by developing a beyond mean-field Popov theory which properly includes the effects of quantum and thermal fluctuations in both the density and spin channels. Results for key thermodynamic quantities, such as the isothermal compressibility and the magnetic susceptibility, are derived from a perturbative calculation of the grand-canonical potential. We find that thermal fluctuations  can play a crucial role on the miscibility condition of a binary mixture, favoring phase separation at finite temperature even if the mixture is soluble at zero temperature, as already  anticipated in a previous work [Ota et al., Phys. Rev. Lett. \textbf{123}, 075301 (2019)]. We further investigate the miscibility condition for binary mixtures in the presence of asymmetry in the intra-species interactions, as well as in the masses of the two components. Furthermore, we discuss the superfluid behavior of the mixture and the temperature dependence of the Andreev-Bashkin effect.
\end{abstract}

\maketitle




\section{Introduction} \label{Sec:Intro}

The equation of state (EOS) of classical or quantum fluids characterizes completely the thermodynamic behavior of the system, by providing unique informations about the fundamental properties of the fluids at finite temperature~\cite{book,Huang}, such as their behavior at the phase transition, the role of quantum statistics and the effects of the interatomic forces. For instance, in liquid ${}^4 \mathrm{He}$, the observation of the celebrated superfluid lambda point was achieved from the measurement of the specific heat~\cite{Lipa2003}. Half century later, the same lambda transition was observed in the context of the unitary Fermi gas~\cite{Ku2012}, by extracting the EOS of the homogeneous gas from a measurement carried out on a trapped system. This methodology, based on the local density approximation, has been successfully used in obtaining the EOS of two-dimensional Bose~\cite{Desbuquois2014,Yefsah2011} and Fermi gases~\cite{Boettcher2016,Fenech2016,Makhalov2014}. As for the three-dimensional Bose gas, the zero-temperature EOS has been probed experimentally in Ref.~\cite{Navon2011}, and the role of quantum fluctuations giving rise to beyond mean-field effects has been verified. However, a complete determination of the EOS at finte temperature for the homogeneous gas is still lacking, the main difficulties arising from the absence of universal description, and the sharp change in the density profile of the trapped gas as one crosses the transition point, requiring therefore high precision measurements~\cite{Nascimbene2010b,Mordini2020}.
\par
On the theoretical side, the simplest mean-field Hartree-Fock (HF) theory has been widely used to describe the equilibrium properties of dilute Bose gases at finite temperature~\cite{book}, and shown to describe experimental data with reasonably good accuracy~\cite{Dalfovo1999,Gerbier2004,Smith2011,Mordini2020}. The satisfactory description of thermodynamics provided by HF theory relies on the weakness of interactions in these systems as well as on the relatively more marginal role played by beyond mean-field effects at finite temperature, including on the thermodynamic behavior near the transition between the superfluid and the normal phase. However, in the last few years, novel experimental techniques allowing for a more precise determination of the EOS have become available. These include the box-like trapping potential~\cite{Gaunt2013,Lopes2017,Chomaz2015,Mukherjee2017}, which allows to probe a homogeneous gas, as well as the development of high resolution imaging techniques~\cite{Ramanathan2012,Mordini2020b}. Besides, since the experimental realization of coherent coupling~\cite{Lin2011,Zhang2012} and the observation of self-bound quantum droplets~\cite{Cabrera2018,Semeghini2018}, there has been a growing interest for mixtures of Bose-Einstein condensates (BECs), for which the finite-temperature behavior still remains an open question. These recent developments all indicate the need for a reliable finite-temperature theory, which allows one to study the thermodynamics of Bose gases in diverse configurations, with the same accuracy up to the critical temperature.
\par
The fundamental elements of HF theory are single-particle excitations. Further improvements accounting for pair excitations are brought about by the Hartree-Fock-Bogoliubov (HFB) theory~\cite{Proukakis2008,Griffin1996}, which is based on non-interacting quasi-particles. The HFB theory takes into account effects of quantum fluctuations including the quantum depletion of the condensate. However, the HFB approach suffers from the presence of an unphysical gap in the excitation spectrum, and many studies have been devoted to the understanding of its origin and ways to overcome it ~\cite{Takano1961,Proukakis1998,Shi1998}. In particular, the pathology of the HFB theory arises from the incorrect treatment of second-order terms in the interaction strength~\cite{Shi1998}, and an improvement of the theory, referred to as the finite-temperature Belieav technique (or Popov theory~\cite{Popov1983}) has been put forward (see {\it e.g} Ref.~\cite{Fedichev1998}). Although Popov theory is known to be the proper theory accounting for leading corrections to the thermodynamic quantities of a weakly interacting Bose gas, only a few works have used this approach to investigate the equilibrium properties of Bose gases at finite temperature~\cite{Capogrosso-Sansone2010}.
\par
The main purpose of this paper is therefore to provide with a straightforward methodology to construct the finite-temperature Popov theory for weakly-interacting Bose gases, which properly takes into account the effects of thermal and quantum fluctuations. We give a derivation of the Popov theory based on the diagonalization of the Hamiltonian in terms of Bogoliubov quasi-particles and of its perturbative solution. An equivalent derivation can be carried out using diagrammatic techniques~\cite{Fedichev1998,Capogrosso-Sansone2010}. For a single-component gas we present our calculations for the condensate density and several thermodynamic quantities, including the isothermal compressibility, which is particularly sensitive to interaction effects. Furthermore, the method can be applied to more complex Bose systems and in this paper we extend the formalism to binary mixtures of BECs. We point out the improvements of the Popov approach with respect to the predictions of Hartree-Fock theory, which turn out to be particularly important in the study of the miscibility of a quantum mixture at finite temperature. For binary condensates, we find that the inclusion of beyond mean-field terms change drastically the thermodynamic behavior, eventually leading to the emergence of new phases, such as the self-bound quantum droplets~\cite{Petrov2015,Cabrera2018,Semeghini2018,Ota2020} and non-trivial phase-separated states~\cite{Ota2019,He2020,Roy2020} as well as the occurrence of collisionless spin drag~\cite{Andreev1975,Nespolo2017}.  
\par
The structure of the paper is as follows. First, in Sec.~\ref{Sec:theory_sc}, we derive the thermodynamic potential for a uniform single-component dilute Bose gas, starting from the grand-canonical Hamiltonian, which we diagonalize by means of the Bogoliubov transformation. We present our numerical results for the single-component condensate density, as well as the chemical potential and the isothermal compressibility. We extend the formalism of Popov theory to the case of two-component mixtures in Sec.~\ref{Sec:theory_mixt} and show our numerical results for the main thermodynamic quantities. We discuss in Sec.~\ref{Sec:Phase Separation} the free energy of the mixture and the miscibility condition also for interaction and mass imbalanced systems, extending the findings of a recent work~\cite{Ota2019}. Finally in Sec.~\ref{Sec:AB effects} we discuss the Andreev-Bashkin effect at finite temperature by calculating explicitly the superfluid densities for the mixture.




\section{Single-component Bose gas: Formalism of Popov theory}\label{Sec:theory_sc}

\subsection{Diagonalization of the Hamiltonian}\label{Sec:diag_sc}

Our starting point is the grand-canonical Hamiltonian for a single component homogeneous Bose gas, in the absence of external potentials. In terms of the single-particle creation and annihilation operators, $\hat{a}^\dagger_\mathbf{k}$ and $\hat{a}_\mathbf{k}$, the Hamiltonian including all two-body collisions takes the form:
\begin{equation} \label{Eq:H_sc}
\hat{H} = \sum_\mathbf{k} \varepsilon_\mathbf{k} \hat{a}_{\mathbf{k}}^\dagger \hat{a}_{\mathbf{k}} + \frac{g}{2V} \sum_{\mathbf{k}, \mathbf{k}', \mathbf{q}} \hat{a}_{\mathbf{k}}^\dagger \hat{a}_{\mathbf{k'+q}}^\dagger \hat{a}_{\mathbf{k'}} \hat{a}_{\mathbf{k+q}}
\end{equation}
where $\varepsilon_\mathbf{k} = \hbar^2 \mathbf{k}^2/(2m)$ is the single-particle kinetic energy. In the above equation we have assumed a point-like interaction between particles $V_\mathrm{int}(\mathbf{r}-\mathbf{r'}) = g \delta (\mathbf{r} - \mathbf{r'})$, with $g$ the interaction coupling constant related to the $s$-wave scattering length $a_s$ by $g = 4 \pi \hbar^2 a_s / m$.
\par
After applying the usual Bogoliubov prescription, which consists in replacing the operators $\hat{a}_{0}$ and $\hat{a}_{0}^\dagger$ with the macroscopic number of particles in the condensate $\sqrt{N_{0}}$, one obtains for the grand-canonical Hamiltonian $\hat{K} = \hat{H} - \mu \hat{N}$, where $\hat{N}=N_0+\sum_{\mathbf{k}}\hat{a}_\mathbf{k}^\dagger \hat{a}_\mathbf{k}$, the result:
\begin{align}\label{Eq:K_sc}
\hat{K} =& \frac{g}{2V} N_{0} ^2 - \frac{g}{V} \tilde{N}^2 - \mu N_{0} + \sum_{\mathbf{k} \neq 0} \left( \varepsilon_\mathbf{k} + 2 g n - \mu \right) \hat{a}_\mathbf{k}^\dagger \hat{a}_\mathbf{k} \nonumber \\
&+ \frac{g}{2V} N_{0} \sum_{\mathbf{k} \neq 0} \left( \hat{a}_\mathbf{k}^\dagger \hat{a}_\mathbf{-k}^\dagger + \hat{a}_\mathbf{k} \hat{a}_\mathbf{-k} \right) \, ,
\end{align}
where we have introduced the number of non-condensed atoms, $\tilde{N} = \langle \hat{N} \rangle-N_0 = \sum_{\mathbf{k} \neq 0} \langle \hat{a}_\mathbf{k}^\dagger \hat{a}_\mathbf{k} \rangle$, and the total atom number density $n = \langle \hat{N} \rangle/V = n_0 + \tilde{n}$. In obtaining Eq.~\eqref{Eq:K_sc}, we have applied a mean-field treatment on the interaction terms involving non-condensate operators $\hat{a}_{\mathbf{k} \neq 0}$, $\hat{a}_{\mathbf{k} \neq 0}^\dagger$ and we neglected higher order contributions. In particular, discarded terms include: cubic products of non-condensate operators and terms of the form $g \tilde{m} \hat{a}^\dagger_\mathbf{k} \hat{a}^\dagger_\mathbf{-k}$ and $g \tilde{m}^2$, where $\tilde{m} = V^{-1} \sum_{\mathbf{k}\neq 0} \langle \hat{a}_\mathbf{k} \hat{a}_\mathbf{-k} \rangle$ is the anomalous density. As one shall see below, the leading order of the anomalous density is linear in $g$ and these terms correspond therefore to contributions beyond second-order. We briefly note that these higher order terms are included in the Hartree-Fock-Bogoliubov (HFB) theory, making a key difference with the present approach, since they yield a gapped excitation spectrum. One should also notice that the last term of Eq.~\eqref{Eq:K_sc} is of order $g^2$, leading to the well-known problem of ultraviolet divergence. This issue, which arises from the approximated treatment of inter-atomic interactions, is conveniently solved by a proper renormalization of the coupling constant~\cite{LandauSP2}: $g \rightarrow g [1 + gV^{-1} \sum_\mathbf{k} 1/(2\varepsilon_\mathbf{k}) ]$.
\par
One can diagonalize Eq.~\eqref{Eq:K_sc} by means of the canonical Bogoliubov transformation:
\begin{align}\label{Eq:Bogo_trans}
\begin{split}
\hat{a}_\mathbf{k} &= u_\mathbf{k} \hat{\alpha}_\mathbf{k} + v_\mathbf{-k}^* \hat{\alpha}_\mathbf{-k}^\dagger \, , \\
\hat{a}^\dagger_\mathbf{k} &= u^*_\mathbf{k} \hat{\alpha}_\mathbf{k}^\dagger + v_\mathbf{-k} \hat{\alpha}_\mathbf{-k} \, .
\end{split}
\end{align}
In the above equations, $\hat{\alpha}_\mathbf{k}$ and $\hat{\alpha}_\mathbf{k}^\dagger$ are the quasi-particle annihilation and creation operators obeying Bose commutation relations. This involves the normalization $\vert u_\mathbf{k} \vert^2 - \vert v_\mathbf{-k} \vert^2 = 1$ for the quasi-particle amplitudes and, after substituting~\eqref{Eq:Bogo_trans} in Eq.~\eqref{Eq:K_sc}, one finds that the off-diagonal terms vanish for the following values of the functions $u_\mathbf{k}$ and $v_\mathbf{k}$:
\begin{equation}\label{Eq:uv_sc}
u_\mathbf{k}, v_\mathbf{-k} = \pm \left( \frac{\varepsilon_\mathbf{k} + \Lambda}{2 \tilde{E}_\mathbf{k}} \pm \frac{1}{2} \right)^{1/2},
\end{equation}
where we have introduced the quantity $\Lambda = 2gn - \mu \geq 0$ for future convenience~\cite{Capogrosso-Sansone2010}, while $\tilde{E}_\mathbf{k} = \sqrt{(\varepsilon_\mathbf{k} + \Lambda)^2 - (gn_0)^2}$ is the Bogoliubov quasi-particle spectrum. Notice that $\Lambda$ corresponds to the shift $\delta\mu=\mu_c-\mu$ of the chemical potential with respect to its value $\mu_c=2gn$, holding at the critical point according to mean-field theory. By means of Eq.~\eqref{Eq:uv_sc}, the Hamiltonian~\eqref{Eq:K_sc} reduces to a pseudo-Hamiltonian describing a gas of non-interacting quasi-particles:
\begin{equation}
\hat{K} = \Omega_0 + \sum_{\mathbf{k} \neq 0} \tilde{E}_\mathbf{k} \hat{\alpha}_\mathbf{k}^\dagger \hat{\alpha}_\mathbf{k} \, ,
\end{equation}
with $\Omega_0$ the thermodynamic potential of the vacuum of quasi-particles:
\begin{equation}\label{Eq:Omega0_sc}
\Omega_0 = g \frac{N_0^2}{2V} - g \frac{\tilde{N}^2}{V} - \mu N_0 + \frac{1}{2} \sum_{\mathbf{k} \neq 0} \left[ \tilde{E}_\mathbf{k} - \varepsilon_\mathbf{k} - \Lambda + \frac{(gn_0)^2}{2\varepsilon_\mathbf{k}} \right] .
\end{equation}
The thermodynamic potential is obtained according to $\Omega = \frac{1}{\beta} \ln Z $, where $Z = \mathrm{Tr} (e^{-\beta \hat{K}})$ is the grand-partition function with inverse thermal energy $\beta = (k_B T)^{-1}$. The trace is taken over the quasi-particle states and one finds:
\begin{equation}\label{Eq:Omega_sc}
\Omega = \Omega_0  + \frac{1}{\beta} \sum_\mathbf{k} \ln \left( 1 - e^{-\beta \tilde{E}_\mathbf{k}} \right) \, .
\end{equation}

\subsection{Equation of state}\label{Sec:theory_SC}
Let us now calculate the chemical potential. In the BEC phase, this is achieved from the saddle point equation $\partial (\Omega/V) / \partial n_0 \vert_{\tilde{n},\mu,T} = 0$ which provides the following result:
\begin{align}\label{Eq:mu_gapped_sc}
\mu &= g n_0 \nonumber \\
&+ \frac{g}{2V} \sum_\mathbf{k} \left\lbrace \frac{2 ( \varepsilon_\mathbf{k} + \Lambda ) - g n_0}{\tilde{E}_\mathbf{k}} [2f(\tilde{E}_\mathbf{k}) + 1] - 2  + \frac{g n_0}{\varepsilon_\mathbf{k}} \right\rbrace ,
\end{align}
where $f(\tilde{E}_\mathbf{k}) = \langle \hat{\alpha}^\dagger_\mathbf{k} \hat{\alpha}_\mathbf{k} \rangle = (e^{\beta \tilde{E}_\mathbf{k}} -1)^{-1}$ is the Bose distribution function of quasi-particles. In principle, the above equation has to be solved self-consistently together with the equation for the non-condensate density:
\begin{equation}\label{Eq:nT_gapped_sc}
\tilde{n} = \frac{1}{2V}\sum_\mathbf{k} \left\lbrace \frac{\varepsilon_\mathbf{k} + \Lambda}{\tilde{E}_\mathbf{k}} [2f(\tilde{E}_\mathbf{k})+1] -1 \right\rbrace \,
\end{equation}
obtained from the extremal condition $\partial (\Omega/V) / \partial \tilde{n} \vert_{n_0,\mu,T} = 0$. However, such procedure is known to exhibit an unphysical gap in the quasi-particle energies~\cite{Griffin1996}. In this work, we follow the methodology of Ref.~\cite{Giorgini2000} and solve perturbatively the coupled equations. This allows one to avoid the problem of the gap and provides the correct leading order correction to the chemical potential. Indeed, Eq.~\eqref{Eq:mu_gapped_sc}, together with Eq.~\eqref{Eq:nT_gapped_sc}, can be expressed as $gn_0 = \Lambda + (\text{higher order terms})$. Thus, to the lowest order in the coupling constant, $gn_0 \simeq \Lambda$ and consequently the Bogoliubov spectrum becomes gapless:
\begin{equation}\label{Eq:E_sc}
E_\mathbf{k} = \sqrt{\varepsilon_\mathbf{k}^2 + 2 \Lambda \varepsilon_\mathbf{k}} \, .
\end{equation}
Inserting this expression in Eq.~\eqref{Eq:nT_gapped_sc}, one finds the leading correction for the non-condensed density:
\begin{equation} \label{Eq:nT_sc}
\tilde{n} = n_T^0 + \left( \frac{m \Lambda}{2 \pi \hbar^2} \right)^{3/2} G (\tau )
\end{equation}
where $n_T^0 = \zeta(3/2)/\lambda_T^3$ is the density of thermal atoms in an ideal Bose gas, with $\zeta(s)$ the Riemann zeta function, and $G(\tau)$ is a dimensionless function of the reduced temperature $\tau = k_B  T / \Lambda$ given by
\begin{equation}\label{Eq:G_sc}
G(\tau) =  \frac{2\sqrt{2}}{3\sqrt{\pi}} + \frac{2}{\sqrt{\pi}} \tau \int_0^\infty dx f(x) ( \sqrt{u-1} - \sqrt{\tau x} ) ,
\end{equation}
with $u = \sqrt{1+(\tau x)^2}$. The corresponding correction to $\mu$ is calculated from Eq.~\eqref{Eq:mu_gapped_sc} by replacing $\tilde{E}_\mathbf{k} \rightarrow E_\mathbf{k}$ and $gn_0 \rightarrow \Lambda$ in the terms in brackets:
\begin{equation}\label{Eq:mu_sc}
\mu = g n + gn_T^0 + g \left( \frac{m \Lambda}{2 \pi \hbar^2} \right)^{3/2} H (\tau) \, ,
\end{equation}
with the dimensionless function defined as:
\begin{equation}\label{Eq:Htau_sc}
H (\tau) =  \frac{8 \sqrt{2}}{3\sqrt{\pi}} + \frac{2}{\sqrt{\pi}} \tau \int_0^\infty d x f (x) \left[ \frac{(u-1)^{3/2}}{u} - \sqrt{\tau x} \right] ,
\end{equation}
where we have used Eq.~\eqref{Eq:nT_sc} to express $n_0=n-\tilde{n}$ as a function of $\Lambda$. Equation~\eqref{Eq:mu_sc} provides the proper leading order beyond mean-field correction to the chemical potential, as a function of the total density $n$ and temperature $T$. This result was first derived by Popov~\cite{Popov1983} in the high-temperature regime (see Eq.~\eqref{Eq:muHT_sc} below), and the same expression~\eqref{Eq:mu_sc} was found in Refs.~\cite{Griffin1996,Capogrosso-Sansone2010} within the finite-temperature extension of the Beliaev diagrammatic techniques, as well as in Ref.~\cite{Giorgini2000} starting from the time-dependent HFB equations. In our work, we therefore refer to this approach as Popov theory.
\par
Equation~\eqref{Eq:mu_sc} can be solved either perturbatively, the second-order expression being obtained by inserting the lowest order expression $\Lambda \simeq \Lambda^0 = g(n-n_T^0)$ in the last term, or self-consistently, from the definition $\Lambda = 2gn - \mu$. %
Although the latter procedure would allow for the calculation of higher order corrections, the validity of these new terms is questionable. Indeed, Eq.~\eqref{Eq:E_sc} assumes $\Lambda = gn_0$ to hold, which is true only at the lowest order in the interaction, while it is an approximation when higher order contributions are included. Solving self-consistently Eq.~\eqref{Eq:mu_sc} is therefore an ad-hoc procedure which \textit{assumes} a gapless spectrum~\eqref{Eq:E_sc}. It is nonetheless insightful to compare the two approaches, and in what follows we will investigate both the \textit{self-consistent} Popov theory where $\Lambda$ is obtained by solving self-consistently Eq.~\eqref{Eq:mu_sc}, and the \textit{second-order} Popov theory where $\Lambda^0 = g (n-n_T^0)$ is used. Actually, within the same accuracy one can also replace $\Lambda$ by $gn_0$ and solve Eq.~\eqref{Eq:nT_sc} self-consistently. The choice of the perturbation parameter is only a matter of convenience, since it gives the same second-order results and differences arise only for higher order terms (which are, \textit{a priori}, unreliable)~\cite{Capogrosso-Sansone2010}. In our work, we have chosen to solve self-consistently in $\Lambda$ since, by construction, it has the same beyond leading order corrections as the chemical potential. As we shall see below, this correspondence provides the correct low-temperature expansion of the chemical potential, as well as the correct lowest order expression for the free energy (see Appendix~\ref{App:F}). The beyond mean-field theory developed in this work is therefore valid as far as the following inequalities are satisfied:
\begin{equation}\label{Eq:inequality}
1 \gg \frac{\Lambda}{k_B T_\mathrm{BEC}} \gg (na^3)^{2/3} \, ,
\end{equation}
with $k_B T_\mathrm{BEC} = 2 \pi \hbar^2 / m \left[n/\zeta(3/2) \right]^{2/3}$ the BEC critical temperature for a non-interacting Bose gas. The first inequality in Eq.~\eqref{Eq:inequality} corresponds to the weakness of the interaction strength (diluteness condition), whereas the second inequality ensures that corrections to thermodynamics arising from critical fluctuations close to the phase transition are sufficiently small~\cite{book}. In other words, our approach fails in describing the region in the close vicinity of the BEC transition, $|T-T_\mathrm{BEC}|/T_\mathrm{BEC} \lesssim n^{1/3}a$. Here $\Lambda$ becomes very small as the chemical potential approaches the value $\mu_c=2gn$ at the critical point.
\par
As for the anomalous density, the expression $\tilde{m}=(1/V) \sum_{\mathbf{k} \neq 0} \langle \hat{a}_\mathbf{k}^\dagger \hat{a}_\mathbf{-k}^\dagger \rangle$ yields together with the gapless spectrum~\eqref{Eq:E_sc}:
\begin{equation} \label{Eq:m_sc}
\tilde{m} = - \frac{1}{V} \sum_\mathbf{k} \frac{\Lambda}{E_\mathbf{k}} \left[ f(E_\mathbf{k}) + \frac{1}{2} \right] \, .
\end{equation}
We notice that, as already mentioned previously, the second term in the right-hand side of Eq.~\eqref{Eq:m_sc} is ultraviolet divergent, and needs to be treated carefully, with a proper renormalization of the coupling constant. Finally, using the two densities Eqs.~\eqref{Eq:nT_sc} and~\eqref{Eq:m_sc}, the chemical potential can be rewritten as:
\begin{equation}\label{Eq:mu2_sc}
\mu = g n_0 + 2 g \tilde{n} + g \tilde{m} \, .
\end{equation}
One can verify that the above expression coincides with Eq.~\eqref{Eq:mu_sc} upon applying the renormalization of the coupling constant $g n_0 + g \tilde{m} \rightarrow g n_0 [1 + gV^{-1} \sum_\mathbf{k} 1/(2\varepsilon_\mathbf{k})] + g\tilde{m}$. 
\par
We now discuss the behavior of $\mu$ in different temperature regimes. First, at zero temperature $H(\tau) = 8 \sqrt{2}/(3 \sqrt{\pi})$, and one obtains
\begin{equation}\label{Eq:muT0_sc}
\mu (T=0) = g n \left( 1 + \frac{32}{3\sqrt{\pi}} \sqrt{na^3} \right) \, ,
\end{equation}
corresponding to the chemical potential calculated by Lee, Huang and Yang~\cite{Lee1957} and accounting for the effects of quantum fluctuations through the second term in the parenthesis. 
\par
At low temperature, $\tau \ll 1$, one can expand the dimensionless function $H(\tau)$ in Eq.~\eqref{Eq:Htau_sc} according to:
\begin{equation}\label{Eq:HtauLT_sc}
H (\tau) \simeq \sqrt{\frac{2}{\pi}} \left[ \frac{8}{3} - \sqrt{\frac{\pi}{2}} \zeta (3/2) \tau ^{3/2} + \frac{\pi^4}{30} \tau^4 \right] \, .
\end{equation}
By inserting this expression in Eq.~\eqref{Eq:mu_sc}, one obtains the low-temperature behavior of the chemical potential:
\begin{equation}\label{Eq:muLT_sc}
\mu \simeq gn + g \left( \frac{m \Lambda}{2 \pi \hbar^2} \right)^{3/2} \sqrt{\frac{2}{\pi}} \left( \frac{8}{3} + \frac{\pi^4}{30} \tau^4 \right) \, ,
\end{equation}
where $\Lambda$ is evaluated at $T=0$ and the $(k_BT)^4$ contribution arises from phonon excitations which are dominant at low temperatures.
Similarly, in the same temperature regime Eq.~\eqref{Eq:nT_sc} provides the result
\begin{equation}\label{Eq:n0LT_sc}
n_0 = n - \left( \frac{m \Lambda}{2 \pi \hbar^2} \right)^{3/2} \frac{2}{\sqrt{\pi}} \left( \frac{\sqrt{2}}{3} + \frac{\pi^2}{6\sqrt{2}} \tau^2 \right) \, ,
\end{equation}
for the condensate density \footnote{Alternatively, one can also replace $\Lambda$ by $gn_0$ in Eqs.~\eqref{Eq:muLT_sc}-\eqref{Eq:n0LT_sc} and solving self-consistently Eq.~\eqref{Eq:n0LT_sc} in $gn_0$. However, this procedure would restrict the validity of Eq.~\eqref{Eq:muLT_sc} to a narrower temperature region with a lower bound given by $k_B T \gg (na^3)^{1/4}$.}.
\par
At high temperature instead, one can neglect in Eqs.~\eqref{Eq:nT_sc} and~\eqref{Eq:mu_sc} the contribution from quantum fluctuations, independent of the reduced temperature $\tau$, and expand the Bose distribution function as $f(E) \simeq (\beta E)^{-1}$. This gives the following results~\cite{Popov1983}:
\begin{gather}
\mu \simeq g (n + n_T^0) - g \frac{2\sqrt{2\pi}}{\lambda_T^3} \sqrt{\beta \Lambda} \, , \label{Eq:muHT_sc} \\
n_0 \simeq n - n_T^0 + \frac{\sqrt{2\pi}}{\lambda_T^3}\sqrt{\beta\Lambda} \, , \label{Eq:n0HT_sc}
\end{gather}
as a function of the parameter $\Lambda$. By choosing the leading order result $\Lambda^0=g(n-n_T^0)$, Eq.~\eqref{Eq:muHT_sc} reduces to the expression $\mu = g(n+n^0_T) -g^{3/2} 2\sqrt{2\pi} \sqrt{\beta(n-n^0_T)}/\lambda_T^3$,   showing that at high temperature the leading correction to the mean-field value $\mu^0= g(n +n^0_T)$  scales like $g^{3/2}$, differently from the $g^{5/2}$ correction accounting for the effects of quantum fluctuations at zero temperature (see Eq.~\eqref{Eq:muT0_sc}). It is insightful to compare the above result with the prediction of HF theory. The HF theory is obtained from the model Hamiltonian Eq.~\eqref{Eq:K_sc} by neglecting the last terms in which annihilation and creation operators appear in pairs~\cite{book}. Then, proceeding in the same way as previously, one finds for the thermal density $\tilde{n}^\mathrm{HF} = g_{3/2} (e^{-\beta \Lambda^\mathrm{HF}}) / \lambda_T^3$, where $g_p(z)$ is the Bose special function, and for the chemical potential $\mu^\mathrm{HF} = g (n + \tilde{n}_\mathrm{HF})$. As for the Popov theory, the HF equations can be solved either self-consistently by using $\Lambda^\mathrm{HF} = 2gn- \mu^\mathrm{HF}$, or up to second-order with $\Lambda^0 = g(n-n_T^0)$. In the high-temperature limit one finds:
\begin{equation}
\mu^\mathrm{HF} \simeq gn + gn_T^0 - g \frac{2\sqrt{\pi}}{\lambda_T^3} \sqrt{\beta \Lambda^\mathrm{HF}} \, .
\end{equation}
Therefore the HF approach provides a qualitatively similar result to Eq.~\eqref{Eq:muHT_sc}, though it underestimates the effects of thermal fluctuations by a factor $\sqrt{2}$. This can be understood by the fact that at high temperatures, large momentum modes $\hbar k \sim \sqrt{2mk_B T}$ contribute the most to the excitations and one can therefore approximate the excitation spectrum Eq.~\eqref{Eq:E_sc} by the single-particle expression $E_\mathbf{k} \simeq \varepsilon_\mathbf{k} + \Lambda$. Let us however notice that while in the HF approach the superfluid density is found to coincide with the condensate fraction $n_s = n_0$, such identity does not hold anymore within the Popov theory (see Appendix~\ref{App:ns}). 
\par
Finally, in the absence of Bose-Einstein condensation ($n_0=0$), the Popov approach reduces to the HF theory, in which $\mu = \mu^\mathrm{IBG} + 2 gn$, with $\mu^\mathrm{IBG}$ the ideal Bose gas chemical potential. Consequently, within the second-order Popov theory the BEC phase transition is predicted to occur at the ideal gas phase transition temperature, $T_\mathrm{BEC}$.


\subsection{Results}\label{Sec:Results1}

\begin{figure}[t!]
\begin{center}
\includegraphics[width=0.8\columnwidth]{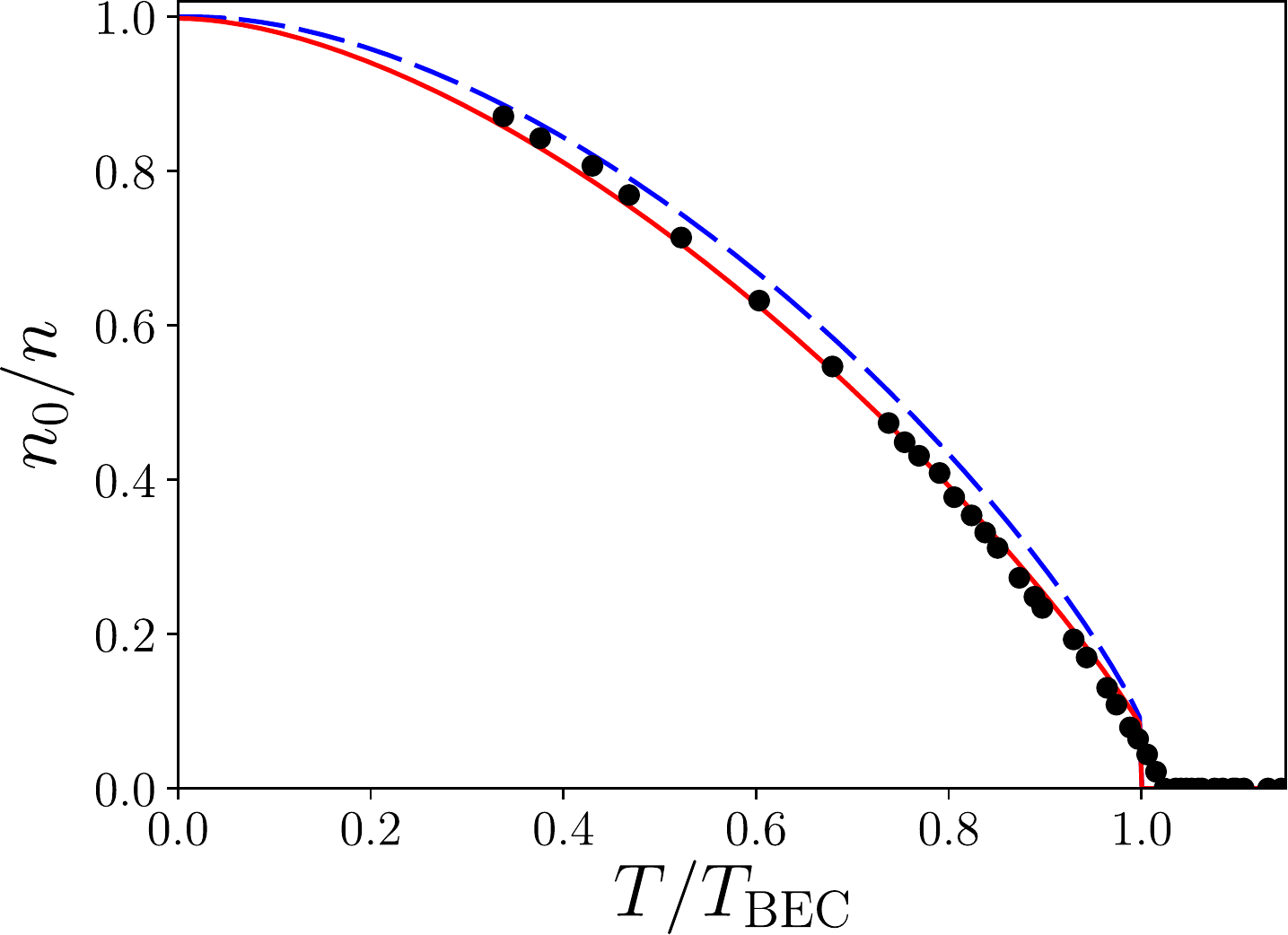}
\includegraphics[width=0.8\columnwidth]{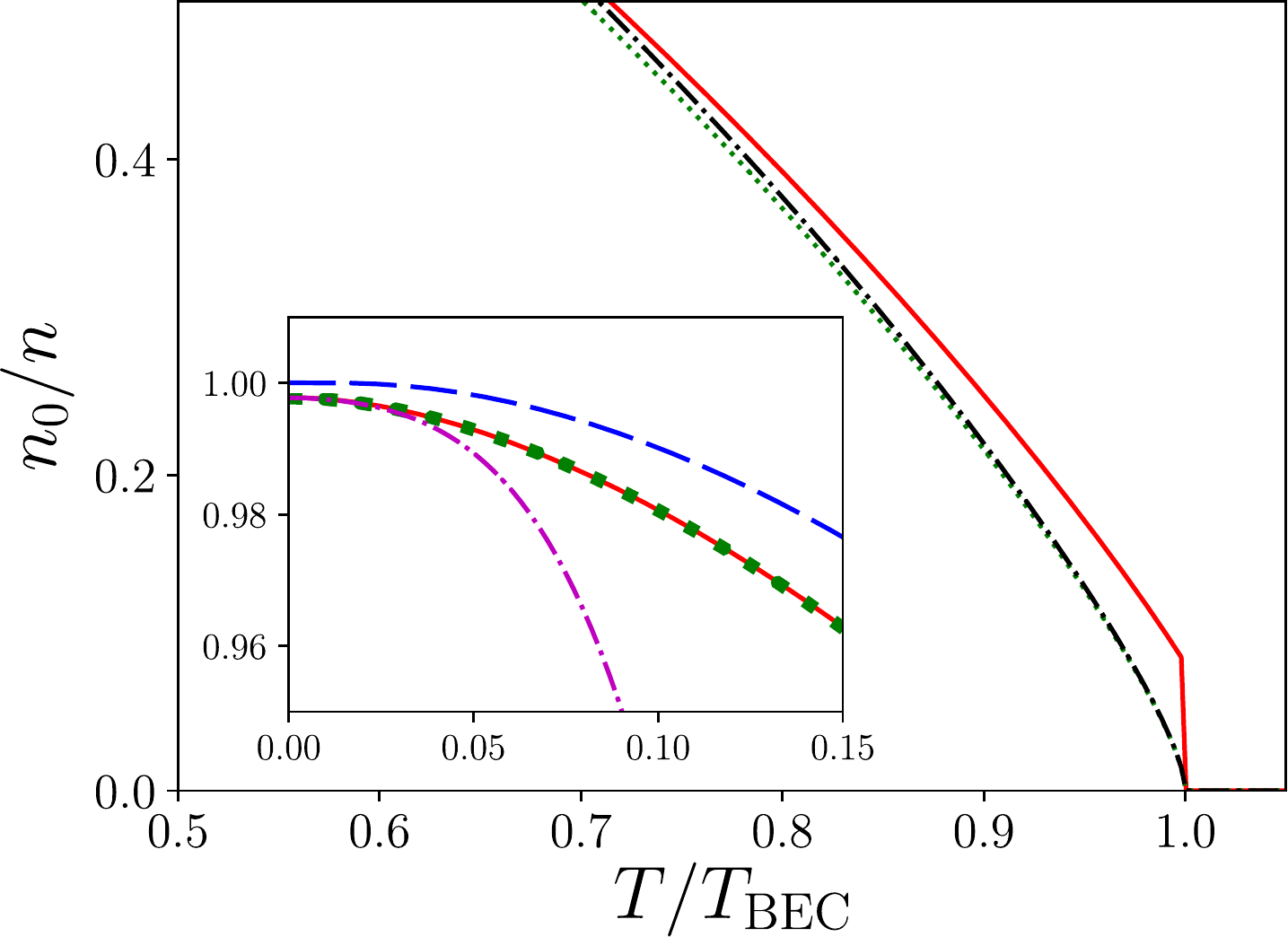}
\caption{Condensate density $n_0 = n - \tilde{n}$ as a function of temperature, calculated for $gn/(k_B T_\mathrm{BEC}) = 0.05$. Upper panel: comparison of different theories. The red solid line is the Popov theory prediction in which Eqs.~\eqref{Eq:nT_sc} and~\eqref{Eq:mu_sc} have been solved self-consistently. The blue dashed line shows the result of the HF theory and the black dots are the predictions from the universal relations of Ref.~\cite{Prokofev2004}. Lower panel: comparison of Popov theory in different limits. Red solid and blue dashed lines: same as upper panel. Green dotted line: Popov theory calculated up to second-order (by using $\Lambda^0 = g (n-n_T^0)$ in Eq.~\eqref{Eq:nT_sc}). The black dot-dashed line in the main figure is the high-temperature expression~\eqref{Eq:n0HT_sc}. The inset shows Popov and HF theory as in the upper and lower panel, whereas the purple dot-dashed line corresponds to the low-temperature expansion~\eqref{Eq:n0LT_sc}.} 
\label{Fig:n0_sc}
\end{center}
\end{figure}

\begin{figure}[t!]
\begin{center}
\includegraphics[width=0.8\columnwidth]{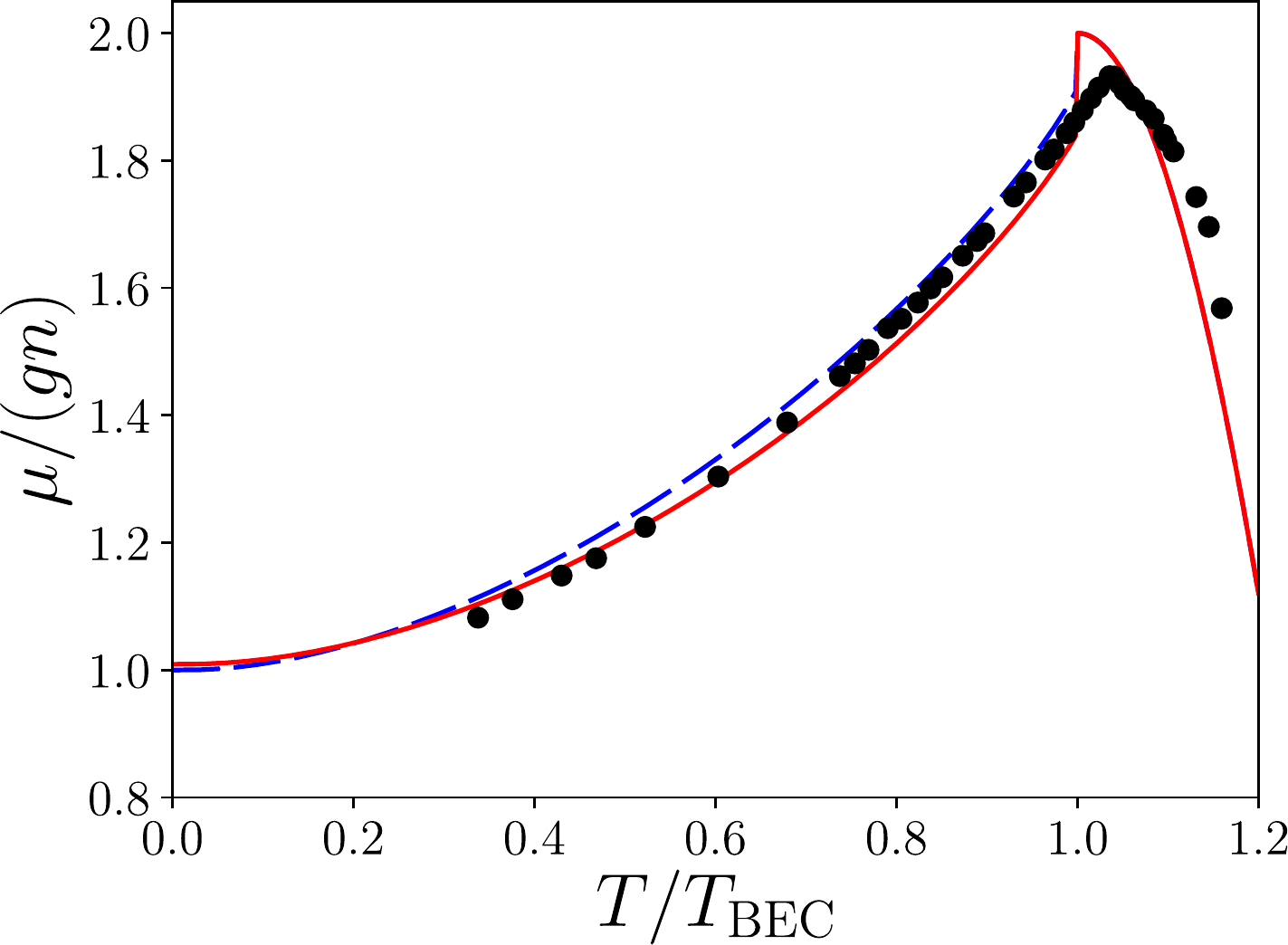}
\includegraphics[width=0.8\columnwidth]{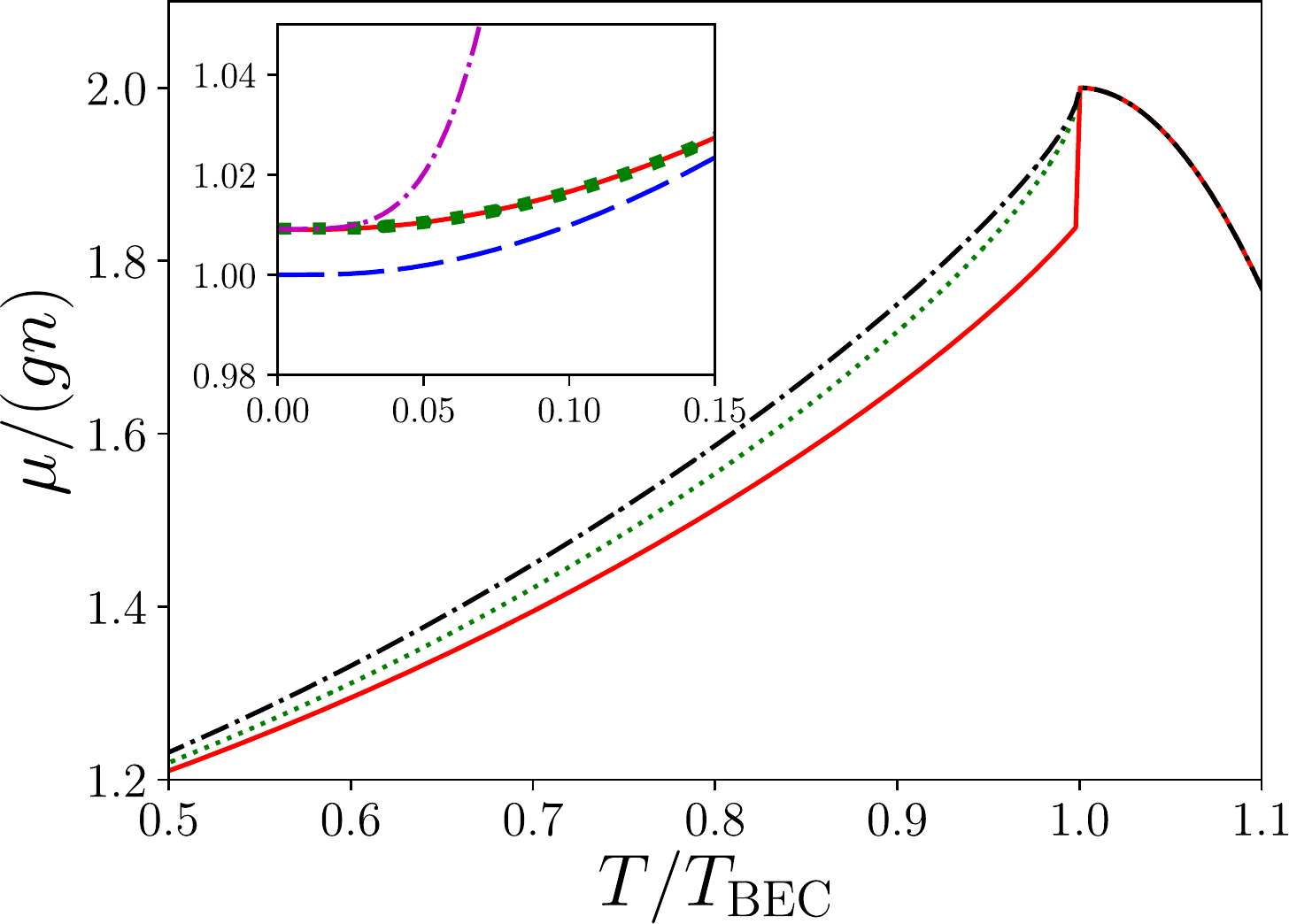}
\caption{Chemical potential as a function of temperature for $gn/(k_B T_\mathrm{BEC}) =0.05$. Line guides are the same as in Fig.~\ref{Fig:n0_sc}.} 
\label{Fig:mu_sc}
\end{center}
\end{figure}

We now discuss the numerical results obtained for some key thermodynamic quantities. Figure~\ref{Fig:n0_sc} shows the condensate density evaluated from Eq.~\eqref{Eq:nT_sc} for the interaction parameter $gn/(k_B T_\mathrm{BEC}) = 0.05$. This choice corresponds to the typical value of the gas parameter $na^3 \sim 10^{-6}$. In the upper panel, we compare the results from the self-consistent Popov and Hartree-Fock theories, together with the predictions from the universal relations. This last approach describes the region in the vicinity of the phase transition, where perturbative theory fails due to strong fluctuations, but a universal description of the weakly interacting Bose gas exists~\cite{Prokofev2001}. In this region, the equation of state depends on a single variable $X = \beta^2 (\mu - \mu_c)/(\hbar^6 m^3 g^2)$, according to $ n-n_c=f(X)$, with $\mu_c$ and $n_c$ the chemical potential and density at the critical point, respectively. Explicit results for the universal function $f$ in 3D were calculated from classical Monte-Carlo simulations in Ref.~\cite{Prokofev2004}. Our calculations show that Popov theory agrees well with the predictions of the universal theory. We briefly note that the unphysical jump of the condensate density in both the self-consistent Popov and HF theories arises from the inclusion of higher order terms~\cite{Shi1998}, and is absent in the second-order Popov approach (see lower panel of Fig.~\ref{Fig:n0_sc}). We also notice that the universal relations of Ref.~\cite{Prokofev2004} are consistent with a small upward shift of the critical temperature arising from many-body effects~\cite{Arnold2001} which is not captured by our perturbative treatment. In the lower panel of Fig.~\ref{Fig:n0_sc} we compare the results of Popov theory in the vicinity of the phase transition. In particular, we see that the second-order Popov result (green dotted line), in which we have used the lowest order expression for the effective chemical potential $\Lambda^0 = g(n-n_T^0)$, agrees with the self-consistent calculation up to the close vicinity of $T_\mathrm{BEC}$.  The inset of Fig.~\ref{Fig:n0_sc} also shows that Popov theory predicts correctly the depletion of the condensate at zero temperature. In Figure~\ref{Fig:mu_sc} we make a similar comparison for the chemical potential. In the lowest order  mean-field description,  where $\mu^0 = g(n + n_T^0) $ the chemical potential is predicted to evolve monotonically from $gn$ at zero temperature to $2gn$ at the critical temperature. The self-consistent Popov theory confirms this picture, although predicting a shift of $\mu$ at $T=0$ due to quantum fluctuations and an unphysical jump at $T_\mathrm{BEC}$.
\par
\par
\begin{figure}[t!]
\begin{center}
\includegraphics[width=0.9\columnwidth]{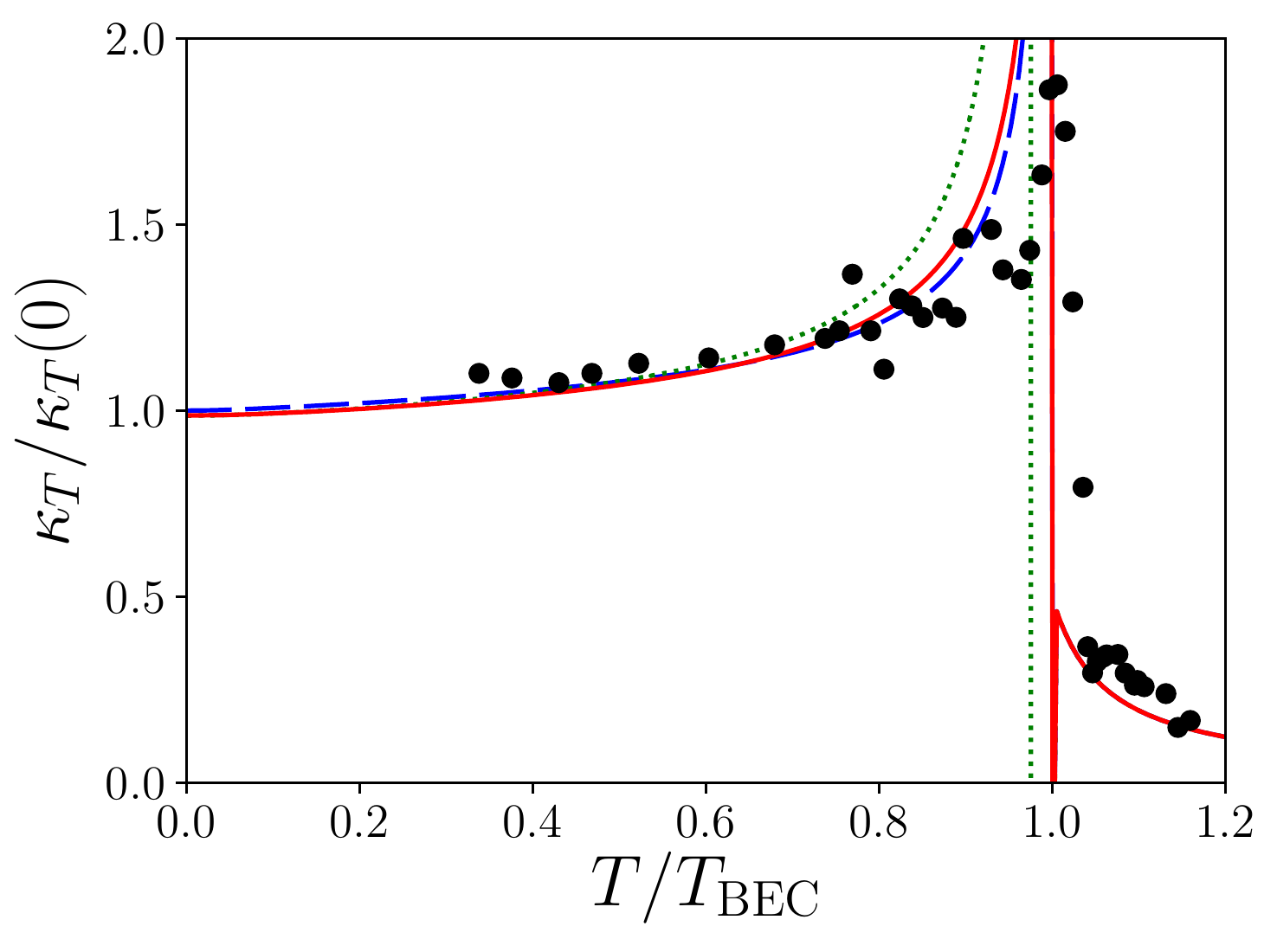}
\caption{Isothermal compressibility as a function of temperature for $gn/(k_BT_\mathrm{BEC})=0.05$. Red solid line: self-consistent Popov theory. Blue dashed line: HF theory. Green dotted line: Popov theory calculated up to second-order. Black dots: prediction from the universal relations of Ref.~\cite{Prokofev2004}.}
\label{Fig:kT_sc}
\end{center}
\end{figure}
For an ideal Bose gas, the isothermal compressibility $\kappa_T = \partial P / \partial n \vert_T$ is predicted to diverge in the BEC phase, and therefore the quantity is expected to be sensitive to the way interaction is treated in the theory. This is shown in Fig.~\ref{Fig:kT_sc}, where one finds that all approaches predict a finite compressibility in the BEC phase, increasing with the temperature. In particular, one finds that the second-order Popov theory (green dotted line) shows a worse agreement with the prediction from the universal relations, compared to self-consistent approaches. This is understood from the fact that in the vicinity of $T_\mathrm{BEC}$, thermal fluctuations become important and beyond second-order terms have non-negligible contributions to the thermodynamic quantities. Although the correctness of the self-consistent Popov theory is questionable in this regime, its solution automatically captures higher order terms. One should point out, however, that our numerical extraction of the isothermal compressibility from the universal relations of Ref.~\cite{Prokofev2004} results in a set of quite scattered values which make the comparison with other theories rather difficult. More precise results for $\kappa_T$ from exact approaches would be useful in order to carry out quantitative comparisons.




\section{Two-component Bose mixtures: Equation of state}\label{Sec:theory_mixt}

\subsection{Diagonalization}

We consider now a uniform mixture of two-component Bose gases, and extend the Popov theory using the same methodology as for the single-component gas. We consider a regime of temperatures and densities where both components are in the condensed phase and we aim to determine thermodynamic quantities of the mixture including beyond mean-field corrections in both the intra-species and the inter-species coupling strength. The Hamiltonian including all point-like interactions takes the form,
\begin{align}
\hat{H} = & \sum_{i = 1, 2} \left( \sum_\mathbf{k} \varepsilon_{i, \mathbf{k}} \hat{a}_{i, \mathbf{k}}^\dagger \hat{a}_{i, \mathbf{k}} + \frac{g_{ii}}{2V} \sum_{\mathbf{k}, \mathbf{k}', \mathbf{q}} \hat{a}_{i, \mathbf{k}}^\dagger \hat{a}_{i, \mathbf{k'+q}}^\dagger \hat{a}_{i, \mathbf{k'}} \hat{a}_{i, \mathbf{k+q}} \right) \nonumber \\ 
&+ \frac{g_{12}}{V} \sum_{\mathbf{k}, \mathbf{k}', \mathbf{q}} \hat{a}_{1, \mathbf{k}}^\dagger \hat{a}_{1, \mathbf{k+q}} \hat{a}_{2, \mathbf{k'+q}}^\dagger \hat{a}_{2, \mathbf{k'}} \, ,
\end{align}
where the subscript $i = \{1, 2\}$ refers to the $i^\mathrm{th}$ component of the mixture. We have further introduced the coupling constants for the intra-species interactions $g_{ii} = 4\pi \hbar^2 a_{ii} / m_i$ in terms of the scattering length $a_{ii}$ and mass $m_i$, as well as for the inter-species interaction $g_{12} = 4 \pi \hbar^2 a_{12}/m_R$, with reduced mass $m_R = 2m_1 m_2/(m_1+m_2)$. By applying the Bogoliubov prescription and replacing $\hat{a}_{i, 0}$ and $\hat{a}_{i, 0}^\dagger$ with the number of particles in the condensate $\sqrt{N_{i,0}}$, one obtains for the grand-canonical Hamiltonian $\hat{K} = \hat{H}-\sum_i \mu_i \hat{N}_i$:
\begin{widetext}
\begin{align}\label{Eq:K_mix}
\hat{K} =& \sum_{\substack{i,j=1,2 \\ i \neq j}} \left[ \frac{g_{ii}}{2V} ( N_{i, 0} ^2 - 2 \tilde{N}_i^2 ) - \mu_i N_{i, 0} + \sum_{\mathbf{k} \neq 0} ( \varepsilon_{i, \mathbf{k}} + 2 g_{ii} n_i + g_{12} n_j - \mu_i ) \hat{a}_{i, \mathbf{k}}^\dagger \hat{a}_{i, \mathbf{k}} + \frac{g_{ii}}{2V} N_{i, 0} \sum_{\mathbf{k} \neq 0} ( \hat{a}_{i, \mathbf{k}}^\dagger \hat{a}_{i, \mathbf{-k}}^\dagger + \hat{a}_{i, \mathbf{k}} \hat{a}_{i, \mathbf{-k}} ) \right] \nonumber \\
&+ \frac{g_{12}}{V} N_{1, 0} N_{2, 0} - \frac{g_{12}}{V} \tilde{N}_1 \tilde{N}_2 + \frac{g_{12}}{V} \sqrt{N_{1, 0} N_{2, 0}} \sum_{\mathbf{k} \neq 0} ( \hat{a}_{1, \mathbf{k}}^\dagger + \hat{a}_{1, \mathbf{-k}} ) ( \hat{a}_{2, \mathbf{-k}}^\dagger + \hat{a}_{2, \mathbf{k}} ) \, ,
\end{align}
\end{widetext}
with $\varepsilon_{i, \mathbf{k}} = \hbar^2 k^2 / (2m_i)$. In the above equation, we have again kept quadratic terms in $\hat{a}_{\mathbf{k} \neq 0}$, $\hat{a}^\dagger_{\mathbf{k} \neq 0}$ up to second-order in the coupling constants, and we neglected quadratic terms in the fluctuations of the non-condensate densities around their mean value, as well as terms proportional to the anomalous densities. The terms in the bracket of Eq.~\eqref{Eq:K_mix} correspond to the single-species Hamiltonian~\eqref{Eq:K_sc} for each component, whereas the last terms contain the interspecies interaction terms. The grand-canonical Hamiltonian Eq.~\eqref{Eq:K_mix} can be diagonalized by means of canonical transformations to uncouple the two components, followed by Bogoliubov transformations, as well as proper renormalization of the coupling constants. The details of the calculation can be found in Appendix~\ref{App:mixt}, and here we show the final result:
\begin{equation}\label{Eq:K_Po_mix}
\hat{K} = \Omega_0 + \sum_{\mathrm{k} \neq 0} \left( E_{+ , \mathbf{k}} \hat{\alpha}_\mathbf{k}^\dagger \hat{\alpha}_\mathbf{k} + E_{- , \mathbf{k}} \hat{\beta}_\mathbf{k}^\dagger \hat{\beta}_\mathbf{k} \right) \, ,
\end{equation}
where $\hat{\alpha}_\mathbf{k}^\dagger$ (resp. $\hat{\beta}_\mathbf{k}^\dagger$) is the creation operator for the quasiparticles in the density (resp. spin) channel, obeying Bose statistics. The expression for the vacuum energy of Bogoliubov quasi-particles $\Omega_0$ is given by Eq.~\eqref{Eq:Omega0_mixt}, and the gapless excitation spectrum of the system reads
\begin{equation}\label{Eq:E_mix}
E_{\pm, \mathbf{k}} = \sqrt{\left( \frac{\nu_1^2 + \nu_2^2}{2} \right) \varepsilon_\mathbf{k}^2 + 2 \varepsilon_\mathbf{k} \Lambda_{\pm, \mathbf{k}}} \, ,
\end{equation}
where we have introduced the kinetic energy in terms of the reduced mass $\varepsilon_\mathbf{k} = \hbar^2 k^2 / (2m_R)$ and the inverse mass ratios $\nu_i = m_R / m_i$. The effective chemical potential $\Lambda_{\pm,\mathbf{k}}$ is therefore associated to the Bogoliubov density and spin sounds, and takes the following expression:
\begin{gather}
\Lambda_{\pm, \mathbf{k}} = \frac{1}{2} \left(  \nu_1 \Lambda_1 + \nu_2 \Lambda_2 \pm \Gamma_\mathbf{k} \right) \, , \label{Eq:Lambda} \\
\Gamma_\mathbf{k} = \sqrt{\left[ \frac{( \nu_1^2 - \nu_2^2 )}{2} \varepsilon_\mathbf{k} + (\nu_1 \Lambda_1 - \nu_2 \Lambda_2) \right]^2 + 4 \bar{g}^2 \nu_1 \Lambda_1 \nu_2 \Lambda_2 } \label{Eq:Gamma} 
\end{gather}
with $\Lambda_1 = 2 g_{11} n_1 + g_{12} n_2 - \mu_1$ and $\Lambda_2$ is obtained by inverting the indexes $(1 \leftrightarrow 2)$. In the above equation, we have also introduced the reduced coupling constant $\bar{g} = g_{12}/\sqrt{g_{11}g_{22}}$.

\subsection{Equation of state}

The chemical potential in each component can be calculated in a similar fashion to the single-component case, by evaluating the saddle point equation $\partial \Omega / \partial n_{i,0} = 0$ and solving it perturbatively \footnote{Similarly to the single-component case, the saddle point equation has to be evaluated from the unperturbed grand-canonical potential with the gapped spectrum, and \textit{not} from Eqs.~\eqref{Eq:K_mix} and~\eqref{Eq:E_mix} where we have assumed the lowest order identity $\Lambda_i = g_{ii}n_i$ to hold. The details of the calculation can be found in Appendix~\ref{App:mixt}.}. We give in Appendix~\ref{App:mixt} the derivation of the equation of state in the most general case, and here we only show the results for the equal masses configuration $m_1 = m_2 = M$. Then, the function $\Gamma_\mathbf{k}$ in Eq.~\eqref{Eq:Gamma} becomes independent of the wave-vector~\cite{Pethick}:
\begin{equation}\label{Eq:Lambda_sym}
\Lambda_\pm = \frac{1}{2} (\Lambda_1 + \Lambda_2 \pm \sqrt{(\Lambda_1 - \Lambda_2)^2 + 4 \bar{g}^2 \Lambda_1 \Lambda_2} ) \, ,
\end{equation}
and one can write the condensate depletion in a form similar to the single-component case:
\begin{equation}\label{Eq:nT_mix_sym}
\tilde{n}_1 = n_T^0 + \sum_\pm \left(\frac{m \Lambda_\pm}{2 \pi \hbar^2}\right)^{3/2} G_\pm (\tau_\pm , l) \, ,
\end{equation}
where the dimensionless function depends now on the reduced temperature $\tau_\pm = k_B T / \Lambda_\pm$, and we have introduced the ratio $l = \Lambda_2/\Lambda_1$ of effective chemical potentials,
\begin{align}
& G_\pm(\tau_\pm , l) = \frac{1}{2} \left(1 \pm \frac{1-l}{\sqrt{(1-l)^2+4\bar{g}^2l}}\right) \nonumber \\
&\times \left[ \frac{2\sqrt{2}}{3\sqrt{\pi}} + \frac{2}{\sqrt{\pi}} \tau_\pm \int_0^\infty dx f(x) \left( \sqrt{u_\pm - 1} -  \sqrt{\tau_\pm x} \right) \right] \, ,
\end{align}
with $u_\pm = \sqrt{1 + \tau_\pm^2 x^2}$. The condensate depletion $\tilde{n}_2$ in the second component is instead obtained from Eq.~\eqref{Eq:nT_mix_sym}, replacing $l$ by $1/l = \Lambda_1 / \Lambda_2$. For the chemical potential one finds $(1 \leftrightarrow 2)$,
\begin{equation}\label{Eq:mu_mix_sym}
\mu_1 = g_{11} (n_1 + n_T^0) + g_{12} n_2 + g_{11} \sum_\pm \left(\frac{m \Lambda_\pm}{2 \pi \hbar^2}\right)^{3/2} H_\pm (\tau_\pm, l) \,
\end{equation}
where by $(1 \leftrightarrow 2)$ we also mean $(l \leftrightarrow 1/l)$, and the dimensionless function is given by:
\begin{align}\label{Eq:H_mix}
&H_\pm(\tau_\pm, l) = \frac{1}{2} \left[1 \pm \frac{1 + (2 \bar{g}^2 -1) l}{\sqrt{(1 - l)^2 + 4 \bar{g}^2 l}}\right]\nonumber \\
&\times \left[ \frac{8\sqrt{2}}{3\sqrt{\pi}} + \frac{2}{\sqrt{\pi}}  \tau_\pm \int_0^\infty dx f(x) \left( \frac{(u_\pm -1)^{3/2}}{u_\pm} - \sqrt{\tau_\pm x} \right) \right] .
\end{align}
As in the single-component case, the above equations can be solved either self-consistently or perturbatively, the second-order expression being obtained by inserting the leading order result $\Lambda_i^0 = g_{ii} (n_i - n_T^0)$ for the effective chemical potential. We can verify that from Eqs.~\eqref{Eq:nT_mix_sym} and~\eqref{Eq:mu_mix_sym} one retrieves the single-component result Eqs.~\eqref{Eq:nT_sc} and~\eqref{Eq:mu_sc} respectively, when putting $\Lambda_2 = 0$.
\par
In analogy to the single component gas, one can define anomalous densities involving two creation or annihilations operators. In particular, the binary system possesses two additional anomalous pair densities,
\begin{equation}
\tilde{n}_{12} = \frac{1}{V} \sum_\mathbf{k} \left\langle \hat{a}^\dagger_{1, \mathbf{k}} \hat{a}_{2, \mathbf{k}} \right\rangle \, , \quad \tilde{m}_{12} = \frac{1}{V} \sum_\mathbf{k} \left\langle \hat{a}_{1, \mathbf{k}} \hat{a}_{2, \mathbf{-k}} \right\rangle \, ,
\end{equation}
describing processes where, due to the presence of the condensate reservoir, particles are exchanged or pairing correlations emerge between the two components. Using the newly introduced densities, the chemical potential in the equal masses case can be conveniently written in the form
\begin{equation}\label{Eq:mu_mixt_densities}
\mu_1 = g_{11} (n_1 + \tilde{n}_1 + \tilde{m}_1) + g_{12} n_2 + g_{12} \sqrt{\frac{g_{11} \Lambda_2}{g_{22} \Lambda_1}} \left( \tilde{n}_{12} + \tilde{m}_{12} \right) \, .
\end{equation}
For future purpose, it is insightful to compare the above expression with the HF prediction. Similarly to the single-component case, the chemical potential within HF theory is obtained by neglecting the terms in Eq.~\eqref{Eq:K_mix} in which the annihilation and creation operators appear in pairs. One readily finds ($1 \leftrightarrow 2$):
\begin{equation}\label{Eq:mu_mixt_HF}
\mu_1^\mathrm{HF} = g_{11} (n_1 + \tilde{n}_1^\mathrm{HF}) + g_{12} n_2 \, ,
\end{equation}
where $\tilde{n}_1^\mathrm{HF} = g_{3/2} \left( e^{-\beta \Lambda_1^\mathrm{HF}} \right) / \lambda_{1,T}^3$ is the HF density of thermal atoms, with $\Lambda_1^\mathrm{HF}=2g_{11}n_1+g_{12}n_2-\mu_1^\mathrm{HF} $. Equation~\eqref{Eq:mu_mixt_HF} clearly shows that in HF theory, beyond mean-field effects appear only in the intra-species interaction terms, the inter-species coupling being considered to the lowest linear order.
\par
At zero temperature, one finds the following expression for the quantum depletion
\begin{align}
n_{1,0} =& n_1 \Bigg\lbrace 1 - \frac{4}{3\sqrt{\pi}} \sqrt{n_1 a_{11}^3} \sum_\pm \left( 1 \pm \frac{1-l}{\sqrt{(1-l)^2+4\bar{g}^2l}} \right) \nonumber \\
& \times  \left[\frac{1}{2} \left(1+l \pm \sqrt{(1-l)^2+4\bar{g}^2l} \right) \right]^{3/2} \Bigg\rbrace \, .
\end{align}
As for the chemical potential, one finds instead
\begin{align}\label{Eq:mu_mixt_T0}
&\mu_1 (T=0) = g_{11} n_1 + g_{12} n_2  \nonumber \\
&+  \frac{16}{3\sqrt{\pi}} g_{11}n_1 \sqrt{n_1 a_{11}^3} \sum_\pm \left( 1 \pm \frac{1+(2\bar{g}^2-1)l}{\sqrt{(1-l)^2+4\bar{g}^2l}} \right) \nonumber \\
& \times  \left[\frac{1}{2} \left(1+l \pm \sqrt{(1-l)^2+4\bar{g}^2l} \right) \right]^{3/2}  \, ,
\end{align}
which corresponds to the chemical potential evaluated from the LHY energy functional in Ref.~\cite{Petrov2015}.
\par
At temperature $k_B T \gg \mu_i(T=0)$, the Bose distribution function can be expanded in the same way as in the single-component case, yielding the following expression for the chemical potential:
\begin{align}\label{Eq:mu_mixt_HT}
\mu_1 &\simeq g_{11} (n_1 + n_T^0) + g_{12} n_2 - g_{11} \frac{2\sqrt{2\pi}}{\lambda_T^3} \nonumber \\
&\times \sum_\pm \frac{1}{2} \left(1 \pm \frac{1 + (2 \bar{g}^2 -1) l}{\sqrt{(1 - l)^2 + 4 \bar{g}^2 l}}\right) \sqrt{\beta \Lambda_\pm} \, .
\end{align}
It is worth noticing that the HF theory in the same temperature regime predicts the chemical potential to behave like
\begin{equation}\label{Eq:mu_mixt_HT_HF}
\mu_1^\mathrm{HF} \simeq g_{11} (n_1 + n_T^0) + g_{12} n_2 - g_{11} \frac{2\sqrt{\pi}}{\lambda_T^3} \sqrt{\beta \Lambda_1^\mathrm{HF}} \, ,
\end{equation} 
as one can easily verify using Eq.~(\ref{Eq:mu_mixt_HF}). In the next section, we study how the difference in the last terms of Eqs.~(\ref{Eq:mu_mixt_HT}), (\ref{Eq:mu_mixt_HT_HF}) affects the calculation of the thermodynamic quantities. 

\subsection{Results}

\begin{figure}[t!]
\begin{center}
\includegraphics[width=0.8\columnwidth]{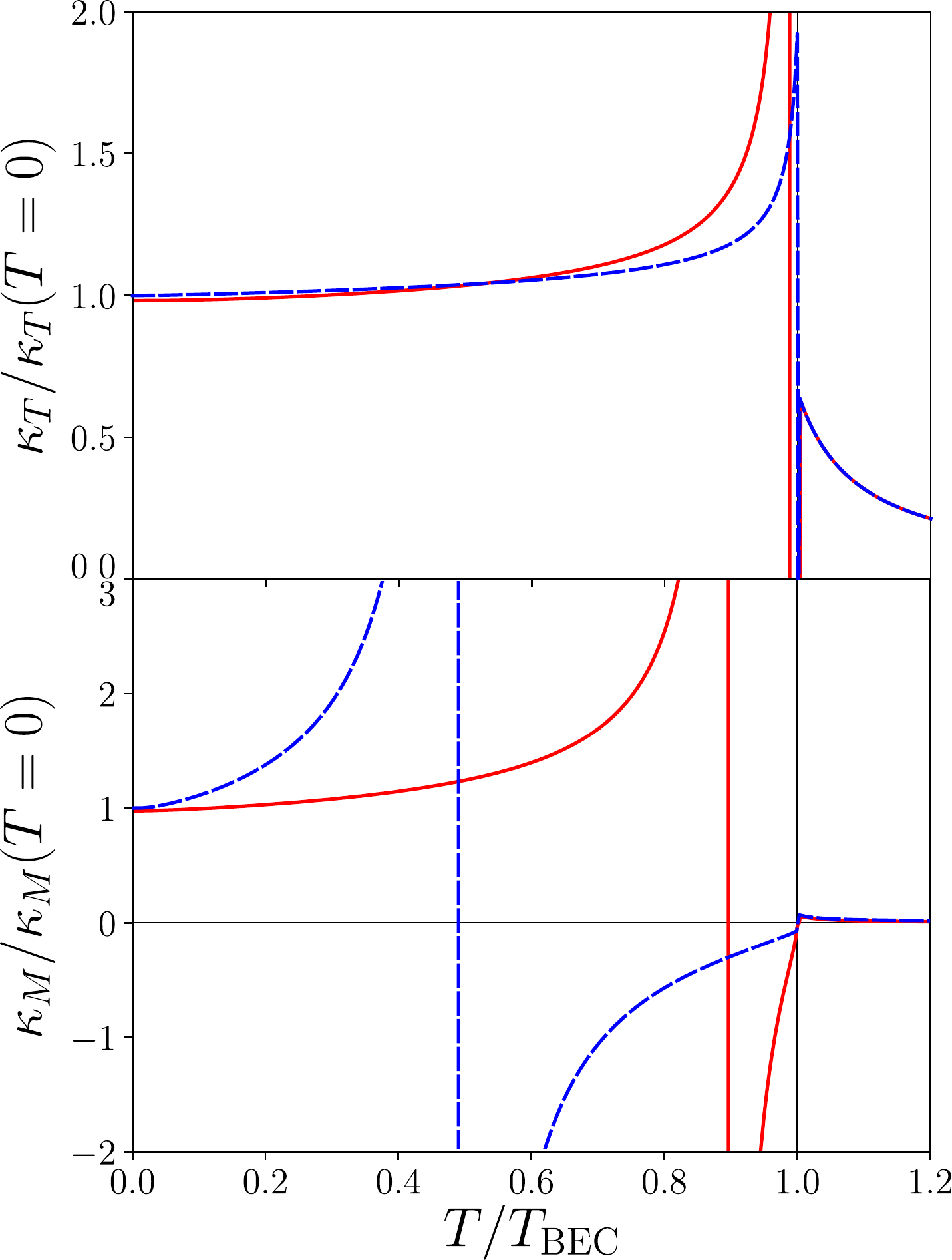}
\caption{Isothermal compressibility (a) and spin susceptibility (b) in Eq.~\eqref{Eq:kappa} for binary mixtures of Bose gases, with interaction parameters $gn/(k_B T_\mathrm{BEC})=0.1$ and $\delta g/g=0.07$. The blue dashed and the red solid lines are the predictions of HF theory and second-order Popov theory, respectively. Both quantities are normalized to the mean-field $T=0$ values, $\kappa_{T, M}(T=0) = 2/(g \pm g_{12})$.} 
\label{Fig:k_mixt}
\end{center}
\end{figure}

We now discuss the numerical results for the mixture of two weakly interacting Bose gases, obtained within the second-order Popov theory (using $\Lambda_i^0 = g_{ii}(n_i - n_T^0)$ for the perturbation parameter). Let us first consider the symmetric configuration in which $n_1 = n_2$, $m_1 = m_2 = M$ and $g_{11} = g_{22} = g$. We further consider the system to be near the miscible-unmiscible transition, with $(g - g_{12})/ g = 0.07$. Such situation can for instance be found in mixtures of sodium atoms~\cite{Bienaime2016,Fava2018}. Figure~\ref{Fig:k_mixt} shows the isothermal compressibility $\kappa_T$ and the spin susceptibility $\kappa_M$, as a function of temperature $T/T_\mathrm{BEC}$ with $k_BT_\mathrm{BEC} = 2\pi\hbar^2/M [n/(2\zeta(3/2))]^{2/3}$, where $n=(N_1 + N_2)/V$ is the total atom density. These quantities are defined from the chemical potential~\eqref{Eq:mu_mix_sym} as:
\begin{equation}\label{Eq:kappa}
\kappa_{T(M)} = \left[\frac{\partial (\mu_1 \pm \mu_2)}{\partial (n_1 \pm n_2)}\right]^{-1}_T \, .
\end{equation}
In the upper panel of Fig.~\ref{Fig:k_mixt}, one can see that both the second-order Popov theory and the HF theory predict essentially the same behavior for the compressibility, similar to the single-component gas (see Fig.~\ref{Fig:kT_sc}).
Remarkably, the susceptibility predicted by the HF theory shown in the lower panel of Fig.~\ref{Fig:k_mixt} exhibits a divergent behavior at $T \simeq 0.5 T_\mathrm{BEC}$, thereby signalling the onset of a magnetic dynamical instability~\cite{Ota2019}. The origin of this instability can be understood if one writes the analytical expression for the spin susceptibility, obtained from the high-temperature expression~\eqref{Eq:mu_mixt_HT_HF} for the HF chemical potential:
\begin{equation}\label{Eq:kappaM_HF}
2 \left( \kappa_M^\mathrm{HF} \right)^{-1} \simeq \delta g - g^{3/2} \frac{\sqrt{\pi}}{\lambda_T^3} \sqrt{\frac{\beta}{n-2n_T^0}} \, . 
\end{equation}
The onset of the dynamical instability in the HF description is due to the last $g^{3/2}$-term in Eq.~\eqref{Eq:kappaM_HF}, arising from interaction driven thermal fluctuations. As the temperature increases, beyond mean-field effects are enhanced, eventually leading to a divergent behavior of $\kappa_M^\mathrm{HF}$ at finite temperature. However, as shown by the red solid line in the lower panel of Fig.~\ref{Fig:k_mixt}, we find that the spin susceptibility predicted by the Popov theory deviates strongly from the HF calculation. In order to understand the major differences provided by the two approaches, we derive the high-temperature analytical expression of the spin susceptibility, now calculated within the Popov approach Eq.~\eqref{Eq:mu_mixt_HT}. We find:
\begin{align}
2 \left( \kappa_M\right)^{-1} \simeq & \, \delta g - g^{3/2} \frac{\delta g}{g_{12}} \frac{2 \sqrt{\pi}}{\lambda_T^3} \sqrt{\frac{\beta}{n-2n_T^0}}  \nonumber \\
& \times \left[ \left( 1 + \frac{g_{12}}{g} \right)^{3/2} - \left( 1 + \frac{g_{12}}{g} \right) \sqrt{\frac{\delta g}{g}} \right] \, . \label{Eq:kappaM_Po}
\end{align}
In contrast to the HF prediction Eq.~\eqref{Eq:kappaM_HF}, the Popov approach gives rise to contributions proportional to $\delta g$ also for the beyond mean-field terms (second term in the right-hand side of Eq.~\eqref{Eq:kappaM_Po}). A careful comparison between Eqs.~\eqref{Eq:mu_mixt_densities} and~\eqref{Eq:mu_mixt_HT} reveals that the emergence of such beyond mean-field terms in $g_{12}$ is due to the inclusion in Popov theory of effects involving the mixed anomalous densities $\tilde{n}_{12}$ and $\tilde{m}_{12}$.




\section{Phase-separation in two-component mixtures}\label{Sec:Phase Separation}

We now discuss the phenomenon of phase-separation in the mixture of weakly interacting BECs~\cite{Hall1998,Papp2008}. Recently, it has been found in Ref.~\cite{Ota2019} that a mixture initially miscible at zero-temperature can undergo a phase-separation as one increases the temperature, as a result of interaction driven thermal fluctuations. In what follows, we analyze the onset of phase-separation for the Bose mixtures in diverse configurations.

\subsection{Homogeneous symmetric mixtures}

Let us first consider the case of a uniform and symmetric mixture in a box of volume $V$. The onset of such phase transition can be conveniently assessed from an analysis of the Helmoltz free energy $F = \Omega + \sum_i \mu_i n_i$. Proceeding in the same way as for the single-component gas, one finds from Eqs.~\eqref{Eq:K_Po_mix} and \eqref{Eq:mu_mix_sym} the following second-order expression for the free energy of the mixture in the mixed state (see Appendix~\ref{App:F}):
\begin{align} \label{Eq:F_mixt_mis}
\frac{F}{V} =& \frac{g}{2} \left( n_1^2 + n_2 ^2 \right) +g_{12} n_1 n_2 \nonumber \\
&+ g \frac{\zeta(3/2)^2}{\lambda_T^6} + \frac{1}{\beta V} \sum_\pm  \sum_\mathbf{k} \ln \left( 1 - e^{-\beta E_{\pm , \mathbf{k}}^0} \right) \nonumber  \\
&+ \left( \frac{M}{2\pi\hbar^2} \right)^{3/2} \frac{4}{15 \sqrt{\pi}} \sum_\pm \left( 2 \Lambda_\pm^0 \right)^{5/2} \, ,
\end{align}
where $E_{\pm,\mathbf{k}}^0$ and $\Lambda_\pm^0$ are the lowest order expressions, evaluated from Eq.~\eqref{Eq:Lambda_sym} using $\Lambda_i^0$. As for the phase-separated state, since we consider a uniform system, the mixture is prone to separate into two domains ($\mathcal{A}, \mathcal{B}$) of equal volume $V/2$, conserving the total density $n^\mathcal{A} = n^\mathcal{B} = n$, but with opposite magnetization $m^\mathcal{A} = -m^\mathcal{B}=m$. The two domains are in equilibrium when both the pressure ($P^\mathcal{A} = P^\mathcal{B}$) and the chemical potential ($\mu_i^\mathcal{A} = \mu_i^\mathcal{B}$) equilibrium conditions are satisfied. While the equilibrium condition for the pressure is always satisfied for the symmetric configuration, the chemical potential equilibrium at a given temperature is found to be fulfilled at a specific value of the magnetization only. In particular, the equilibrium magnetization must satisfy $m > n-2\zeta(3/2)/\lambda_T^3$, thus corresponding to a regime where in each domain one of the two components is in the normal phase. For such a configuration the Popov free energy in each domain is given by:
\begin{align}\label{Eq:F_mixt_sep}
\frac{F}{V} =& \frac{g}{2} \left( n_1^2 + 2 n_2^2  + \frac{\zeta(3/2)^2}{\lambda_T^6} \right) +g_{12} n_1 n_2 + \mu_2^\mathrm{IBG} n_2 \nonumber \\
&+ \left( \frac{M}{2\pi\hbar^2} \right)^{3/2} \frac{4}{15 \sqrt{\pi}} \left( 2 \Lambda_1^0 \right)^{5/2} \nonumber \\
&+ \frac{1}{\beta V} \sum_\mathbf{k} \ln \left( 1 - e^{-\beta E_\mathbf{k}^0} \right) \nonumber \\ 
&+ \frac{1}{\beta V} \sum_\mathbf{k} \ln  \left( 1 - e^{-\beta (\varepsilon_\mathbf{k} - \mu_2^\mathrm{IBG} )} \right) \, ,
\end{align}
where we have chosen $n_2$ to be the minority component in the normal phase. The ideal Bose gas chemical potential $\mu_2^\mathrm{IBG}$ is defined through the relationship $n_2 = g_{3/2}(e^{\beta \mu_2^\mathrm{IBG}})/\lambda_T^3$, with $g_p(z)$ the usual Bose special function~\cite{book}. As for the majority component in the condensed phase, it is now described by the quasi-particle energy $E_\mathbf{k}^0 = \sqrt{\varepsilon_\mathbf{k}^2 + 2 \varepsilon_\mathbf{k} \Lambda_1^0}$.\par
\begin{figure}[t]
\begin{center}
\includegraphics[width=0.9\columnwidth]{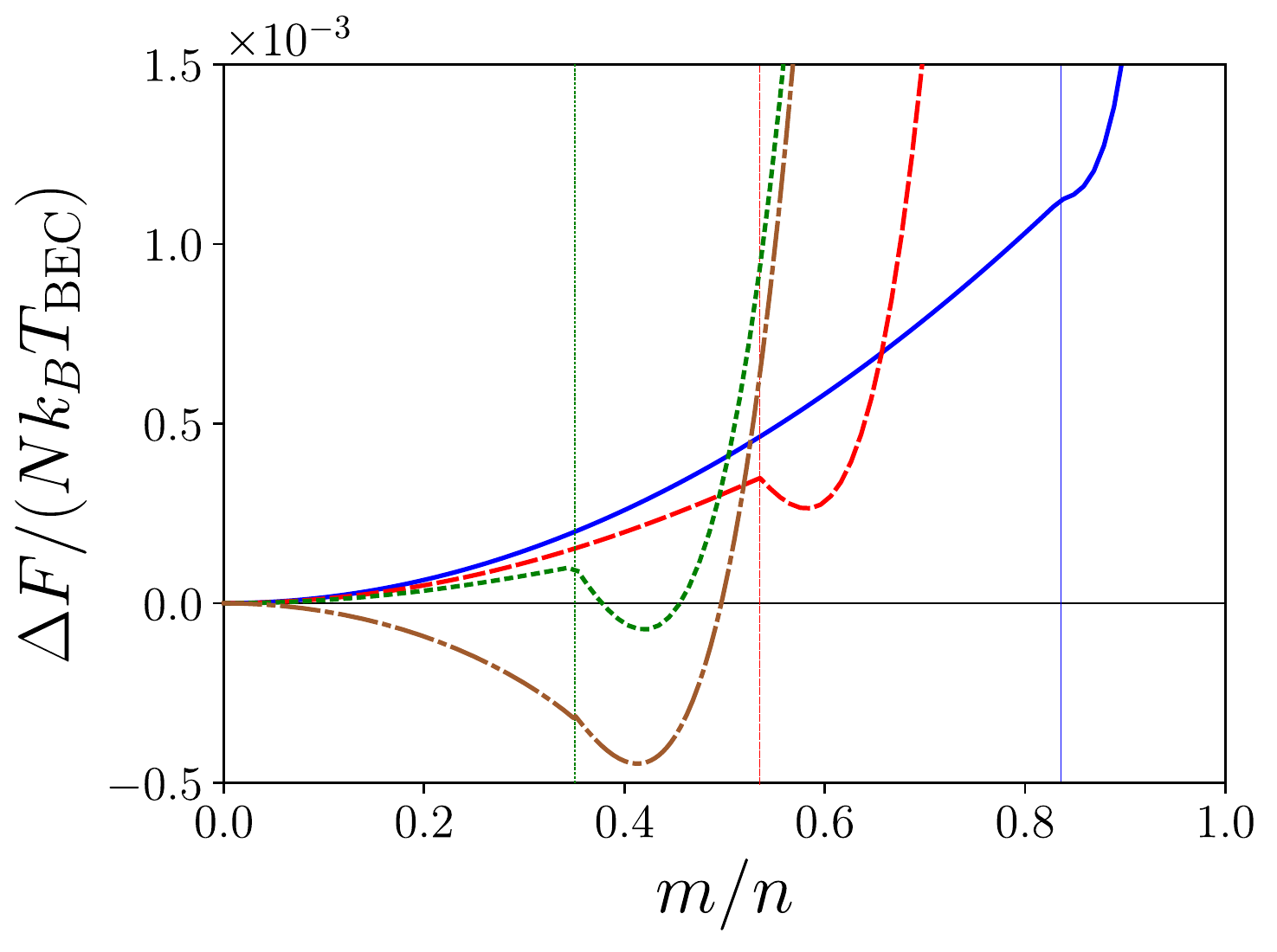}
\caption{Difference of free energies between the miscible state ($m=0$) and the phase-separated state described in the main text, calculated within the Popov theory for $gn/(k_B T_\mathrm{BEC})=0.1$ and $\delta g/g=0.07$. Blue solid line: $T<T^*$, red dashed line: $T^*<T<T_M$, green dotted line $T>T_M$. The brown dot-dashed line is the HF theory result for $T > T_M$. The vertical lines indicate the critical magnetization $m=n-2 \zeta(3/2)/\lambda_T^3$ above which the minority component is purely thermal.} 
\label{Fig:F_mis}
\end{center}
\end{figure}
\par
Figure~\ref{Fig:F_mis} shows the calculated free energy as a function of the magnetization density, for different values of temperature. At low temperature, the free energy is a monotonously increasing function (see blue solid line), with a unique minimum at zero magnetization, corresponding to the mixed state. At a given temperature hereafter called $T^*$, a second minimum starts to develop in the region where the minority component is purely thermal, $m > n-2\zeta(3/2)/\lambda_T^3$ (red dashed line). As already stressed, the emergence of such metastable state corresponds to the fulfillment of the chemical potential equilibrium between the two domains. An analytical expression for the temperature $T^*$ can be obtained from Eq.~\eqref{Eq:F_mixt_mis}, by employing the high temperature $k_B T \gg gn$ expansion for the Bose distribution function:
\begin{equation}\label{Eq:T*}
\frac{T^*}{T_\mathrm{BEC}} \simeq \frac{\delta g}{g} \frac{\zeta(3/2)}{\sqrt{2 \pi}} \sqrt{\frac{k_B T_\mathrm{BEC}}{gn}} \, .
\end{equation}
By further increasing the temperature, the energy of the metastable state decreases, eventually reaching the same energy as the unpolarized state, therefore signaling the onset of a first order phase transition. Hereafter we use the notation $T_M$ to denote this magnetic phase transition temperature, above which the mixed state is energetically unstable with respect to the phase-separated state (green dotted line in Fig.~\ref{Fig:F_mis}). The new equilibrium phase predicted by Popov theory is hence characterized by a full space separation of the Bose-Einstein condensed components of the two atomic species, their thermal components remaining instead mixed, with a finite magnetization. We briefly note that HF theory predicts a similar behavior for the free energy~\cite{Schaeybroeck2013}, but with a dynamical instability, associated to the divergence of the spin susceptibility Eq.~\eqref{Eq:kappaM_HF}. This is shown as the brown dot-dashed line in Fig.~\ref{Fig:F_mis}, where the curvature of the free energy at $m = 0$ becomes negative above $T_M$.
\par
\begin{figure}[t]
\begin{center}
\includegraphics[width=0.9\columnwidth]{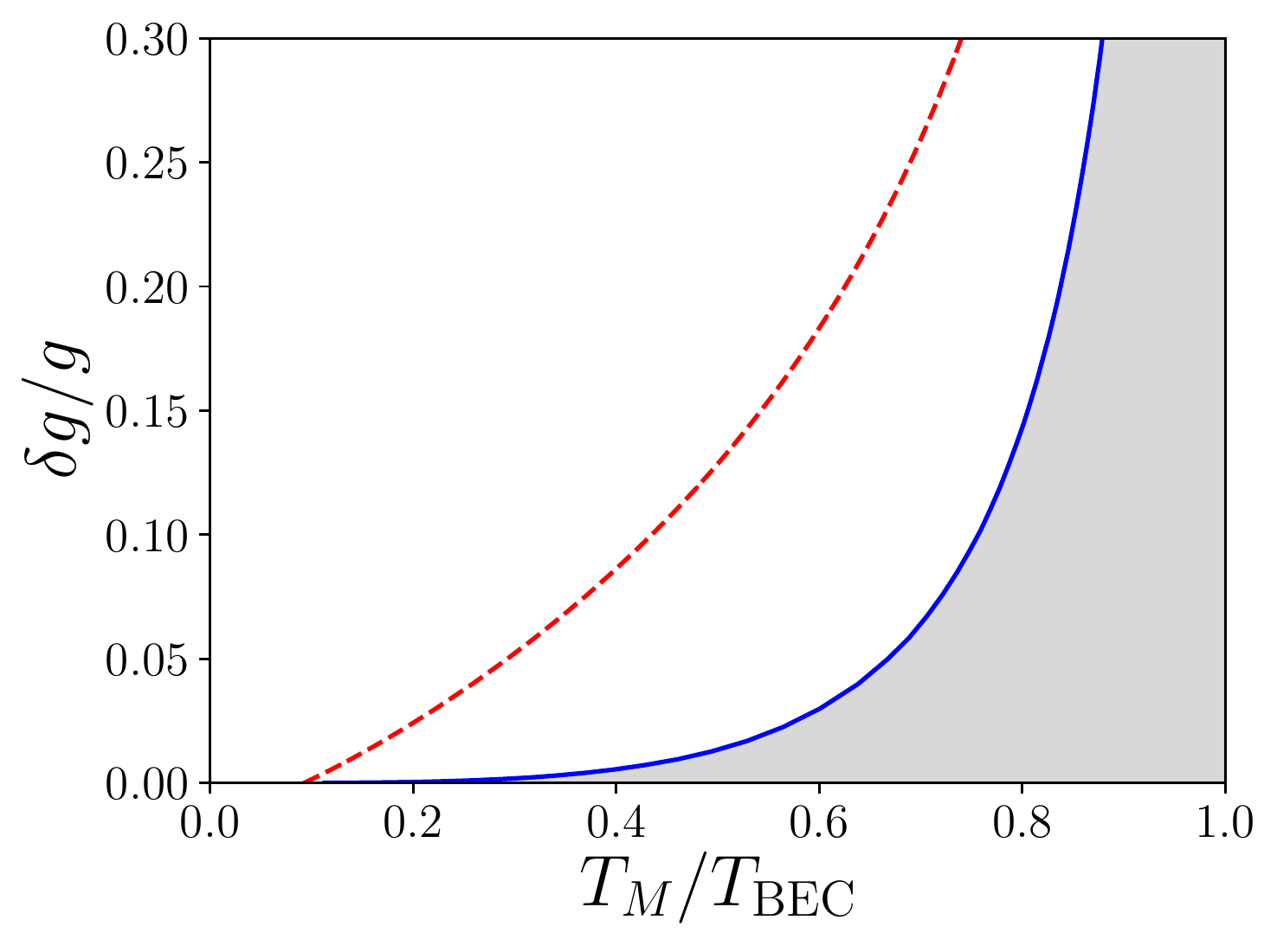}
\caption{Phase diagram for binary condensates with $gn/(k_B T_\mathrm{BEC}) = 0.1$. The blue solid and the red dashed lines are the phase transition temperature $T_M$, and characteristic temperature $T^*$, respectively. The gray area corresponds to the regime of phase-separation.} 
\label{Fig:phaseDiag}
\end{center}
\end{figure}

To summarize, we show in Fig.~\ref{Fig:phaseDiag} the phase diagram of the two-component Bose mixture, by plotting the characteristic temperature $T^*$, providing the onset of a minimum in the free energy with $m \ne 0$, and the phase transition temperature $T_M$, as a function of $\delta g/g$. For the sodium mixture where $\delta g /g=0.07$, we find that the phase-separated state appears as a metastable state at $T^*=0.36 T_\mathrm{BEC}$, while the phase transition occurs at $T_M = 0.71 T_\mathrm{BEC}$. We briefly note that as $\delta g / g \rightarrow 0$, $T^*$ tends to a finite value ($ \simeq 0.1 T_\mathrm{BEC}$), as a consequence of quantum fluctuations, in contrast to Eq.~\eqref{Eq:T*} which only holds if $T^* \gg gn/k_B$. We also find that the phase separated state disappears slightly above the critical temperature $T_\mathrm{BEC}$. At this temperature, the mixture becomes again miscible with both components in the normal phase. We notice that phase separation is the mechanism through which BEC occurs in a symmetric mixture of Bose gases. In fact, instead of being realized simultaneously at the same temperature in both components, the conditions for BEC are attained separately in the two domains of the phase separated state. Only below the temperature $T_M$, the homogeneous and symmetric Bose condensed phase of the mixture emerges as the true equilibrium state. The situation is best understood in terms of the symmetries of the Hamiltonian. In fact, the symmetric mixture enjoys a $U(1)\times U(1) \times Z_2$ symmetry, where $U(1)\times U(1)$ is the gauge symmetry associated to each component, and $Z_2$ is referred to the invariance of the system in respect to the exchange of particles $(1 \leftrightarrow 2)$. Therefore the homogeneous phase with BEC would correspond to the breaking of the $U(1)\times U(1)$ symmetry while the state remains $Z_2$ symmetric. Instead of this picture, the actual scenario is that $U(1) \times Z_2$ is broken in the phase separated state, while each domain remains $U(1)$ symmetric with respect to the minority component. Finally, below $T_M$ the $Z_2$ symmetry is restored.
\par
\begin{figure}[t]
\begin{center}
\includegraphics[width=0.9\columnwidth]{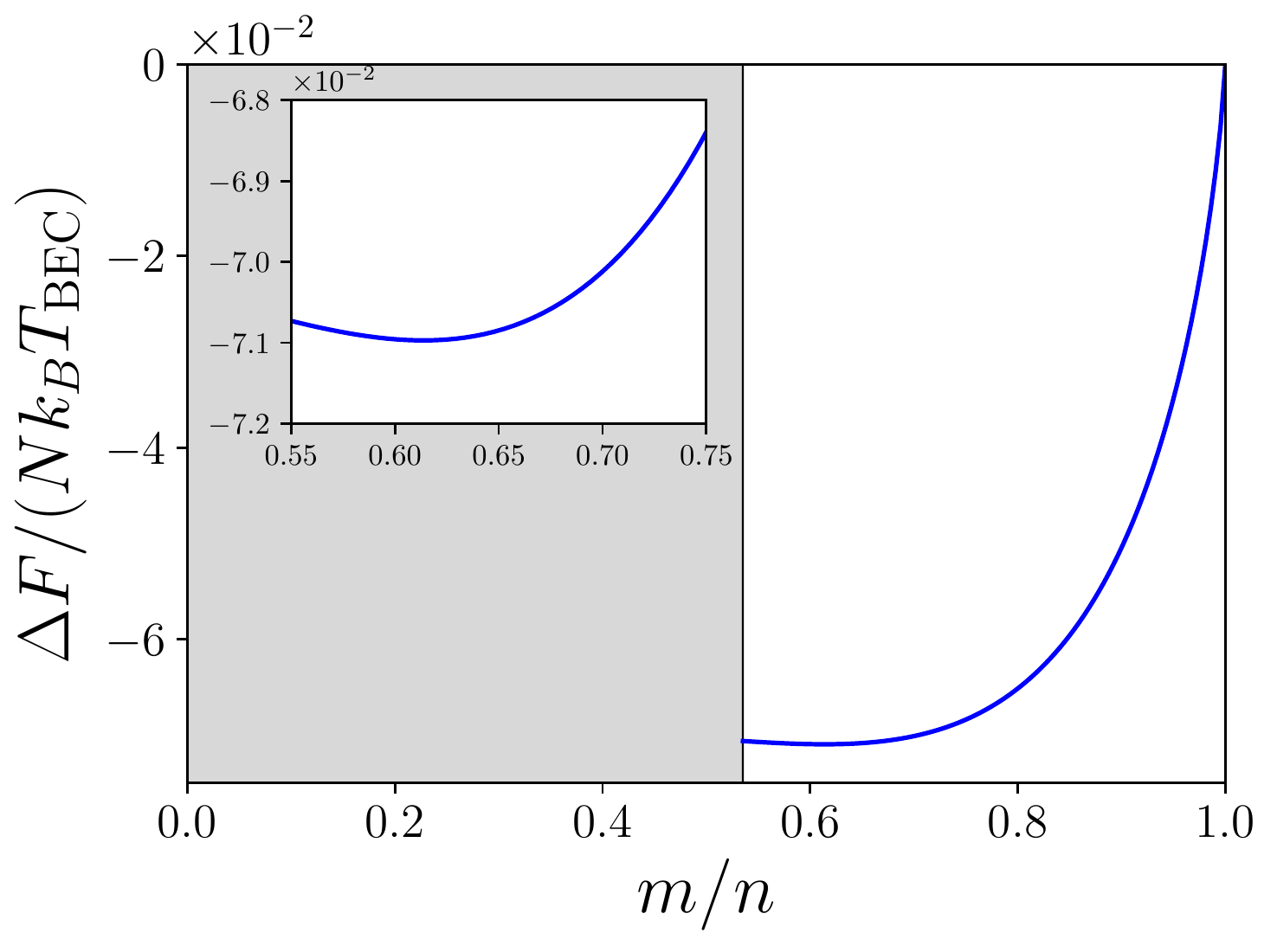}
\caption{Difference of free energies between the fully polarized state ($m=n$) and the phase-separated state described in the main text, for $gn/(k_B T_\mathrm{BEC})=0.1$ and $\delta g/g=-0.07$, calculated at $T=0.6T_\mathrm{BEC}$. The gray shaded region $m < n-2 \zeta(3/2)/\lambda_T^3$ corresponds to the region in which the system is dynamically unstable, with a complex excitation spectrum. Inset: emphasis on the minimum of free energy.}
\label{Fig:F_separated}
\end{center}
\end{figure}

So far, we have restricted our discussion to mixtures satisfying the miscibility criterion at zero-temperature: $g_{12} \leq g$. However, the free energy analysis used above suggests that a similar phase-separation mechanism can take place even when the gas is phase-separated at $T=0$. Indeed, let us consider the situation in which $g_{12}>g$. Then, the spin susceptibility Eq.~\eqref{Eq:kappaM_Po} as well as the square of the spin sound speed Eq.~\eqref{Eq:Lambda_sym} is negative, implying an imaginary Bogoliubov excitation spectrum in the long wave-length limit. These are signatures of dynamical instability, associated to the occurrence of a phase-separation. Now, in the particular case discussed so far, where the two condensates are phase-separated, the spin channel in the Bogoliubov excitation~\eqref{Eq:Lambda_sym} vanishes, and the system is well described by Eq.~\eqref{Eq:F_mixt_sep}, regardless the values of $g$ and $g_{12}$. In Fig.~\ref{Fig:F_separated} we show the behavior of the free energy as a function of the magnetization density, for $\delta g / g = -0.07$, at $T=0.6T_\mathrm{BEC}$.
Actually in the regime where $\delta g < 0$, we find that for any small but finite temperature, a minimum of the free energy appears at $m < n$. Although one can not evaluate the free energy in the region where $m < n - 2 \zeta(3/2)/\lambda_T^3$ (shaded region in Fig.~\ref{Fig:F_separated}) because of the complex excitation spectrum, we expect that a complete phase-separation of the two gases ($m=n$) is made possible only at zero-temperature, and any small but finite temperature is responsible for the mixing of the non-condensed parts. Furthermore, the mixture might be phase-separated in the absence of BEC too, provided that $g_{12} \gg g$.

\subsection{Trapped symmetric mixtures}

In the previous section, we have considered the homogeneous mixture in a uniform potential. However, for the experimental purpose, it is important to assess how the physics of phase-separation is modified in presence of a confining trap. This can be conveniently assessed if we work in the grand-canonical ensemble, and use the local density approximation (LDA)~\cite{Dalfovo1999,Ku2012}. For fixed chemical potentials $(\mu_1, \mu_2)$, four possible configurations arise, according to our previous discussion: both components can be in the BEC phase (BEC1-BEC2), or in the normal phase (N1-N2), and the majority component is in the BEC phase while the minority one is in the normal phase (BEC1-N2 and BEC2-N1).
\begin{figure}[t!]
\begin{center}
\includegraphics[width=0.6\columnwidth]{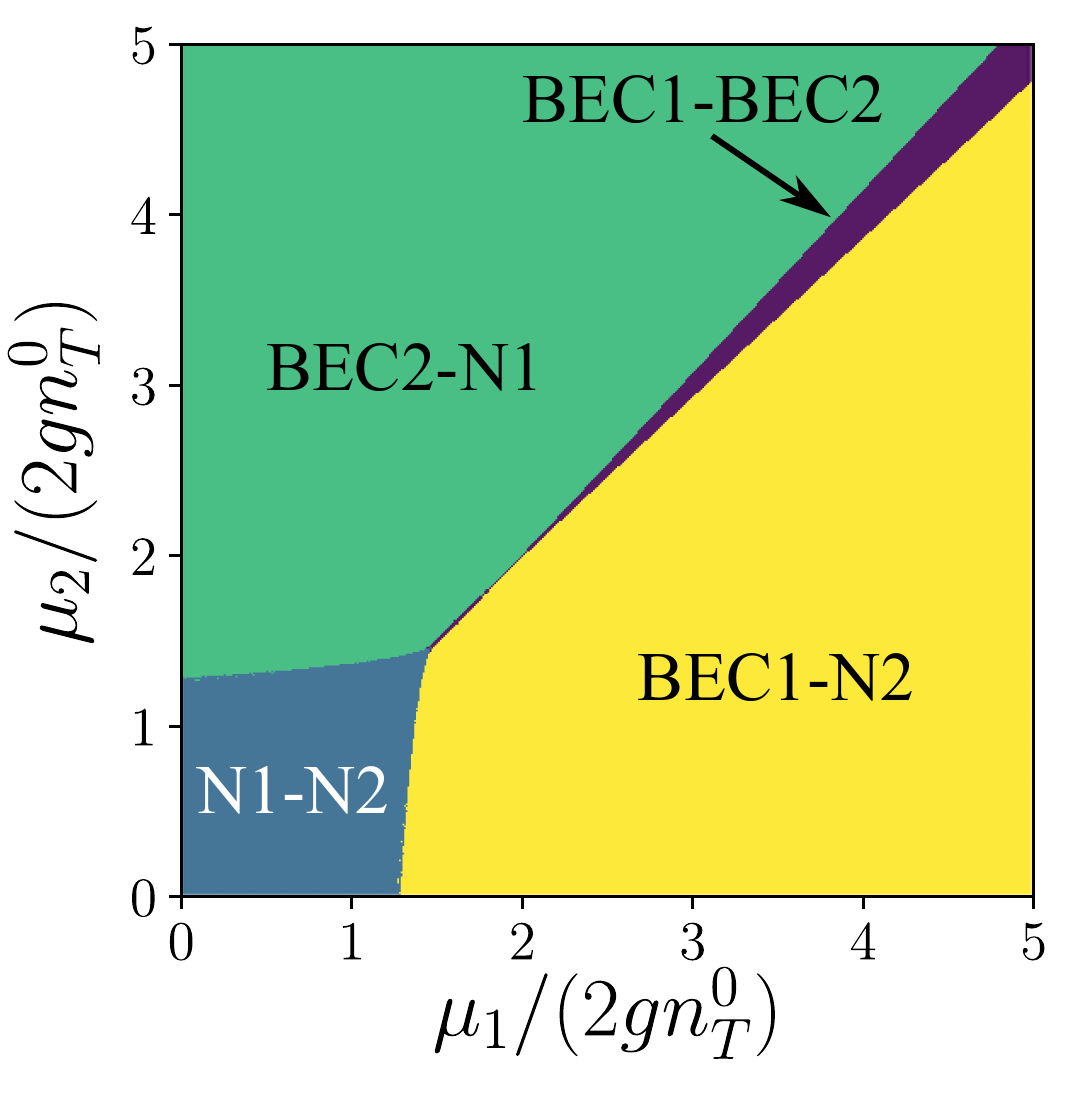}
\caption{Grand-canonical phase diagram for binary condensates, with $gn_T^0 / (k_B T) = 0.05$ and $\delta g / g = 0.07$. The four regions correspond to: both components in the BEC phase (BEC), both in the normal phase (N), component 1 in the BEC phase and component 2 in the normal phase (BEC1-N2), and vice-versa (BEC2-N1).} 
\label{Fig:phaseDiag_GC}
\end{center}
\end{figure}
In Fig.~\ref{Fig:phaseDiag_GC}, we show the grand-canonical phase diagram for the symmetric mixture as a function of chemical potentials, obtained by comparing the thermodynamic energy $\Omega / V$ of these four configurations and searching for the energetically favourable state. The diagram is obtained within the second-order Popov theory, for a fixed value of temperature $g n_T^0 / (k_B T) = 0.05$ and $\delta g / g = 0.07$. Within LDA, the inhomogeneous gas is described as a set of locally homogeneous subsystems, with local chemical potential $\mu_i (\mathbf{r}) = \mu_i - V_{\mathrm{ext},i}(\mathbf{r})$. For an isotropic harmonic trap, $V_{\mathrm{ext},i}(\mathbf{r}) = m_i \omega_i^2 r^2 / 2$, and in the symmetric case where both components feel the same potential, the density profile in the trap is obtained by following the linear curve $\mu_1 = \mu_2 - (\mu_1^0 - \mu_2^0)$, with $\mu_i^0 = \mu_i (r = 0)$, on the phase diagram. Looking closely to Fig.~\ref{Fig:phaseDiag_GC}, one finds that the mixture is miscible at every position of the trap for $\mu_1^0 = \mu_2^0$ only, and an imbalance in the chemical potentials leads to the appearance of a region in which the two BECs do not coexist. We briefly note that a similar phase diagram has been obtained within the HF framework in Ref.~\cite{Schaeybroeck2013}, although predicting the existence of a tricritical point, arising from the divergence of the magnetic susceptibility.

\subsection{Homogeneous asymmetric mixtures}

Finally, let us address the problem of mass and interaction imbalance. For this purpose, we restrict ourselves to the HF framework, since we have seen that this approach provides qualitatively similar results to the Popov theory and is numerically less demanding. In the case of imbalanced mixtures, the system does not separate into two domains of same volume anymore, and one needs to properly solve the pressure and chemical potential equilibrium conditions. The pressure in a given domain $\mathcal{A}$ is given within the HF theory by:
\begin{align}
P^\mathcal{A} =& \sum_{i=1 , 2} \left[\frac{1}{\beta \lambda_{i, T}^3 } g_{5/2} (z_i^\mathcal{A}) + g_{ii} (n_i^\mathcal{A})^2 - \frac{g_{ii}}{2} (n_{i, 0}^\mathcal{A})^2\right] \nonumber \\
& + g_{12} n_1^\mathcal{A} n_2^\mathcal{A} \, ,
\end{align}
with $z_1^\mathcal{A} = e^{\mu_1^\mathcal{A} - 2 g_{11} n_1^\mathcal{A} - g_{12} n_2^\mathcal{A}}$ $(1 \leftrightarrow 2)$. As for the chemical potential, its expression is given in Eq.~\eqref{Eq:mu_mixt_HF}. Solving these equilibrium equations, together with the overall condition $N_1 = N_2$ for the total numbers of particles, one obtains at a given temperature the equilibrium densities for each component in each domain. In the same way as for the symmetric case, the comparison of free energy at equilibrium with the one of the miscible mixture allows for the determination of the critical temperature $T_M$ where the phase-separated state becomes energetically favourable.  
\par
\begin{figure}[t!]
\begin{center}
\includegraphics[width=0.8\columnwidth]{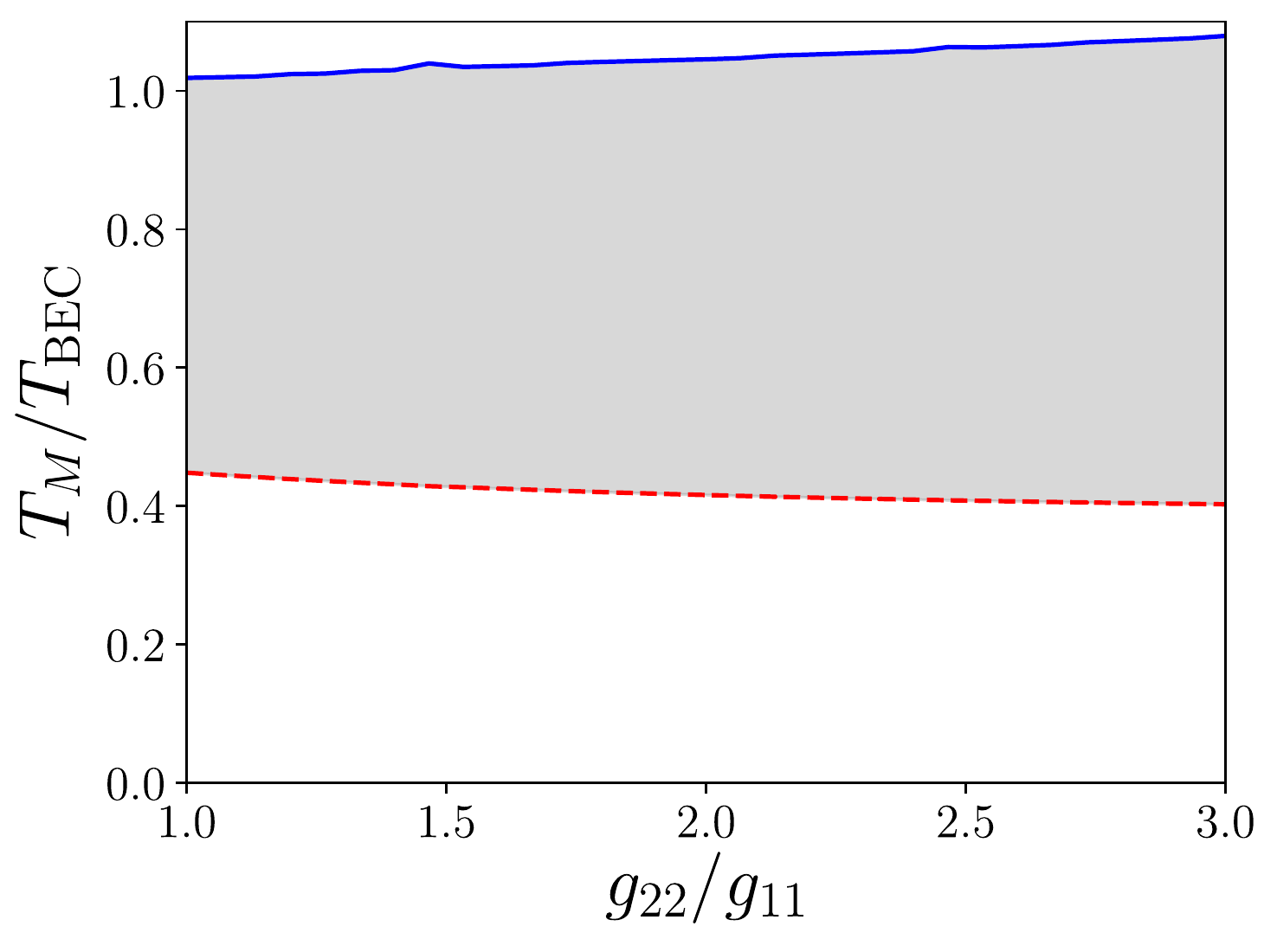}
\includegraphics[width=0.8\columnwidth]{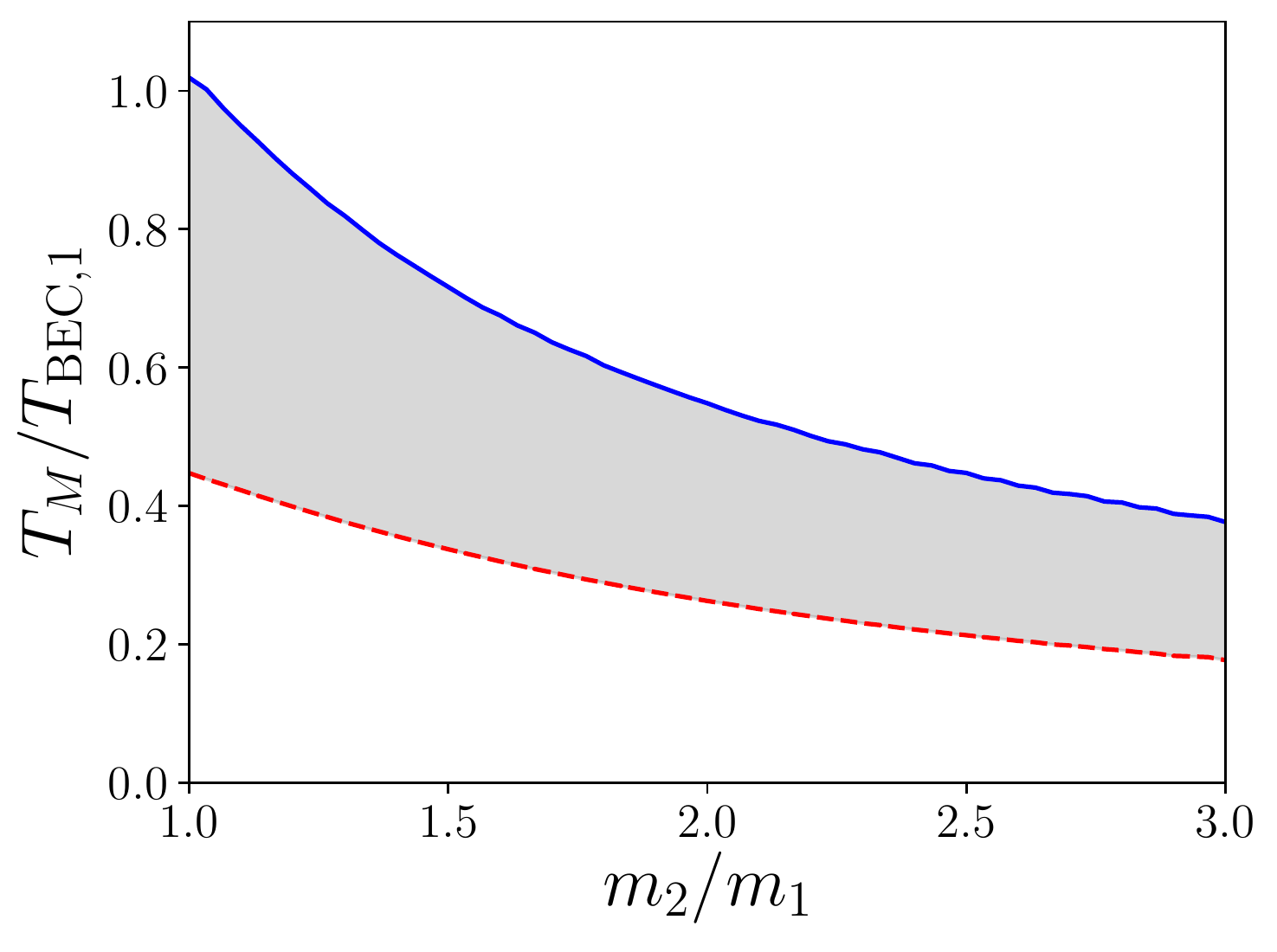}
\caption{Magnetic phase transition temperatures for the imbalanced mixture of Bose gases, with $g_{11} n / (k_B T_\mathrm{BEC}) = 0.1$ and $\bar{g} = 1.07$. Upper panel: as a function of interaction imbalance $g_{22}/g_{11}$, with equal masses $m_1 = m_2$. Lower panel: as a function of mass imbalance $m_2/m_1$, with equal intra-species interaction $g_{11} = g_{22}$.}
\label{Fig:T_M}
\end{center}
\end{figure}
\par
In the upper panel of Fig.~\ref{Fig:T_M}, we show the calculated $T_M$, as a function of the coupling constant ratio $g_{22} / g_{11}$, with $m_1=m_2$ and $\bar{g}=1.07$. We find that the phase-separation is not very sensitive to the interaction imbalance, with a phase-separated region (shown as gray shaded area) practically independent of the value of $g_{22}/g_{11}$. The lower panel of Fig.~\ref{Fig:T_M} shows instead the dependence of $T_M$ on the mass imbalance $m_2 / m_1$ for $g_{11} = g_{22}$ and $\bar{g}=1.07$. The temperature on this plot is normalized to the critical temperature of the light component $T_{\mathrm{BEC},1} = 2 \pi \hbar^2 / m_1 (n/2\zeta(3/2))^{2/3}$. In contrast to the previous case, the region in which the phase-separated state is favorable is found to shrink as one increases $m_2/m_1$. This is understood from the fact that the BEC critical temperature for the heavy component scales as $1/m_2$, thus lowering the upper bound for the phase-separation (blue solid line in Fig.~\ref{Fig:T_M}), which corresponds essentially to the BEC critical temperature.




\section{Superfluid density of two component mixtures}\label{Sec:AB effects}

We finally discuss the superfluid densities in binary Bose gases. As a peculiarity of superfluid mixtures, the coupling between the two atomic components will be responsible for an entrainment effect, known as the Andreev-Bashkin effect~\cite{Andreev1975}. The superfluid current in each component is coupled through a drag term ($1 \leftrightarrow 2$)~\cite{Nespolo2017}:
\begin{equation}
m_1 \mathbf{j}_1 = \rho_{1,n} \mathbf{v}_n + \rho_{1, s} \mathbf{v}_{1,s} + \rho_{12} \mathbf{v}_{2, s} \, ,
\end{equation}
where we have introduced the normal component velocity $\mathbf{v}_n$ as well as the superfluid velocities $\mathbf{v}_{i, s}$. The normal component $\rho_{i,n}$ and superfluid component $\rho_{i,s}$ in each atomic species is normalized according to
\begin{equation}
\rho_{i} = m_i n_i = \rho_{i,n} + \rho_{i,s} + \rho_{12} \, ,
\end{equation}
with $\rho_{12} = \rho_{21}$ the "drag" density. In this work, we use the methodology developed in Ref.~\cite{Romito2019}, which extends the linear response formalism used for the single-component gas to atomic mixtures (see Appendix~\ref{App:ns}). In this framework, the normal densities as well as the drag density are given by:
\begin{gather}
\rho_{i,n} = \frac{m_i^2}{V} \lim_{\mathbf{q} \to 0} \chi^\perp_{\mathbf{j}_i \mathbf{j}_i} (\mathbf{q}) - \rho_{12} \, , \label{Eq:rho_n_mixt} \\
\rho_{12} = - \frac{m_1 m_2}{V} \lim_{\mathbf{q} \to 0} \chi^\perp_{\mathbf{j}_1 \mathbf{j}_2}(\mathbf{q}) \, , \label{Eq:rho12_mixt}
\end{gather}
where $\chi^\perp_{\mathbf{j}_i \mathbf{j}_j}$ is the transverse component of the current density response function for the mixtures
\begin{align}\label{Eq:chi_mixt}
\chi_{\mathbf{j}_i \mathbf{j}_j} (\mathbf{q}) =& \frac{1}{Q} \sum_{m,n} e^{-\beta E_m} \left(\frac{\langle n | \hat{\mathbf{j}}_i^\dagger(\mathbf{q})|m \rangle \langle m | \hat{\mathbf{j}}_j(\mathbf{q})|n \rangle}{E_n - E_m + i\eta} \right. \nonumber \\
&- \left. \frac{\langle n | \hat{\mathbf{j}}_i(\mathbf{q})|m \rangle \langle m | \hat{\mathbf{j}}_j^\dagger(\mathbf{q})|n \rangle}{E_m - E_n + i\eta} \right) \, ,
\end{align}
whereas the current density operator in each component $\hat{\mathbf{j}}_i$ is given by Eq.~\eqref{Eq:j_sc}, with the proper corresponding creation and annihilation operators $\hat{a}^\dagger_{i, \mathbf{k}}$ and $\hat{a}_{i, \mathbf{k}}$. The calculation of the transverse response function follows essentially the same steps as the single-component case.
\par
Let us focus primarily on the collisionless drag $\rho_{12}$. After expressing the single-particle creation and annihilation operators in the quasi-particle basis by means of Eqs.~\eqref{Eq:cano_tans_mixt} and~\eqref{Eq:Bogo_trans_mixt}, one finds that the matrix elements product in Eq.~\eqref{Eq:chi_mixt} has three categories of non-vanishing contributions in the limit of long wavelengths ($\mathbf{q}\to0$). The first one corresponds to quasiparticle excitation-annihilation matrix elements in the single spin or density channel (e.g. $\langle n | \hat{\alpha}^\dagger \hat{\alpha} | m \rangle \langle m | \hat{\alpha}^\dagger \hat{\alpha} | n \rangle $), and is analogous to the single-component result Eq.~\eqref{Eq:chiT_sc}:
\begin{equation}
\chi^\perp_{\mathbf{j}_1 \mathbf{j}_2} \Bigr|_\mathrm{Single} = -\frac{1}{3} \frac{\hbar^2}{m_1 m_2} \sum_\mathbf{k}  k^2 \lambda_\mathbf{k}^2 (z_\mathbf{k}^2 - w_\mathbf{k}^2) \sum_\pm \frac{\partial f_{\pm,\mathbf{k}}}{\partial E_{\pm,\mathbf{k}}} \,
\end{equation}
where $\lambda_\mathbf{k}$, $w_\mathbf{k}$ and $z_\mathbf{k}$ are given by Eq.~\eqref{Eq:qpf_amp_mixt}, and we have introduced the short-hand notation $f_{\pm,\mathbf{k}} = f(E_{\pm, \mathbf{k}})$. The second contribution arises from multi-channel quasiparticle excitation-annihilation matrix elements (e.g. $\langle n | \hat{\alpha}^\dagger \hat{\beta} | m \rangle \langle m | \hat{\beta}^\dagger \hat{\alpha} | n \rangle $):
\begin{align}
\chi^\perp_{\mathbf{j}_1 \mathbf{j}_2} \Bigr|_\mathrm{Multi, 1} = & \frac{2}{3} \frac{\hbar^2}{m_1 m_2} \sum_\mathbf{k}  k^2 \frac{\lambda_\mathbf{k}^2 (z_\mathbf{k}^2 - w_\mathbf{k}^2)}{4 E_{+, \mathbf{k}}E_{-,\mathbf{k}}} \nonumber \\
&\times \frac{\left(E_{+,\mathbf{k}} + E_{-,\mathbf{k}}\right)^2}{E_{+,\mathbf{k}} - E_{-,\mathbf{k}}} ( f_{+,\mathbf{k}} - f_{-,\mathbf{k}} ) \, .
\end{align}
Finally, the last contribution comes from anomalous multi-channel excitations (e.g. $\langle n | \hat{\alpha}^\dagger \hat{\beta}^\dagger | m \rangle \langle m | \hat{\alpha} \hat{\beta} | n \rangle $):
\begin{align}
\chi^\perp_{\mathbf{j}_1 \mathbf{j}_2} \Bigr|_\mathrm{Multi, 2} =&  -\frac{2}{3} \frac{\hbar^2}{m_1 m_2} \sum_\mathbf{k}  k^2 \frac{\lambda_\mathbf{k}^2 (z_\mathbf{k}^2 - w_\mathbf{k}^2)}{4 E_{+, \mathbf{k}}E_{-,\mathbf{k}}} \nonumber \\
&\times \frac{\left(E_{+,\mathbf{k}} - E_{-,\mathbf{k}}\right)^2}{E_{+,\mathbf{k}} + E_{-,\mathbf{k}}} ( 1 + f_{+,\mathbf{k}} + f_{-,\mathbf{k}} ) ,
\end{align}
which remains finite also at $T=0$~\cite{Romito2019}. Summing up the three contributions we find from Eq.~\eqref{Eq:rho12_mixt}:
\begin{align}\label{Eq:rho12_Popov}
\rho_{12} =& \frac{4}{3} \frac{\sqrt{m_1 m_2}}{V} \sum_\mathbf{k} \frac{\bar{g}^2 \Lambda_1 \Lambda_2 (\varepsilon_{1,\mathbf{k}}\varepsilon_{2,\mathbf{k}})^{3/2}}{E_{+,\mathbf{k}} E_{-,\mathbf{k}}} \left[ \frac{1+f_{+,\mathbf{k}} + f_{-,\mathbf{k}}}{(E_{+,\mathbf{k}} + E_{-,\mathbf{k}})^3} \right. \nonumber \\
&\left. - \frac{f_{+,\mathbf{k}} - f_{-,\mathbf{k}}}{(E_{+,\mathbf{k}} - E_{-,\mathbf{k}})^3} + \frac{2 E_{+,\mathbf{k}} E_{-,\mathbf{k}}}{(E_{+,\mathbf{k}}^2- E_{-,\mathbf{k}}^2)^2} \sum_\pm \frac{\partial f_{\pm,\mathbf{k}}}{\partial E_{\pm,\mathbf{k}}}  \right] \, .
\end{align}
The above result is valid for any temperature up to the close vicinity of the critical point. In particular, in the low-temperature regime where $k_B T \ll \mu_i(T=0)$, one can safely replace $\Lambda_i$ by the zero-temperature expression $g_{ii}n_i$. In this way, we retrieve the result of Ref.~\cite{Fil2005}, obtained by calculating the lowest order change in the free energy of the mixture due to a finite superfluid velocity. As for the normal density, one can carry out a similar development starting from Eq.~\eqref{Eq:rho_n_mixt}, and find ($1 \leftrightarrow 2$)
\begin{align}
\rho_{1,n} =& - \frac{1}{3} \frac{m_1}{V} \sum_\mathbf{k} \varepsilon_{1,\mathbf{k}} \left[\left(\frac{\partial f_{+,\mathbf{k}}}{\partial E_{+,\mathbf{k}}} + \frac{\partial f_{-,\mathbf{k}}}{\partial E_{-,\mathbf{k}}}\right) \right. \nonumber \\
&\left. + \frac{E_{1,\mathbf{k}}^2 - E_{2,\mathbf{k}}^2}{(E_{+,\mathbf{k}}^2 - E_{-,\mathbf{k}}^2)} \left(\frac{\partial f_{+,\mathbf{k}}}{\partial E_{+,\mathbf{k}}} - \frac{\partial f_{-,\mathbf{k}}}{\partial E_{-,\mathbf{k}}}\right)\right] \;.
\end{align}
\par
\begin{figure}[t!]
\begin{center}
\includegraphics[width=0.9\columnwidth]{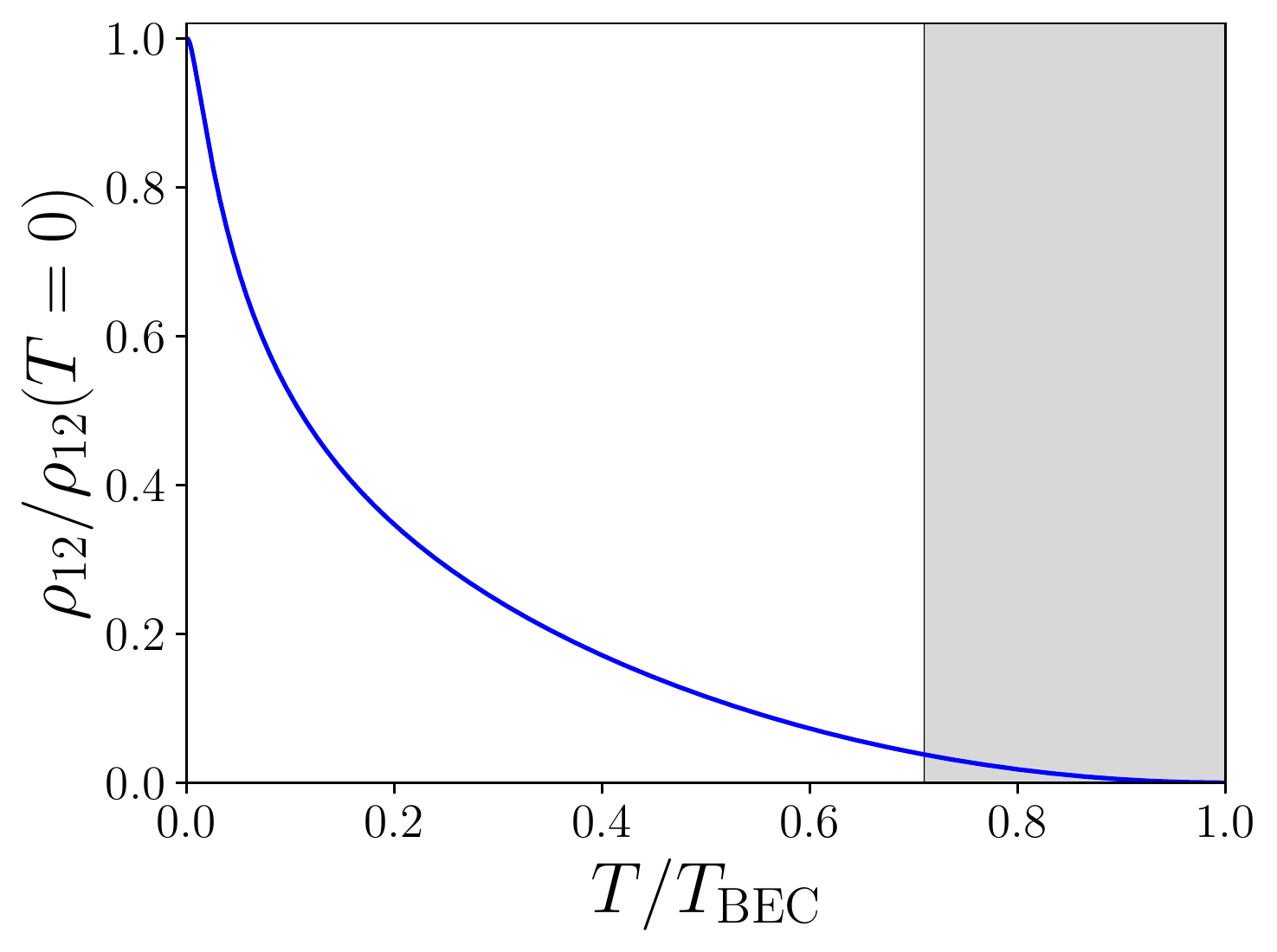}
\caption{Temperature dependence of the collisionless drag $\rho_{12}$, normalized to the zero-temperature value Eq.~\eqref{Eq:rho12_T0}. The drag is calculated for the $^{23}\mathrm{Na}$ symmetric mixture, with $gn / (k_B T_\mathrm{BEC}) = 0.1$ and $\delta g / g = 0.07$. The shaded region corresponds to the temperature regime where the miscible mixture is energetically unstable with respect to phase-separation.} 
\label{Fig:rho12}
\end{center}
\end{figure}

Figure~\ref{Fig:rho12} shows the temperature dependence of the superfluid drag, calculated for a symmetric mixture, such as ${}^{23}\mathrm{Na}$, with $g_{11}=g_{22}=g$ and $m_1 = m_2 =m$. Our results extend to finite temperature the calculations of Ref.~\cite{Romito2019} and generalize the findings of Ref.~\cite{Fil2005} which were restricted to the regime $k_BT\ll\mu$. We also notice that according to what we have discussed in Sec.~\ref{Sec:Phase Separation}, the mixture becomes unstable with respect to phase-separation for $T>T_M$, leading to a vanishing collisionless drag. This is shown by the shaded region in Fig.~\ref{Fig:rho12}. 
In the particular case of equal masses, the zero-temperature value for the drag can be evaluated analytically by turning the momentum sum in Eq.~\eqref{Eq:rho12_Popov} into an integral, and one finds~\cite{Fil2005}
\begin{equation}\label{Eq:rho12_T0}
\rho_{12}(T=0) = mn_1\sqrt{n_1a_{11}^3}F(\bar{g}, l) \, ,
\end{equation}
with $\bar{g} = g_{12}/\sqrt{g_{11}g_{22}}$ and $l = g_{22} n_2 / (g_{11} n_1)$. The dimensionless function on the right hand side of the above equation is given by
\begin{widetext}
\begin{equation}
F(\bar{g}, l) = \frac{128\sqrt{2}}{45\sqrt{\pi}} \frac{\bar{g}^2 l  \left(1+l+3 \sqrt{l(1-\bar{g}^2)}\right)}{\left(\sqrt{1+l+\sqrt{(1-l)^2+4\bar{g}^2l}} + \sqrt{1+l-\sqrt{(1-l)^2+4\bar{g}^2l}}\right)^{3}} \,.
\end{equation}
\end{widetext}
As discussed in Ref.~\cite{Fil2005}, this function displays a weak dependence on the parameters $\bar{g}$ and $l$ yielding values in the range $0.7\le F\le0.8$.
\par
The present study shows that the Andreev-Bashkin effect is most important at zero temperature. On the other hand, it is known that at $T=0$ the spin speed of sound is closely related to both the magnetic susceptibility and the drag density \cite{Parisi2017}. Experimentally one could therefore observe the Andreev-Bashkin effect from an independent measurement of the susceptibility and the spin sound velocity~\cite{Kim2020,Roy2020}.



\section{Conclusion}

In conclusion, we have developed the beyond mean-field Popov theory for the systems of weakly interacting Bose gases. Our derivation is based on simple theoretical tools, and can be applied to a variety of problems involving BEC. We have illustrated the approach by deriving the Popov theory for the mixtures of BECs, which includes the effects of thermal and quantum fluctuations in both the density and spin channels. As a result, we have extended our previous study on the magnetic phase transition at finite temperature~\cite{Ota2019} to the case of trapped systems, as well as in presence of interaction and mass imbalances. Our numerical results show that, in experiments with trapped systems, a miscible mixture can exhibit a spatial region in which the two BECs do not coexist. On the other hand, we found that in general, the presence of an asymmetry reduces the temperature window in which the phase-separated state is energetically favorable. Finally, we have calculated the temperature-dependence of the collisionless drag, by means of linear response theory combined to the Popov approach. Important open issues concern the propagation of sound in these polarized domains, the possible emergence of a similar magnetic phase transition in two dimensions, and the structure of the interface between different domains.

\begin{acknowledgments}
We are indebted to Sandro Stringari for many stimulating discussions and suggestions during the preparation of this work. We also thank Donato Romito for useful comments. This project has received funding from the EU Horizon 2020 research and innovation programme under grant agreement No. 641122 QUIC, and by Provincia Autonoma di Trento.
\end{acknowledgments}




\appendix




\section{Free energy}\label{App:F}

For the study of the phase diagram of Bose mixtures, it is useful to evaluate the leading order beyond mean-field corrections to the Helmoltz free energy. Since the calculation is essentially the same between the single-component gas and the binary mixture, we focus in what follows to the single-component case. Let us assume for this purpose a gapless spectrum, as given by Eq.~\eqref{Eq:E_sc}. Then, the Helmoltz free energy in the BEC phase is given from Eq.~\eqref{Eq:Omega_sc} according to $F = \Omega + \mu N$:
\begin{align}\label{Eq:F_Po_1}
\frac{F}{V} =& \frac{g}{2} n_0^2 - g \tilde{n}^2 + \mu \tilde{n} + \frac{1}{\beta V} \sum_\mathbf{k} \ln \left(1 - e^{-\beta E_\mathbf{k}}\right) \nonumber \\
&+ \frac{1}{2V} \sum_{\mathbf{k} \neq 0} \left( E_\mathbf{k} - \varepsilon_\mathbf{k} - \Lambda + \frac{\Lambda^2}{2\varepsilon_\mathbf{k}} \right) \, .
\end{align}
Using the results in Eqs.~\eqref{Eq:nT_sc} and~\eqref{Eq:mu_sc} respectively for $\tilde{n}$ and $\mu$ the free energy can also be expressed as
\begin{align}\label{Eq:F_Po_2}
\frac{F}{V} =& \frac{g}{2} n^2 + \frac{g}{2} {n_T^0}^2 + gn_T^0 \left(\frac{m\Lambda}{2\pi \hbar^2}\right)^{3/2}H(\tau) \nonumber \\
&+ \frac{1}{\beta V} \sum_\mathbf{k} \ln \left(1 - e^{-\beta E_\mathbf{k}}\right) + \frac{16\sqrt{2}}{15\sqrt{\pi}} \left(\frac{m}{2\pi \hbar^2}\right)^{3/2} \Lambda^{5/2} \, ,
\end{align}
where contributions of order higher than $\Lambda^{5/2}$ have been neglected. The last sum over thermal excitations requires a careful expansion in terms of the parameter 
 $\Lambda = \Lambda^0 - g \left(\frac{m\Lambda}{2\pi \hbar^2}\right)^{3/2}H(\tau)$. One finds
\begin{align}
\frac{1}{\beta V} \sum_\mathbf{k} \ln \left(1 - e^{-\beta E_\mathbf{k}}\right) \simeq & \frac{1}{\beta V} \sum_\mathbf{k} \ln \left(1 - e^{-\beta E^0_\mathbf{k}}\right) \nonumber \\
&-g n_T^0 \left(\frac{m\Lambda}{2\pi \hbar^2}\right)^{3/2}H(\tau) \, ,
\end{align}
so that the last term cancels with the corresponding one in Eq.~\eqref{Eq:F_Po_2}. In the above expression $E^0_\mathbf{k}=\sqrt{\varepsilon_\mathbf{k}^2+2\Lambda^0\varepsilon_\mathbf{k}}$ denotes the quasi-particle spectrum~\eqref{Eq:E_sc} to the lowest order in $\Lambda$. It is worth noticing that such cancellation would not happen, if one chooses $gn_0 = \Lambda^0 - g^{5/2} \left(\frac{mn_0}{2\pi \hbar^2}\right)^{3/2} G(\tau)$ instead of $\Lambda$ for the perturbation parameter, leading to an incorrect expression for the free energy.
Finally, the full expression of the free energy including the leading order corrections in the interaction coupling reads:
\begin{align}\label{Eq:F_Po}
\frac{F}{V} \simeq & \frac{g}{2} (n^2 + {n_T^0}^2) + \frac{1}{\beta V} \sum_\mathbf{k} \ln \left(1 - e^{-\beta E^0_\mathbf{k}}\right) \nonumber \\
& + \frac{16\sqrt{2}}{15\sqrt{\pi}} \left(\frac{m}{2\pi \hbar^2}\right)^{3/2} (\Lambda^0)^{5/2} \, .
\end{align}
One can verify that by taking the derivative of Eq.~\eqref{Eq:F_Po} with respect to $n$, we recover the expression for the chemical potential Eq.~\eqref{Eq:mu_sc}, when evaluated using $\Lambda^0$. The derivation of the free energy for the mixtures follow the same line of calculations.




\section{Superfluid density}\label{App:ns}

The calculation of the superfluid density for both the single-component and binary mixtures of Bose gases can be achieved in a convenient way by means of linear response theory. For a single-component Bose gas, the normal component mass density $\rho_n = m n_n = m(n - n_s)$ is calculated according to~\cite{book}
\begin{equation}\label{Eq:rho_n_sc}
\rho_n = \frac{m^2}{V} \lim_{\mathbf{q} \to 0} \chi^\perp_\mathbf{j} (\mathbf{q}) \,
\end{equation}
which relates $\rho_n$ to the transverse component (denoted by $\perp$, with $\mathbf{q} \cdot \mathbf{j} = 0$) of the current response function, defined as
\begin{align}\label{Eq:chi_sc}
\chi_\mathbf{j} (\mathbf{q}) =& \frac{1}{Q} \sum_{m,n} e^{-\beta E_m} \left[ \frac{\langle n | \hat{\mathbf{j}}^\dagger(\mathbf{q})|m \rangle \langle m | \hat{\mathbf{j}}(\mathbf{q})|n \rangle}{E_n - E_m + i\eta} \right. \nonumber \\
&- \left. \frac{\langle n | \hat{\mathbf{j}}(\mathbf{q})|m \rangle \langle m | \hat{\mathbf{j}}^\dagger(\mathbf{q})|n \rangle}{E_m - E_n + i\eta} \right] \, ,
\end{align}
where $Q=\sum_m e^{-\beta E_m}$ is the canonical partition function, $|n\rangle$ and $E_n$ are the eigenstates and eigenvalues of the Hamiltonian~\eqref{Eq:H_sc} and $\eta$ a small positive value. The current density operator reads
\begin{equation}\label{Eq:j_sc}
\hat{\mathbf{j}}(\mathbf{q}) = \frac{\hbar}{2m} \sum_{\mathbf{k}} \left(2 \mathbf{k} + \mathbf{q}\right) \hat{a}^\dagger_{\mathbf{k}}\hat{a}_{\mathbf{k}+\mathbf{q}} \, .
\end{equation}
The matrix elements of the current density operator in Eq.~\eqref{Eq:chi_sc} can be calculated straightforwardly if one expresses $\hat{a}_\mathbf{k}$ and $\hat{a}^\dagger_\mathbf{k}$ in terms of the quasi-particle creation and annihilation operators in Eq.~\eqref{Eq:Bogo_trans}. Indeed, the matrix element $\langle m | \hat{\alpha}^\dagger_\mathbf{k} \hat{\alpha}_\mathbf{k'}|n \rangle$ yields non-vanishing contribution only for $E_n - E_m = E_\mathbf{k'} - E_\mathbf{k}$.  Choosing the vector $\mathbf{q}$ along the $z$-direction, the transverse response is provided by the $x$-component of the current density, and one finds evaluating the trace:
\begin{align}\label{Eq:chiT_sc}
\chi_{j_x}^\perp(\mathbf{q}) =& -\frac{\hbar^2}{m^2} \sum_\mathbf{k} k_x^2 \left(u_\mathbf{k} u_\mathbf{k+q}-v_\mathbf{k} v_\mathbf{k+q} \right)^2 \nonumber \\
& \times \frac{f(E_{\mathbf{k}+\mathbf{q}}) - f(E_\mathbf{k})}{E_{\mathbf{k} + \mathbf{q}} - E_\mathbf{k}} \, .
\end{align}
Taking the limit $\mathbf{q} \to 0$ as well as making use of the isotropy of the system, one finally obtains from Eqs.~\eqref{Eq:rho_n_sc} and~\eqref{Eq:chiT_sc}:
\begin{equation}
\rho_n = -\frac{\hbar^2}{3V} \sum_\mathbf{k} k^2 \frac{\partial f(E_\mathbf{k})}{\partial E_\mathbf{k}} \, ,
\end{equation}
retrieving the well-known Landau formula. Using the Popov excitation spectrum Eq.~\eqref{Eq:E_sc}, one gets the expression
\begin{equation}
\rho_n = \frac{1}{m\lambda_T^3} \frac{4}{3\sqrt{\pi}} \frac{1}{\sqrt{\tau}} \int_0^\infty dx x \frac{(u-1)^{3/2}}{u} \frac{f(x)}{1-e^{-x}} \, ,
\end{equation}
where $u=\sqrt{1+(\tau x)^2}$ and $\tau=k_BT/\Lambda$ as in Eqs.~\eqref{Eq:G_sc} and~\eqref{Eq:Htau_sc}. The present approach is extended to the two-component mixtures in Sec.~\ref{Sec:AB effects}, allowing for the calculation of the collisionless drag.

\section{Two component Bose mixtures: General case}\label{App:mixt}

In order to diagonalize the grans-canonical Hamiltonian of the binary mixture Eq.~\eqref{Eq:K_mix}, one first applies the following canonical transformations:
\begin{align}\label{Eq:cano_tans_mixt}
\begin{split}
\hat{a}'_{1, \mathbf{k}} &= \lambda_\mathbf{k} \hat{a}_{1, \mathbf{k}} + z_\mathbf{k} \hat{a}_{2, \mathbf{k}} + w_\mathbf{k} \hat{a}_{2, -\mathbf{k}}^\dagger  \, , \\
\hat{a}'_{2, \mathbf{k}} &= \lambda_\mathbf{k} \hat{a}_{2, \mathbf{k}} - z_\mathbf{k} \hat{a}_{1, \mathbf{k}} + w_\mathbf{k} \hat{a}_{1, -\mathbf{k}}^\dagger  \, .
\end{split}
\end{align}
The new operators must obey the Bose commutation relation, and one finds that this is satisfied if $\lambda_\mathbf{k}^2 + z_\mathbf{k}^2 - w_\mathbf{k}^2 = 1$. Then the terms involving the coupling of both components, in which the operators $\hat{a}'_1$ and $\hat{a}'_2$ appear in pairs, vanish if the weight functions take the expressions:
\begin{equation}
\lambda_\mathbf{k}^2 = \frac{1}{2} \left[ 1 + \frac{\tilde{E}_{1, \mathbf{k}}^2 - \tilde{E}_{2, \mathbf{k}}^2}{\sqrt{(\tilde{E}_{1, \mathbf{k}}^2 - \tilde{E}_{2, \mathbf{k}}^2)^2 + 16 g_{12}^2 \tilde{\varepsilon}_{1, \mathbf{k}} \tilde{\varepsilon}_{2, \mathbf{k}} n_{1, 0} n_{2, 0}}} \right] \nonumber 
\end{equation}
\begin{align}\label{Eq:qpf_amp_mixt}
z_\mathbf{k}^2 (w_\mathbf{k}^2) &= \frac{(\tilde{\varepsilon}_{1, \mathbf{k}} \pm \tilde{\varepsilon}_{2, \mathbf{k}})^2}{8 \tilde{\varepsilon}_{1, \mathbf{k}} \tilde{\varepsilon}_{2, \mathbf{k}}} \nonumber \\
& \times \left[ 1 - \frac{\tilde{E}_{1, \mathbf{k}}^2 - \tilde{E}_{2, \mathbf{k}}^2}{\sqrt{(\tilde{E}_{1, \mathbf{k}}^2 - \tilde{E}_{2, \mathbf{k}}^2)^2 + 16 g_{12}^2 \tilde{\varepsilon}_{1, \mathbf{k}} \tilde{\varepsilon}_{2, \mathbf{k}} n_{1, 0} n_{2, 0}}} \right]
\end{align}
where we have introduced the "gapped" kinetic energy $\tilde{\varepsilon}_{1, \mathbf{k}} = \varepsilon_{1, \mathbf{k}} + \Lambda_1 - g_{11}n_{1,0}$ with effective chemical potential $\Lambda_1 = 2 g_{11} n_1 + g_{12} n_2 - \mu_1 $ and ($1 \leftrightarrow 2$), as well as the single-component excitation spectrum $\tilde{E}_{i, \mathbf{k}} = \sqrt{(\varepsilon_{i, \mathbf{k}} + \Lambda_i)^2 - (g_{ii} n_{i, 0})^2}$. Henceforth, Eq.~\eqref{Eq:K_mix} reduces to the sum of two uncoupled Hamiltonian, which can be diagonalized by means of the Bogoliubov transformation Eq.~\eqref{Eq:Bogo_trans} applied to $(\hat{a}_{1, \mathbf{k}}^{'\dagger} , \hat{a}'_{1, \mathbf{k}})$ and $(\hat{a}_{2, \mathbf{k}}^{'\dagger} , \hat{a}'_{2, \mathbf{k}})$, respectively:
\begin{align}\label{Eq:Bogo_trans_mixt}
\begin{split}
\hat{a}'_{1, \mathbf{k}} &= u_{+,\mathbf{k}} \hat{\alpha}_\mathbf{k} + v_{+,\mathbf{-k}}^* \hat{\alpha}_\mathbf{-k}^\dagger \, , \\
\hat{a}'_{2, \mathbf{k}} &= u_{-,\mathbf{k}} \hat{\beta}_\mathbf{k} + v_{-,\mathbf{-k}}^* \hat{\beta}_\mathbf{-k}^\dagger \, .
\end{split}
\end{align}
The off-diagonal terms are found to vanish for the following values of the quasi-particle amplitudes:
\begin{align}\label{Eq:uv_mixt}
\begin{split}
u_{+,\mathbf{k}}, v_{+, \mathbf{k}} = \pm \frac{1}{2} \left( \sqrt{\frac{\varepsilon_{1, \mathbf{k}}}{E_{+,\mathbf{k}}}} \pm \sqrt{\frac{E_{+,\mathbf{k}}}{\varepsilon_{1, \mathbf{k}}}} \right) \, , \\
u_{-,\mathbf{k}}, v_{-, \mathbf{k}} = \pm \frac{1}{2} \left( \sqrt{\frac{\varepsilon_{2, \mathbf{k}}}{E_{-,\mathbf{k}}}} \pm \sqrt{\frac{E_{-,\mathbf{k}}}{\varepsilon_{2, \mathbf{k}}}} \right) \, .
\end{split}
\end{align}

Finally, the grand-canonical Hamiltonian in the diagonalized form is expressed as:
\begin{equation}
\hat{K} = \Omega_0 + \sum_{\mathrm{k} \neq 0} \left( \tilde{E}_{+ , \mathbf{k}} \hat{\alpha}_\mathbf{k}^\dagger \hat{\alpha}_\mathbf{k} + \tilde{E}_{- , \mathbf{k}} \hat{\beta}_\mathbf{k}^\dagger \hat{\beta}_\mathbf{k} \right)
\end{equation}
where $\hat{\alpha}_\mathbf{k}^\dagger$ (resp. $\hat{\beta}_\mathbf{k}^\dagger$) is the creation operator for the quasiparticles in the density (resp. spin) channel, obeying Bose statistics. The excitation spectrum of the system reads
\begin{align}\label{Eq:E_mix_gapped}
\tilde{E}_{\pm,\mathbf{k}} =& \frac{1}{\sqrt{2}} \left[\tilde{E}_{1, \mathbf{k}}^2 + \tilde{E}_{2, \mathbf{k}}^2 \right. \nonumber \\
& \left. \pm \sqrt{(\tilde{E}_{1, \mathbf{k}}^2 - \tilde{E}_{2, \mathbf{k}}^2)^2 + 16 g_{12}^2 \tilde{\varepsilon}_{1, \mathbf{k}} \tilde{\varepsilon}_{2, \mathbf{k}} n_{1, 0} n_{2, 0}} \right]^{1/2}
\end{align}
and the vacuum energy of Bogoliubov quasi-particles is given by
\begin{widetext}
\begin{align}\label{Eq:Omega0_mixt}
\Omega_0 =& \sum_{i=1, 2} \left[ \frac{g_{ii}}{2V} \left( N_{i, 0} ^2 -2 \tilde{N}_i^2 \right) - \mu_i N_{i, 0} \right] 
+ \frac{g_{12}}{V} N_{1, 0} N_{2, 0} - \frac{g_{12}}{V} \tilde{N}_1 \tilde{N}_2  \nonumber \\
&+ \frac{1}{2} \sum_\mathbf{k \neq 0} \left( \tilde{E}_{+, \mathbf{k}} - \varepsilon_{1, \mathbf{k}} - \Lambda_1 \right) + \left( \tilde{E}_{-, \mathbf{k}} - \varepsilon_{2, \mathbf{k}} - \Lambda_2 \right)
 + \frac{1}{\hbar^2 k^2} \sum_{i=1, 2} \left[ m_i (g_{ii} n_{i, 0})^2 + 2 m_R g_{12}^2 n_{1, 0} n_{2, 0} \right] \, .
\end{align}
\end{widetext}
In the above expression, the first terms correspond to the mean-field contribution, whereas the terms summed over the wave-vector account for the quantum fluctuations. In particular, the last terms arise from the renormalization of the coupling constants $g_{ii} \rightarrow g_{ii} [1 + g_{ii}/V \sum_\mathbf{k} m_i/(\hbar k)^2 ]$ and $g_{12} \rightarrow g_{12} [1 + g_{12}/V \sum_\mathbf{k} m_R/(\hbar k)^2 ]$~\cite{Larsen1963}.  
\par
The chemical potential in each component can be calculated in a similar fashion to the single-component case, by evaluating the saddle point equation $\partial \Omega / \partial n_{i,0} = 0$ and solving it perturbatively. We naturally find that $g_{ii}n_{i,0} = \Lambda_i + (\text{higher order terms})$, providing the gapless excitation spectrum $E_{i, \mathbf{k}} = \sqrt{\varepsilon_{i, \mathbf{k}}^2 + 2 \Lambda_i \varepsilon_{i, \mathbf{k}}}$ and $\tilde{\varepsilon}_{i, \mathbf{k}} \rightarrow \varepsilon_{i, \mathbf{k}}$ upon replacing $g_{ii}n_{i,0}$ by $\Lambda_i$ in Eq.~\eqref{Eq:E_mix_gapped}.
It is worth noticing that the excitation spectrum Eq.~\eqref{Eq:E_mix_gapped} can also be written as~\cite{Timmermans1998}
\begin{equation}
E_{\pm, \mathbf{k}} = \sqrt{\left( \frac{\nu_1^2 + \nu_2^2}{2} \right) \varepsilon_\mathbf{k}^2 + 2 \varepsilon_\mathbf{k} \Lambda_{\pm, \mathbf{k}}} \, ,
\end{equation}
where we have introduced the kinetic energy in terms of the reduced mass $\varepsilon_\mathbf{k} = \hbar^2 k^2 / (2m_R)$ and the inverse mass ratios $\nu_i = m_R / m_i$. The effective chemical potential $\Lambda_{\pm,\mathbf{k}}$ is therefore associated to the Bogoliubov density and spin sounds, and takes the following expression:
\begin{equation}
\Lambda_{\pm, \mathbf{k}} = \frac{1}{2} \left(  \nu_1 \Lambda_1 + \nu_2 \Lambda_2 \pm \Gamma_\mathbf{k} \right) \, ,
\end{equation}
\begin{equation}
\Gamma_\mathbf{k} = \sqrt{\left[ \frac{( \nu_1^2 - \nu_2^2 )}{2} \varepsilon_\mathbf{k} + (\nu_1 \Lambda_1 - \nu_2 \Lambda_2) \right]^2 + 4 \bar{g}^2 \nu_1 \Lambda_1 \nu_2 \Lambda_2 }
\end{equation}
where we have introduced the reduced coupling constant $\bar{g} = g_{12}/\sqrt{g_{11}g_{22}}$. Finally, the condensate density $n_{i,0} = n_i - V^{-1} \sum_{\mathbf{k} \neq 0} \langle \hat{a}^\dagger_{i, \mathbf{k}} \hat{a}_{i, \mathbf{k}} \rangle$ is given by ($1 \leftrightarrow 2$):
\begin{align}\label{Eq:n0_mixt}
n_{1,0} = n_1 - & \frac{1}{4V} \sum_\mathbf{k} \sum_\pm \left(1 \pm \frac{E_{1, \mathbf{k}}^2 - E_{2, \mathbf{k}}^2}{2 \varepsilon_\mathbf{k} \Gamma_\mathbf{k}} \right) \nonumber \\
&\times \left[ \frac{\varepsilon_{1, \mathbf{k}}^2+ E_{\pm, \mathbf{k}}^2}{2 \varepsilon_{1, \mathbf{k}} E_{\pm, \mathbf{k}}} \left( 2 f(E_{\pm, \mathbf{k}}) + 1 \right) - 1 \right] \,
\end{align}
and the chemical potential:
\begin{align}\label{Eq:mu_mixt}
\mu_1 =& g_{11} n_1 + g_{12} n_2 + \frac{g_{11}}{2V} \sum_\mathbf{k} \frac{1}{\varepsilon_\mathbf{k}} \left[ \frac{1}{\nu_1} \Lambda_1 + \bar{g} \Lambda_2 \right] \nonumber \\
&+ \frac{g_{11}}{4V} \sum_\mathbf{k} \sum_\pm \left( 1 \pm \frac{E_{1, \mathbf{k}}^2 - E_{2, \mathbf{k}}^2 + 4 \bar{g}^2 \Lambda_2 \varepsilon_{2, \mathbf{k}}}{2 \varepsilon_\mathbf{k} \Gamma_\mathbf{k}} \right) \nonumber \\
& \times \left[ \frac{\varepsilon_{1, \mathbf{k}}}{E_{\pm, \mathbf{k}}} \left( 2 f(E_{\pm, \mathbf{k}}) +1 \right) - 1 \right] \nonumber \\
& \, .
\end{align}
As in the single-component case, the above equation can be solved either self-consistently or perturbatively, the second-order expression being obtained by inserting the leading order result $\Lambda_i^0 = g_{ii} (n_i - n_{i,T}^0)$ for the effective chemical potential, where $n_{i,T}^0=\zeta (3/2)/\lambda_{i,T}^3$ is the ideal gas thermal density with species dependent thermal de Broglie wavelength $\lambda_{i,T} = \sqrt{2\pi\hbar^2/(m_i k_B T)}$.
\par
The equation of state for the equal mass case Eqs.~\eqref{Eq:nT_mix_sym}-\eqref{Eq:H_mix} in the main text are obtained from Eqs.~\eqref{Eq:n0_mixt} and~\eqref{Eq:mu_mixt} by putting $m_1 = m_2 = M$




\bibliography{bibPopov}

\begin{thebibliography}{60}%
\makeatletter
\providecommand \@ifxundefined [1]{%
 \@ifx{#1\undefined}
}%
\providecommand \@ifnum [1]{%
 \ifnum #1\expandafter \@firstoftwo
 \else \expandafter \@secondoftwo
 \fi
}%
\providecommand \@ifx [1]{%
 \ifx #1\expandafter \@firstoftwo
 \else \expandafter \@secondoftwo
 \fi
}%
\providecommand \natexlab [1]{#1}%
\providecommand \enquote  [1]{``#1''}%
\providecommand \bibnamefont  [1]{#1}%
\providecommand \bibfnamefont [1]{#1}%
\providecommand \citenamefont [1]{#1}%
\providecommand \href@noop [0]{\@secondoftwo}%
\providecommand \href [0]{\begingroup \@sanitize@url \@href}%
\providecommand \@href[1]{\@@startlink{#1}\@@href}%
\providecommand \@@href[1]{\endgroup#1\@@endlink}%
\providecommand \@sanitize@url [0]{\catcode `\\12\catcode `\$12\catcode
  `\&12\catcode `\#12\catcode `\^12\catcode `\_12\catcode `\%12\relax}%
\providecommand \@@startlink[1]{}%
\providecommand \@@endlink[0]{}%
\providecommand \url  [0]{\begingroup\@sanitize@url \@url }%
\providecommand \@url [1]{\endgroup\@href {#1}{\urlprefix }}%
\providecommand \urlprefix  [0]{URL }%
\providecommand \Eprint [0]{\href }%
\providecommand \doibase [0]{http://dx.doi.org/}%
\providecommand \selectlanguage [0]{\@gobble}%
\providecommand \bibinfo  [0]{\@secondoftwo}%
\providecommand \bibfield  [0]{\@secondoftwo}%
\providecommand \translation [1]{[#1]}%
\providecommand \BibitemOpen [0]{}%
\providecommand \bibitemStop [0]{}%
\providecommand \bibitemNoStop [0]{.\EOS\space}%
\providecommand \EOS [0]{\spacefactor3000\relax}%
\providecommand \BibitemShut  [1]{\csname bibitem#1\endcsname}%
\let\auto@bib@innerbib\@empty
\bibitem [{\citenamefont {Pitaevskii}\ and\ \citenamefont
  {Stringari}(2016)}]{book}%
  \BibitemOpen
  \bibfield  {author} {\bibinfo {author} {\bibfnamefont {L.}~\bibnamefont
  {Pitaevskii}}\ and\ \bibinfo {author} {\bibfnamefont {S.}~\bibnamefont
  {Stringari}},\ }\href@noop {} {\emph {\bibinfo {title} {Bose-Einstein
  Condensation and Superfluidity}}}\ (\bibinfo  {publisher} {Oxford University
  Press},\ \bibinfo {year} {2016})\BibitemShut {NoStop}%
\bibitem [{\citenamefont {Huang}(1963)}]{Huang}%
  \BibitemOpen
  \bibfield  {author} {\bibinfo {author} {\bibfnamefont {K.}~\bibnamefont
  {Huang}},\ }\href@noop {} {\emph {\bibinfo {title} {{Statistical
  Mechanics}}}}\ (\bibinfo  {publisher} {Wiley},\ \bibinfo {address} {New
  York},\ \bibinfo {year} {1963})\BibitemShut {NoStop}%
\bibitem [{\citenamefont {Lipa}\ \emph {et~al.}(2003)\citenamefont {Lipa},
  \citenamefont {Nissen}, \citenamefont {Stricker}, \citenamefont {Swanson},\
  and\ \citenamefont {Chui}}]{Lipa2003}%
  \BibitemOpen
  \bibfield  {author} {\bibinfo {author} {\bibfnamefont {J.~A.}\ \bibnamefont
  {Lipa}}, \bibinfo {author} {\bibfnamefont {J.~A.}\ \bibnamefont {Nissen}},
  \bibinfo {author} {\bibfnamefont {D.~A.}\ \bibnamefont {Stricker}}, \bibinfo
  {author} {\bibfnamefont {D.~R.}\ \bibnamefont {Swanson}}, \ and\ \bibinfo
  {author} {\bibfnamefont {T.~C.~P.}\ \bibnamefont {Chui}},\ }\href {\doibase
  10.1103/PhysRevB.68.174518} {\bibfield  {journal} {\bibinfo  {journal} {Phys.
  Rev. B}\ }\textbf {\bibinfo {volume} {68}},\ \bibinfo {pages} {174518}
  (\bibinfo {year} {2003})}\BibitemShut {NoStop}%
\bibitem [{\citenamefont {Ku}\ \emph {et~al.}(2012)\citenamefont {Ku},
  \citenamefont {Sommer}, \citenamefont {Cheuk},\ and\ \citenamefont
  {Zwierlein}}]{Ku2012}%
  \BibitemOpen
  \bibfield  {author} {\bibinfo {author} {\bibfnamefont {M.~J.~H.}\
  \bibnamefont {Ku}}, \bibinfo {author} {\bibfnamefont {A.~T.}\ \bibnamefont
  {Sommer}}, \bibinfo {author} {\bibfnamefont {L.~W.}\ \bibnamefont {Cheuk}}, \
  and\ \bibinfo {author} {\bibfnamefont {M.~W.}\ \bibnamefont {Zwierlein}},\
  }\href {\doibase 10.1126/science.1214987} {\bibfield  {journal} {\bibinfo
  {journal} {Science}\ }\textbf {\bibinfo {volume} {335}},\ \bibinfo {pages}
  {563} (\bibinfo {year} {2012})}\BibitemShut {NoStop}%
\bibitem [{\citenamefont {Desbuquois}\ \emph {et~al.}(2014)\citenamefont
  {Desbuquois}, \citenamefont {Yefsah}, \citenamefont {Chomaz}, \citenamefont
  {Weitenberg}, \citenamefont {Corman}, \citenamefont {Nascimb\`ene},\ and\
  \citenamefont {Dalibard}}]{Desbuquois2014}%
  \BibitemOpen
  \bibfield  {author} {\bibinfo {author} {\bibfnamefont {R.}~\bibnamefont
  {Desbuquois}}, \bibinfo {author} {\bibfnamefont {T.}~\bibnamefont {Yefsah}},
  \bibinfo {author} {\bibfnamefont {L.}~\bibnamefont {Chomaz}}, \bibinfo
  {author} {\bibfnamefont {C.}~\bibnamefont {Weitenberg}}, \bibinfo {author}
  {\bibfnamefont {L.}~\bibnamefont {Corman}}, \bibinfo {author} {\bibfnamefont
  {S.}~\bibnamefont {Nascimb\`ene}}, \ and\ \bibinfo {author} {\bibfnamefont
  {J.}~\bibnamefont {Dalibard}},\ }\href {\doibase
  10.1103/PhysRevLett.113.020404} {\bibfield  {journal} {\bibinfo  {journal}
  {Phys. Rev. Lett.}\ }\textbf {\bibinfo {volume} {113}},\ \bibinfo {pages}
  {020404} (\bibinfo {year} {2014})}\BibitemShut {NoStop}%
\bibitem [{\citenamefont {Yefsah}\ \emph {et~al.}(2011)\citenamefont {Yefsah},
  \citenamefont {Desbuquois}, \citenamefont {Chomaz}, \citenamefont
  {G\"unter},\ and\ \citenamefont {Dalibard}}]{Yefsah2011}%
  \BibitemOpen
  \bibfield  {author} {\bibinfo {author} {\bibfnamefont {T.}~\bibnamefont
  {Yefsah}}, \bibinfo {author} {\bibfnamefont {R.}~\bibnamefont {Desbuquois}},
  \bibinfo {author} {\bibfnamefont {L.}~\bibnamefont {Chomaz}}, \bibinfo
  {author} {\bibfnamefont {K.~J.}\ \bibnamefont {G\"unter}}, \ and\ \bibinfo
  {author} {\bibfnamefont {J.}~\bibnamefont {Dalibard}},\ }\href {\doibase
  10.1103/PhysRevLett.107.130401} {\bibfield  {journal} {\bibinfo  {journal}
  {Phys. Rev. Lett.}\ }\textbf {\bibinfo {volume} {107}},\ \bibinfo {pages}
  {130401} (\bibinfo {year} {2011})}\BibitemShut {NoStop}%
\bibitem [{\citenamefont {Boettcher}\ \emph {et~al.}(2016)\citenamefont
  {Boettcher}, \citenamefont {Bayha}, \citenamefont {Kedar}, \citenamefont
  {Murthy}, \citenamefont {Neidig}, \citenamefont {Ries}, \citenamefont {Wenz},
  \citenamefont {Z\"urn}, \citenamefont {Jochim},\ and\ \citenamefont
  {Enss}}]{Boettcher2016}%
  \BibitemOpen
  \bibfield  {author} {\bibinfo {author} {\bibfnamefont {I.}~\bibnamefont
  {Boettcher}}, \bibinfo {author} {\bibfnamefont {L.}~\bibnamefont {Bayha}},
  \bibinfo {author} {\bibfnamefont {D.}~\bibnamefont {Kedar}}, \bibinfo
  {author} {\bibfnamefont {P.~A.}\ \bibnamefont {Murthy}}, \bibinfo {author}
  {\bibfnamefont {M.}~\bibnamefont {Neidig}}, \bibinfo {author} {\bibfnamefont
  {M.~G.}\ \bibnamefont {Ries}}, \bibinfo {author} {\bibfnamefont {A.~N.}\
  \bibnamefont {Wenz}}, \bibinfo {author} {\bibfnamefont {G.}~\bibnamefont
  {Z\"urn}}, \bibinfo {author} {\bibfnamefont {S.}~\bibnamefont {Jochim}}, \
  and\ \bibinfo {author} {\bibfnamefont {T.}~\bibnamefont {Enss}},\ }\href
  {\doibase 10.1103/PhysRevLett.116.045303} {\bibfield  {journal} {\bibinfo
  {journal} {Phys. Rev. Lett.}\ }\textbf {\bibinfo {volume} {116}},\ \bibinfo
  {pages} {045303} (\bibinfo {year} {2016})}\BibitemShut {NoStop}%
\bibitem [{\citenamefont {Fenech}\ \emph {et~al.}(2016)\citenamefont {Fenech},
  \citenamefont {Dyke}, \citenamefont {Peppler}, \citenamefont {Lingham},
  \citenamefont {Hoinka}, \citenamefont {Hu},\ and\ \citenamefont
  {Vale}}]{Fenech2016}%
  \BibitemOpen
  \bibfield  {author} {\bibinfo {author} {\bibfnamefont {K.}~\bibnamefont
  {Fenech}}, \bibinfo {author} {\bibfnamefont {P.}~\bibnamefont {Dyke}},
  \bibinfo {author} {\bibfnamefont {T.}~\bibnamefont {Peppler}}, \bibinfo
  {author} {\bibfnamefont {M.~G.}\ \bibnamefont {Lingham}}, \bibinfo {author}
  {\bibfnamefont {S.}~\bibnamefont {Hoinka}}, \bibinfo {author} {\bibfnamefont
  {H.}~\bibnamefont {Hu}}, \ and\ \bibinfo {author} {\bibfnamefont {C.~J.}\
  \bibnamefont {Vale}},\ }\href {\doibase 10.1103/PhysRevLett.116.045302}
  {\bibfield  {journal} {\bibinfo  {journal} {Phys. Rev. Lett.}\ }\textbf
  {\bibinfo {volume} {116}},\ \bibinfo {pages} {045302} (\bibinfo {year}
  {2016})}\BibitemShut {NoStop}%
\bibitem [{\citenamefont {Makhalov}\ \emph {et~al.}(2014)\citenamefont
  {Makhalov}, \citenamefont {Martiyanov},\ and\ \citenamefont
  {Turlapov}}]{Makhalov2014}%
  \BibitemOpen
  \bibfield  {author} {\bibinfo {author} {\bibfnamefont {V.}~\bibnamefont
  {Makhalov}}, \bibinfo {author} {\bibfnamefont {K.}~\bibnamefont
  {Martiyanov}}, \ and\ \bibinfo {author} {\bibfnamefont {A.}~\bibnamefont
  {Turlapov}},\ }\href {\doibase 10.1103/PhysRevLett.112.045301} {\bibfield
  {journal} {\bibinfo  {journal} {Phys. Rev. Lett.}\ }\textbf {\bibinfo
  {volume} {112}},\ \bibinfo {pages} {045301} (\bibinfo {year}
  {2014})}\BibitemShut {NoStop}%
\bibitem [{\citenamefont {Navon}\ \emph {et~al.}(2011)\citenamefont {Navon},
  \citenamefont {Piatecki}, \citenamefont {G{\"{u}}nter}, \citenamefont {Rem},
  \citenamefont {Nguyen}, \citenamefont {Chevy}, \citenamefont {Krauth},\ and\
  \citenamefont {Salomon}}]{Navon2011}%
  \BibitemOpen
  \bibfield  {author} {\bibinfo {author} {\bibfnamefont {N.}~\bibnamefont
  {Navon}}, \bibinfo {author} {\bibfnamefont {S.}~\bibnamefont {Piatecki}},
  \bibinfo {author} {\bibfnamefont {K.}~\bibnamefont {G{\"{u}}nter}}, \bibinfo
  {author} {\bibfnamefont {B.}~\bibnamefont {Rem}}, \bibinfo {author}
  {\bibfnamefont {T.~C.}\ \bibnamefont {Nguyen}}, \bibinfo {author}
  {\bibfnamefont {F.}~\bibnamefont {Chevy}}, \bibinfo {author} {\bibfnamefont
  {W.}~\bibnamefont {Krauth}}, \ and\ \bibinfo {author} {\bibfnamefont
  {C.}~\bibnamefont {Salomon}},\ }\href {\doibase
  10.1103/PhysRevLett.107.135301} {\bibfield  {journal} {\bibinfo  {journal}
  {Phys. Rev. Lett.}\ }\textbf {\bibinfo {volume} {107}},\ \bibinfo {pages}
  {135301} (\bibinfo {year} {2011})}\BibitemShut {NoStop}%
\bibitem [{\citenamefont {Nascimb{\`{e}}ne}\ \emph {et~al.}(2010)\citenamefont
  {Nascimb{\`{e}}ne}, \citenamefont {Navon}, \citenamefont {Chevy},\ and\
  \citenamefont {Salomon}}]{Nascimbene2010b}%
  \BibitemOpen
  \bibfield  {author} {\bibinfo {author} {\bibfnamefont {S.}~\bibnamefont
  {Nascimb{\`{e}}ne}}, \bibinfo {author} {\bibfnamefont {N.}~\bibnamefont
  {Navon}}, \bibinfo {author} {\bibfnamefont {F.}~\bibnamefont {Chevy}}, \ and\
  \bibinfo {author} {\bibfnamefont {C.}~\bibnamefont {Salomon}},\ }\href
  {\doibase 10.1088/1367-2630/12/10/103026} {\bibfield  {journal} {\bibinfo
  {journal} {New J. Phys.}\ }\textbf {\bibinfo {volume} {12}},\ \bibinfo
  {pages} {103026} (\bibinfo {year} {2010})}\BibitemShut {NoStop}%
\bibitem [{\citenamefont {Mordini}\ \emph
  {et~al.}(2020{\natexlab{a}})\citenamefont {Mordini}, \citenamefont
  {Trypogeorgos}, \citenamefont {Farolfi}, \citenamefont {Wolswijk},
  \citenamefont {Stringari}, \citenamefont {Lamporesi},\ and\ \citenamefont
  {Ferrari}}]{Mordini2020}%
  \BibitemOpen
  \bibfield  {author} {\bibinfo {author} {\bibfnamefont {C.}~\bibnamefont
  {Mordini}}, \bibinfo {author} {\bibfnamefont {D.}~\bibnamefont
  {Trypogeorgos}}, \bibinfo {author} {\bibfnamefont {A.}~\bibnamefont
  {Farolfi}}, \bibinfo {author} {\bibfnamefont {L.}~\bibnamefont {Wolswijk}},
  \bibinfo {author} {\bibfnamefont {S.}~\bibnamefont {Stringari}}, \bibinfo
  {author} {\bibfnamefont {G.}~\bibnamefont {Lamporesi}}, \ and\ \bibinfo
  {author} {\bibfnamefont {G.}~\bibnamefont {Ferrari}},\ }\href {\doibase
  10.1103/PhysRevLett.125.150404} {\bibfield  {journal} {\bibinfo  {journal}
  {Phys. Rev. Lett.}\ }\textbf {\bibinfo {volume} {125}},\ \bibinfo {pages}
  {150404} (\bibinfo {year} {2020}{\natexlab{a}})}\BibitemShut {NoStop}%
\bibitem [{\citenamefont {Dalfovo}\ \emph {et~al.}(1999)\citenamefont
  {Dalfovo}, \citenamefont {Giorgini}, \citenamefont {Pitaevskii},\ and\
  \citenamefont {Stringari}}]{Dalfovo1999}%
  \BibitemOpen
  \bibfield  {author} {\bibinfo {author} {\bibfnamefont {F.}~\bibnamefont
  {Dalfovo}}, \bibinfo {author} {\bibfnamefont {S.}~\bibnamefont {Giorgini}},
  \bibinfo {author} {\bibfnamefont {L.~P.}\ \bibnamefont {Pitaevskii}}, \ and\
  \bibinfo {author} {\bibfnamefont {S.}~\bibnamefont {Stringari}},\ }\href
  {\doibase 10.1103/RevModPhys.71.463} {\bibfield  {journal} {\bibinfo
  {journal} {Rev. Mod. Phys.}\ }\textbf {\bibinfo {volume} {71}},\ \bibinfo
  {pages} {463} (\bibinfo {year} {1999})}\BibitemShut {NoStop}%
\bibitem [{\citenamefont {Gerbier}\ \emph {et~al.}(2004)\citenamefont
  {Gerbier}, \citenamefont {Thywissen}, \citenamefont {Richard}, \citenamefont
  {Hugbart}, \citenamefont {Bouyer},\ and\ \citenamefont
  {Aspect}}]{Gerbier2004}%
  \BibitemOpen
  \bibfield  {author} {\bibinfo {author} {\bibfnamefont {F.}~\bibnamefont
  {Gerbier}}, \bibinfo {author} {\bibfnamefont {J.~H.}\ \bibnamefont
  {Thywissen}}, \bibinfo {author} {\bibfnamefont {S.}~\bibnamefont {Richard}},
  \bibinfo {author} {\bibfnamefont {M.}~\bibnamefont {Hugbart}}, \bibinfo
  {author} {\bibfnamefont {P.}~\bibnamefont {Bouyer}}, \ and\ \bibinfo {author}
  {\bibfnamefont {A.}~\bibnamefont {Aspect}},\ }\href {\doibase
  10.1103/PhysRevA.70.013607} {\bibfield  {journal} {\bibinfo  {journal} {Phys.
  Rev. A}\ }\textbf {\bibinfo {volume} {70}},\ \bibinfo {pages} {013607}
  (\bibinfo {year} {2004})}\BibitemShut {NoStop}%
\bibitem [{\citenamefont {Smith}\ \emph {et~al.}(2011)\citenamefont {Smith},
  \citenamefont {Campbell}, \citenamefont {Tammuz},\ and\ \citenamefont
  {Hadzibabic}}]{Smith2011}%
  \BibitemOpen
  \bibfield  {author} {\bibinfo {author} {\bibfnamefont {R.~P.}\ \bibnamefont
  {Smith}}, \bibinfo {author} {\bibfnamefont {R.~L.~D.}\ \bibnamefont
  {Campbell}}, \bibinfo {author} {\bibfnamefont {N.}~\bibnamefont {Tammuz}}, \
  and\ \bibinfo {author} {\bibfnamefont {Z.}~\bibnamefont {Hadzibabic}},\
  }\href {\doibase 10.1103/PhysRevLett.106.250403} {\bibfield  {journal}
  {\bibinfo  {journal} {Phys. Rev. Lett.}\ }\textbf {\bibinfo {volume} {106}},\
  \bibinfo {pages} {250403} (\bibinfo {year} {2011})}\BibitemShut {NoStop}%
\bibitem [{\citenamefont {Gaunt}\ \emph {et~al.}(2013)\citenamefont {Gaunt},
  \citenamefont {Schmidutz}, \citenamefont {Gotlibovych}, \citenamefont
  {Smith},\ and\ \citenamefont {Hadzibabic}}]{Gaunt2013}%
  \BibitemOpen
  \bibfield  {author} {\bibinfo {author} {\bibfnamefont {A.~L.}\ \bibnamefont
  {Gaunt}}, \bibinfo {author} {\bibfnamefont {T.~F.}\ \bibnamefont
  {Schmidutz}}, \bibinfo {author} {\bibfnamefont {I.}~\bibnamefont
  {Gotlibovych}}, \bibinfo {author} {\bibfnamefont {R.~P.}\ \bibnamefont
  {Smith}}, \ and\ \bibinfo {author} {\bibfnamefont {Z.}~\bibnamefont
  {Hadzibabic}},\ }\href {\doibase 10.1103/PhysRevLett.110.200406} {\bibfield
  {journal} {\bibinfo  {journal} {Phys. Rev. Lett.}\ }\textbf {\bibinfo
  {volume} {110}},\ \bibinfo {pages} {200406} (\bibinfo {year}
  {2013})}\BibitemShut {NoStop}%
\bibitem [{\citenamefont {Lopes}\ \emph {et~al.}(2017)\citenamefont {Lopes},
  \citenamefont {Eigen}, \citenamefont {Navon}, \citenamefont {Cl{\'{e}}ment},
  \citenamefont {Smith},\ and\ \citenamefont {Hadzibabic}}]{Lopes2017}%
  \BibitemOpen
  \bibfield  {author} {\bibinfo {author} {\bibfnamefont {R.}~\bibnamefont
  {Lopes}}, \bibinfo {author} {\bibfnamefont {C.}~\bibnamefont {Eigen}},
  \bibinfo {author} {\bibfnamefont {N.}~\bibnamefont {Navon}}, \bibinfo
  {author} {\bibfnamefont {D.}~\bibnamefont {Cl{\'{e}}ment}}, \bibinfo {author}
  {\bibfnamefont {R.~P.}\ \bibnamefont {Smith}}, \ and\ \bibinfo {author}
  {\bibfnamefont {Z.}~\bibnamefont {Hadzibabic}},\ }\href {\doibase
  10.1103/PhysRevLett.119.190404} {\bibfield  {journal} {\bibinfo  {journal}
  {Phys. Rev. Lett.}\ }\textbf {\bibinfo {volume} {119}},\ \bibinfo {pages}
  {190404} (\bibinfo {year} {2017})}\BibitemShut {NoStop}%
\bibitem [{\citenamefont {Chomaz}\ \emph {et~al.}(2015)\citenamefont {Chomaz},
  \citenamefont {Corman}, \citenamefont {Bienaim{\'{e}}}, \citenamefont
  {Desbuquois}, \citenamefont {Weitenberg}, \citenamefont {Nascimb{\`{e}}ne},
  \citenamefont {Beugnon},\ and\ \citenamefont {Dalibard}}]{Chomaz2015}%
  \BibitemOpen
  \bibfield  {author} {\bibinfo {author} {\bibfnamefont {L.}~\bibnamefont
  {Chomaz}}, \bibinfo {author} {\bibfnamefont {L.}~\bibnamefont {Corman}},
  \bibinfo {author} {\bibfnamefont {T.}~\bibnamefont {Bienaim{\'{e}}}},
  \bibinfo {author} {\bibfnamefont {R.}~\bibnamefont {Desbuquois}}, \bibinfo
  {author} {\bibfnamefont {C.}~\bibnamefont {Weitenberg}}, \bibinfo {author}
  {\bibfnamefont {S.}~\bibnamefont {Nascimb{\`{e}}ne}}, \bibinfo {author}
  {\bibfnamefont {J.}~\bibnamefont {Beugnon}}, \ and\ \bibinfo {author}
  {\bibfnamefont {J.}~\bibnamefont {Dalibard}},\ }\href {\doibase
  10.1038/ncomms7162} {\bibfield  {journal} {\bibinfo  {journal} {Nat.
  Commun.}\ }\textbf {\bibinfo {volume} {6}},\ \bibinfo {pages} {6162}
  (\bibinfo {year} {2015})}\BibitemShut {NoStop}%
\bibitem [{\citenamefont {Mukherjee}\ \emph {et~al.}(2017)\citenamefont
  {Mukherjee}, \citenamefont {Yan}, \citenamefont {Patel}, \citenamefont
  {Hadzibabic}, \citenamefont {Yefsah}, \citenamefont {Struck},\ and\
  \citenamefont {Zwierlein}}]{Mukherjee2017}%
  \BibitemOpen
  \bibfield  {author} {\bibinfo {author} {\bibfnamefont {B.}~\bibnamefont
  {Mukherjee}}, \bibinfo {author} {\bibfnamefont {Z.}~\bibnamefont {Yan}},
  \bibinfo {author} {\bibfnamefont {P.~B.}\ \bibnamefont {Patel}}, \bibinfo
  {author} {\bibfnamefont {Z.}~\bibnamefont {Hadzibabic}}, \bibinfo {author}
  {\bibfnamefont {T.}~\bibnamefont {Yefsah}}, \bibinfo {author} {\bibfnamefont
  {J.}~\bibnamefont {Struck}}, \ and\ \bibinfo {author} {\bibfnamefont {M.~W.}\
  \bibnamefont {Zwierlein}},\ }\href {\doibase 10.1103/PhysRevLett.118.123401}
  {\bibfield  {journal} {\bibinfo  {journal} {Phys. Rev. Lett.}\ }\textbf
  {\bibinfo {volume} {118}},\ \bibinfo {pages} {123401} (\bibinfo {year}
  {2017})}\BibitemShut {NoStop}%
\bibitem [{\citenamefont {Ramanathan}\ \emph {et~al.}(2012)\citenamefont
  {Ramanathan}, \citenamefont {Muniz}, \citenamefont {Wright}, \citenamefont
  {Anderson}, \citenamefont {Phillips}, \citenamefont {Helmerson},\ and\
  \citenamefont {Campbell}}]{Ramanathan2012}%
  \BibitemOpen
  \bibfield  {author} {\bibinfo {author} {\bibfnamefont {A.}~\bibnamefont
  {Ramanathan}}, \bibinfo {author} {\bibfnamefont {S.~R.}\ \bibnamefont
  {Muniz}}, \bibinfo {author} {\bibfnamefont {K.~C.}\ \bibnamefont {Wright}},
  \bibinfo {author} {\bibfnamefont {R.~P.}\ \bibnamefont {Anderson}}, \bibinfo
  {author} {\bibfnamefont {W.~D.}\ \bibnamefont {Phillips}}, \bibinfo {author}
  {\bibfnamefont {K.}~\bibnamefont {Helmerson}}, \ and\ \bibinfo {author}
  {\bibfnamefont {G.~K.}\ \bibnamefont {Campbell}},\ }\href {\doibase
  10.1063/1.4747163} {\bibfield  {journal} {\bibinfo  {journal} {Rev. Sci.
  Instrum}\ }\textbf {\bibinfo {volume} {83}},\ \bibinfo {pages} {83119}
  (\bibinfo {year} {2012})}\BibitemShut {NoStop}%
\bibitem [{\citenamefont {Mordini}\ \emph
  {et~al.}(2020{\natexlab{b}})\citenamefont {Mordini}, \citenamefont
  {Trypogeorgos}, \citenamefont {Wolswijk}, \citenamefont {Lamporesi},\ and\
  \citenamefont {Ferrari}}]{Mordini2020b}%
  \BibitemOpen
  \bibfield  {author} {\bibinfo {author} {\bibfnamefont {C.}~\bibnamefont
  {Mordini}}, \bibinfo {author} {\bibfnamefont {D.}~\bibnamefont
  {Trypogeorgos}}, \bibinfo {author} {\bibfnamefont {L.}~\bibnamefont
  {Wolswijk}}, \bibinfo {author} {\bibfnamefont {G.}~\bibnamefont {Lamporesi}},
  \ and\ \bibinfo {author} {\bibfnamefont {G.}~\bibnamefont {Ferrari}},\ }\href
  {\doibase 10.1364/oe.397567} {\bibfield  {journal} {\bibinfo  {journal} {Opt.
  Express}\ }\textbf {\bibinfo {volume} {28}},\ \bibinfo {pages} {29408}
  (\bibinfo {year} {2020}{\natexlab{b}})}\BibitemShut {NoStop}%
\bibitem [{\citenamefont {Lin}\ \emph {et~al.}(2011)\citenamefont {Lin},
  \citenamefont {Jim{\'{e}}nez-Garc{\'{i}}a},\ and\ \citenamefont
  {Spielman}}]{Lin2011}%
  \BibitemOpen
  \bibfield  {author} {\bibinfo {author} {\bibfnamefont {Y.-J.}\ \bibnamefont
  {Lin}}, \bibinfo {author} {\bibfnamefont {K.}~\bibnamefont
  {Jim{\'{e}}nez-Garc{\'{i}}a}}, \ and\ \bibinfo {author} {\bibfnamefont
  {I.~B.}\ \bibnamefont {Spielman}},\ }\href {\doibase 10.1038/nature09887}
  {\bibfield  {journal} {\bibinfo  {journal} {Nature}\ }\textbf {\bibinfo
  {volume} {471}},\ \bibinfo {pages} {83} (\bibinfo {year} {2011})}\BibitemShut
  {NoStop}%
\bibitem [{\citenamefont {Zhang}\ \emph {et~al.}(2012)\citenamefont {Zhang},
  \citenamefont {Ji}, \citenamefont {Chen}, \citenamefont {Zhang},
  \citenamefont {Du}, \citenamefont {Yan}, \citenamefont {Pan}, \citenamefont
  {Zhao}, \citenamefont {Deng}, \citenamefont {Zhai}, \citenamefont {Chen},\
  and\ \citenamefont {Pan}}]{Zhang2012}%
  \BibitemOpen
  \bibfield  {author} {\bibinfo {author} {\bibfnamefont {J.-Y.}\ \bibnamefont
  {Zhang}}, \bibinfo {author} {\bibfnamefont {S.-C.}\ \bibnamefont {Ji}},
  \bibinfo {author} {\bibfnamefont {Z.}~\bibnamefont {Chen}}, \bibinfo {author}
  {\bibfnamefont {L.}~\bibnamefont {Zhang}}, \bibinfo {author} {\bibfnamefont
  {Z.-D.}\ \bibnamefont {Du}}, \bibinfo {author} {\bibfnamefont
  {B.}~\bibnamefont {Yan}}, \bibinfo {author} {\bibfnamefont {G.-S.}\
  \bibnamefont {Pan}}, \bibinfo {author} {\bibfnamefont {B.}~\bibnamefont
  {Zhao}}, \bibinfo {author} {\bibfnamefont {Y.-J.}\ \bibnamefont {Deng}},
  \bibinfo {author} {\bibfnamefont {H.}~\bibnamefont {Zhai}}, \bibinfo {author}
  {\bibfnamefont {S.}~\bibnamefont {Chen}}, \ and\ \bibinfo {author}
  {\bibfnamefont {J.-W.}\ \bibnamefont {Pan}},\ }\href {\doibase
  10.1103/PhysRevLett.109.115301} {\bibfield  {journal} {\bibinfo  {journal}
  {Phys. Rev. Lett.}\ }\textbf {\bibinfo {volume} {109}},\ \bibinfo {pages}
  {115301} (\bibinfo {year} {2012})}\BibitemShut {NoStop}%
\bibitem [{\citenamefont {Cabrera}\ \emph {et~al.}(2018)\citenamefont
  {Cabrera}, \citenamefont {Tanzi}, \citenamefont {Sanz}, \citenamefont
  {Naylor}, \citenamefont {Thomas}, \citenamefont {Cheiney},\ and\
  \citenamefont {Tarruell}}]{Cabrera2018}%
  \BibitemOpen
  \bibfield  {author} {\bibinfo {author} {\bibfnamefont {C.~R.}\ \bibnamefont
  {Cabrera}}, \bibinfo {author} {\bibfnamefont {L.}~\bibnamefont {Tanzi}},
  \bibinfo {author} {\bibfnamefont {J.}~\bibnamefont {Sanz}}, \bibinfo {author}
  {\bibfnamefont {B.}~\bibnamefont {Naylor}}, \bibinfo {author} {\bibfnamefont
  {P.}~\bibnamefont {Thomas}}, \bibinfo {author} {\bibfnamefont
  {P.}~\bibnamefont {Cheiney}}, \ and\ \bibinfo {author} {\bibfnamefont
  {L.}~\bibnamefont {Tarruell}},\ }\href {\doibase 10.1126/science.aao5686}
  {\bibfield  {journal} {\bibinfo  {journal} {Science}\ }\textbf {\bibinfo
  {volume} {359}},\ \bibinfo {pages} {301} (\bibinfo {year}
  {2018})}\BibitemShut {NoStop}%
\bibitem [{\citenamefont {Semeghini}\ \emph {et~al.}(2018)\citenamefont
  {Semeghini}, \citenamefont {Ferioli}, \citenamefont {Masi}, \citenamefont
  {Mazzinghi}, \citenamefont {Wolswijk}, \citenamefont {Minardi}, \citenamefont
  {Modugno}, \citenamefont {Modugno}, \citenamefont {Inguscio},\ and\
  \citenamefont {Fattori}}]{Semeghini2018}%
  \BibitemOpen
  \bibfield  {author} {\bibinfo {author} {\bibfnamefont {G.}~\bibnamefont
  {Semeghini}}, \bibinfo {author} {\bibfnamefont {G.}~\bibnamefont {Ferioli}},
  \bibinfo {author} {\bibfnamefont {L.}~\bibnamefont {Masi}}, \bibinfo {author}
  {\bibfnamefont {C.}~\bibnamefont {Mazzinghi}}, \bibinfo {author}
  {\bibfnamefont {L.}~\bibnamefont {Wolswijk}}, \bibinfo {author}
  {\bibfnamefont {F.}~\bibnamefont {Minardi}}, \bibinfo {author} {\bibfnamefont
  {M.}~\bibnamefont {Modugno}}, \bibinfo {author} {\bibfnamefont
  {G.}~\bibnamefont {Modugno}}, \bibinfo {author} {\bibfnamefont
  {M.}~\bibnamefont {Inguscio}}, \ and\ \bibinfo {author} {\bibfnamefont
  {M.}~\bibnamefont {Fattori}},\ }\href {\doibase
  10.1103/PhysRevLett.120.235301} {\bibfield  {journal} {\bibinfo  {journal}
  {Phys. Rev. Lett.}\ }\textbf {\bibinfo {volume} {120}},\ \bibinfo {pages}
  {235301} (\bibinfo {year} {2018})}\BibitemShut {NoStop}%
\bibitem [{\citenamefont {Proukakis}\ and\ \citenamefont
  {Jackson}(2008)}]{Proukakis2008}%
  \BibitemOpen
  \bibfield  {author} {\bibinfo {author} {\bibfnamefont {N.~P.}\ \bibnamefont
  {Proukakis}}\ and\ \bibinfo {author} {\bibfnamefont {B.}~\bibnamefont
  {Jackson}},\ }\href {\doibase 10.1088/0953-4075/41/20/203002} {\bibfield
  {journal} {\bibinfo  {journal} {J. Phys. B: At. Mol. Opt. Phys.}\ }\textbf
  {\bibinfo {volume} {41}},\ \bibinfo {pages} {203002} (\bibinfo {year}
  {2008})}\BibitemShut {NoStop}%
\bibitem [{\citenamefont {Griffin}(1996)}]{Griffin1996}%
  \BibitemOpen
  \bibfield  {author} {\bibinfo {author} {\bibfnamefont {A.}~\bibnamefont
  {Griffin}},\ }\href {\doibase 10.1103/PhysRevB.53.9341} {\bibfield  {journal}
  {\bibinfo  {journal} {Phys. Rev. B}\ }\textbf {\bibinfo {volume} {53}},\
  \bibinfo {pages} {9341} (\bibinfo {year} {1996})}\BibitemShut {NoStop}%
\bibitem [{\citenamefont {Takano}(1961)}]{Takano1961}%
  \BibitemOpen
  \bibfield  {author} {\bibinfo {author} {\bibfnamefont {F.}~\bibnamefont
  {Takano}},\ }\href {\doibase 10.1103/PhysRev.123.699} {\bibfield  {journal}
  {\bibinfo  {journal} {Physical Review}\ }\textbf {\bibinfo {volume} {123}},\
  \bibinfo {pages} {699} (\bibinfo {year} {1961})}\BibitemShut {NoStop}%
\bibitem [{\citenamefont {Proukakis}\ \emph {et~al.}(1998)\citenamefont
  {Proukakis}, \citenamefont {Morgan}, \citenamefont {Choi},\ and\
  \citenamefont {Burnett}}]{Proukakis1998}%
  \BibitemOpen
  \bibfield  {author} {\bibinfo {author} {\bibfnamefont {N.~P.}\ \bibnamefont
  {Proukakis}}, \bibinfo {author} {\bibfnamefont {S.~A.}\ \bibnamefont
  {Morgan}}, \bibinfo {author} {\bibfnamefont {S.}~\bibnamefont {Choi}}, \ and\
  \bibinfo {author} {\bibfnamefont {K.}~\bibnamefont {Burnett}},\ }\href
  {\doibase 10.1103/PhysRevA.58.2435} {\bibfield  {journal} {\bibinfo
  {journal} {Phys. Rev. A}\ }\textbf {\bibinfo {volume} {58}},\ \bibinfo
  {pages} {2435} (\bibinfo {year} {1998})}\BibitemShut {NoStop}%
\bibitem [{\citenamefont {Shi}\ and\ \citenamefont {Griffin}(1998)}]{Shi1998}%
  \BibitemOpen
  \bibfield  {author} {\bibinfo {author} {\bibfnamefont {H.}~\bibnamefont
  {Shi}}\ and\ \bibinfo {author} {\bibfnamefont {A.}~\bibnamefont {Griffin}},\
  }\href {\doibase https://doi.org/10.1016/S0370-1573(98)00015-5} {\bibfield
  {journal} {\bibinfo  {journal} {Physics Reports}\ }\textbf {\bibinfo {volume}
  {304}},\ \bibinfo {pages} {1 } (\bibinfo {year} {1998})}\BibitemShut
  {NoStop}%
\bibitem [{\citenamefont {Popov}(1983)}]{Popov1983}%
  \BibitemOpen
  \bibfield  {author} {\bibinfo {author} {\bibfnamefont {V.}~\bibnamefont
  {Popov}},\ }\href@noop {} {\emph {\bibinfo {title} {Functional Integrals in
  Quantum Field Theory and Statistical Physics}}}\ (\bibinfo  {publisher} {D.
  Reidel Publishing Company},\ \bibinfo {address} {Dordrecht},\ \bibinfo {year}
  {1983})\BibitemShut {NoStop}%
\bibitem [{\citenamefont {Fedichev}\ and\ \citenamefont
  {Shlyapnikov}(1998)}]{Fedichev1998}%
  \BibitemOpen
  \bibfield  {author} {\bibinfo {author} {\bibfnamefont {P.~O.}\ \bibnamefont
  {Fedichev}}\ and\ \bibinfo {author} {\bibfnamefont {G.~V.}\ \bibnamefont
  {Shlyapnikov}},\ }\href {\doibase 10.1103/PhysRevA.58.3146} {\bibfield
  {journal} {\bibinfo  {journal} {Phys. Rev. A}\ }\textbf {\bibinfo {volume}
  {58}},\ \bibinfo {pages} {3146} (\bibinfo {year} {1998})}\BibitemShut
  {NoStop}%
\bibitem [{\citenamefont {Capogrosso-Sansone}\ \emph
  {et~al.}(2010)\citenamefont {Capogrosso-Sansone}, \citenamefont {Giorgini},
  \citenamefont {Pilati}, \citenamefont {Pollet}, \citenamefont {Prokofev},
  \citenamefont {Svistunov},\ and\ \citenamefont
  {Troyer}}]{Capogrosso-Sansone2010}%
  \BibitemOpen
  \bibfield  {author} {\bibinfo {author} {\bibfnamefont {B.}~\bibnamefont
  {Capogrosso-Sansone}}, \bibinfo {author} {\bibfnamefont {S.}~\bibnamefont
  {Giorgini}}, \bibinfo {author} {\bibfnamefont {S.}~\bibnamefont {Pilati}},
  \bibinfo {author} {\bibfnamefont {L.}~\bibnamefont {Pollet}}, \bibinfo
  {author} {\bibfnamefont {N.}~\bibnamefont {Prokofev}}, \bibinfo {author}
  {\bibfnamefont {B.}~\bibnamefont {Svistunov}}, \ and\ \bibinfo {author}
  {\bibfnamefont {M.}~\bibnamefont {Troyer}},\ }\href {\doibase
  10.1088/1367-2630/12/4/043010} {\bibfield  {journal} {\bibinfo  {journal}
  {New J. Phys.}\ }\textbf {\bibinfo {volume} {12}},\ \bibinfo {pages} {043010}
  (\bibinfo {year} {2010})}\BibitemShut {NoStop}%
\bibitem [{\citenamefont {Petrov}(2015)}]{Petrov2015}%
  \BibitemOpen
  \bibfield  {author} {\bibinfo {author} {\bibfnamefont {D.~S.}\ \bibnamefont
  {Petrov}},\ }\href {\doibase 10.1103/PhysRevLett.115.155302} {\bibfield
  {journal} {\bibinfo  {journal} {Phys. Rev. Lett.}\ }\textbf {\bibinfo
  {volume} {115}},\ \bibinfo {pages} {155302} (\bibinfo {year}
  {2015})}\BibitemShut {NoStop}%
\bibitem [{\citenamefont {Ota}\ and\ \citenamefont
  {Astrakharchik}(2020)}]{Ota2020}%
  \BibitemOpen
  \bibfield  {author} {\bibinfo {author} {\bibfnamefont {M.}~\bibnamefont
  {Ota}}\ and\ \bibinfo {author} {\bibfnamefont {G.~E.}\ \bibnamefont
  {Astrakharchik}},\ }\href {\doibase 10.21468/SciPostPhys.9.2.020} {\bibfield
  {journal} {\bibinfo  {journal} {SciPost Phys.}\ }\textbf {\bibinfo {volume}
  {9}},\ \bibinfo {pages} {020} (\bibinfo {year} {2020})}\BibitemShut {NoStop}%
\bibitem [{\citenamefont {Ota}\ \emph {et~al.}(2019)\citenamefont {Ota},
  \citenamefont {Giorgini},\ and\ \citenamefont {Stringari}}]{Ota2019}%
  \BibitemOpen
  \bibfield  {author} {\bibinfo {author} {\bibfnamefont {M.}~\bibnamefont
  {Ota}}, \bibinfo {author} {\bibfnamefont {S.}~\bibnamefont {Giorgini}}, \
  and\ \bibinfo {author} {\bibfnamefont {S.}~\bibnamefont {Stringari}},\ }\href
  {\doibase 10.1103/PhysRevLett.123.075301} {\bibfield  {journal} {\bibinfo
  {journal} {Phys. Rev. Lett.}\ }\textbf {\bibinfo {volume} {123}},\ \bibinfo
  {pages} {075301} (\bibinfo {year} {2019})}\BibitemShut {NoStop}%
\bibitem [{\citenamefont {He}\ \emph {et~al.}(2020)\citenamefont {He},
  \citenamefont {Gao},\ and\ \citenamefont {Yu}}]{He2020}%
  \BibitemOpen
  \bibfield  {author} {\bibinfo {author} {\bibfnamefont {L.}~\bibnamefont
  {He}}, \bibinfo {author} {\bibfnamefont {P.}~\bibnamefont {Gao}}, \ and\
  \bibinfo {author} {\bibfnamefont {Z.-Q.}\ \bibnamefont {Yu}},\ }\href
  {\doibase 10.1103/PhysRevLett.125.055301} {\bibfield  {journal} {\bibinfo
  {journal} {Phys. Rev. Lett.}\ }\textbf {\bibinfo {volume} {125}},\ \bibinfo
  {pages} {055301} (\bibinfo {year} {2020})}\BibitemShut {NoStop}%
\bibitem [{\citenamefont {{Roy}}\ \emph {et~al.}(2020)\citenamefont {{Roy}},
  \citenamefont {{Ota}}, \citenamefont {{Recati}},\ and\ \citenamefont
  {{Dalfovo}}}]{Roy2020}%
  \BibitemOpen
  \bibfield  {author} {\bibinfo {author} {\bibfnamefont {A.}~\bibnamefont
  {{Roy}}}, \bibinfo {author} {\bibfnamefont {M.}~\bibnamefont {{Ota}}},
  \bibinfo {author} {\bibfnamefont {A.}~\bibnamefont {{Recati}}}, \ and\
  \bibinfo {author} {\bibfnamefont {F.}~\bibnamefont {{Dalfovo}}},\ }\href@noop
  {} {\  (\bibinfo {year} {2020})},\ \Eprint {http://arxiv.org/abs/2010.01933}
  {arXiv:2010.01933} \BibitemShut {NoStop}%
\bibitem [{\citenamefont {Andreev}\ and\ \citenamefont
  {Bashkin}(1975)}]{Andreev1975}%
  \BibitemOpen
  \bibfield  {author} {\bibinfo {author} {\bibfnamefont {A.~F.}\ \bibnamefont
  {Andreev}}\ and\ \bibinfo {author} {\bibfnamefont {E.~P.}\ \bibnamefont
  {Bashkin}},\ }\href@noop {} {\bibfield  {journal} {\bibinfo  {journal} {Zh.
  Eksp. Teor. Fiz}\ }\textbf {\bibinfo {volume} {69}},\ \bibinfo {pages} {319}
  (\bibinfo {year} {1975})}\BibitemShut {NoStop}%
\bibitem [{\citenamefont {Nespolo}\ \emph {et~al.}(2017)\citenamefont
  {Nespolo}, \citenamefont {Astrakharchik},\ and\ \citenamefont
  {Recati}}]{Nespolo2017}%
  \BibitemOpen
  \bibfield  {author} {\bibinfo {author} {\bibfnamefont {J.}~\bibnamefont
  {Nespolo}}, \bibinfo {author} {\bibfnamefont {G.~E.}\ \bibnamefont
  {Astrakharchik}}, \ and\ \bibinfo {author} {\bibfnamefont {A.}~\bibnamefont
  {Recati}},\ }\href {\doibase 10.1088/1367-2630/aa93a0} {\bibfield  {journal}
  {\bibinfo  {journal} {New J. Phys.}\ }\textbf {\bibinfo {volume} {19}},\
  \bibinfo {pages} {125005} (\bibinfo {year} {2017})}\BibitemShut {NoStop}%
\bibitem [{\citenamefont {Lifshitz}\ and\ \citenamefont
  {Pitaevskii}(1981)}]{LandauSP2}%
  \BibitemOpen
  \bibfield  {author} {\bibinfo {author} {\bibfnamefont {E.~M.}\ \bibnamefont
  {Lifshitz}}\ and\ \bibinfo {author} {\bibfnamefont {L.~P.}\ \bibnamefont
  {Pitaevskii}},\ }\href@noop {} {\emph {\bibinfo {title} {Statistical Physics
  Part 2}}}\ (\bibinfo  {publisher} {Pergamon Press},\ \bibinfo {address}
  {Oxford},\ \bibinfo {year} {1981})\BibitemShut {NoStop}%
\bibitem [{\citenamefont {Giorgini}(2000)}]{Giorgini2000}%
  \BibitemOpen
  \bibfield  {author} {\bibinfo {author} {\bibfnamefont {S.}~\bibnamefont
  {Giorgini}},\ }\href {\doibase 10.1103/PhysRevA.61.063615} {\bibfield
  {journal} {\bibinfo  {journal} {Phys. Rev. A}\ }\textbf {\bibinfo {volume}
  {61}},\ \bibinfo {pages} {063615} (\bibinfo {year} {2000})}\BibitemShut
  {NoStop}%
\bibitem [{\citenamefont {Lee}\ \emph {et~al.}(1957)\citenamefont {Lee},
  \citenamefont {Huang},\ and\ \citenamefont {Yang}}]{Lee1957}%
  \BibitemOpen
  \bibfield  {author} {\bibinfo {author} {\bibfnamefont {T.~D.}\ \bibnamefont
  {Lee}}, \bibinfo {author} {\bibfnamefont {K.}~\bibnamefont {Huang}}, \ and\
  \bibinfo {author} {\bibfnamefont {C.~N.}\ \bibnamefont {Yang}},\ }\href
  {\doibase 10.1103/PhysRev.106.1135} {\bibfield  {journal} {\bibinfo
  {journal} {Physical Review}\ }\textbf {\bibinfo {volume} {106}},\ \bibinfo
  {pages} {1135} (\bibinfo {year} {1957})}\BibitemShut {NoStop}%
\bibitem [{Note1()}]{Note1}%
  \BibitemOpen
  \bibinfo {note} {Alternatively, one can also replace $\Lambda $ by $gn_0$ in
  Eqs.~\protect \textup {\hbox {\mathsurround \z@ \protect \normalfont
  (\ignorespaces \ref {Eq:muLT_sc}\unskip \@@italiccorr )}}-\protect \textup
  {\hbox {\mathsurround \z@ \protect \normalfont (\ignorespaces \ref
  {Eq:n0LT_sc}\unskip \@@italiccorr )}} and solving self-consistently
  Eq.~\protect \textup {\hbox {\mathsurround \z@ \protect \normalfont
  (\ignorespaces \ref {Eq:n0LT_sc}\unskip \@@italiccorr )}} in $gn_0$. However,
  this procedure would restrict the validity of Eq.~\protect \textup {\hbox
  {\mathsurround \z@ \protect \normalfont (\ignorespaces \ref
  {Eq:muLT_sc}\unskip \@@italiccorr )}} to a narrower temperature region with a
  lower bound given by $k_B T \gg (na^3)^{1/4}$.}\BibitemShut {Stop}%
\bibitem [{\citenamefont {Prokof'ev}\ \emph {et~al.}(2001)\citenamefont
  {Prokof'ev}, \citenamefont {Ruebenacker},\ and\ \citenamefont
  {Svistunov}}]{Prokofev2001}%
  \BibitemOpen
  \bibfield  {author} {\bibinfo {author} {\bibfnamefont {N.}~\bibnamefont
  {Prokof'ev}}, \bibinfo {author} {\bibfnamefont {O.}~\bibnamefont
  {Ruebenacker}}, \ and\ \bibinfo {author} {\bibfnamefont {B.}~\bibnamefont
  {Svistunov}},\ }\href {\doibase 10.1103/PhysRevLett.87.270402} {\bibfield
  {journal} {\bibinfo  {journal} {Phys. Rev. Lett.}\ }\textbf {\bibinfo
  {volume} {87}},\ \bibinfo {pages} {270402} (\bibinfo {year}
  {2001})}\BibitemShut {NoStop}%
\bibitem [{\citenamefont {Prokof'ev}\ \emph {et~al.}(2004)\citenamefont
  {Prokof'ev}, \citenamefont {Ruebenacker},\ and\ \citenamefont
  {Svistunov}}]{Prokofev2004}%
  \BibitemOpen
  \bibfield  {author} {\bibinfo {author} {\bibfnamefont {N.}~\bibnamefont
  {Prokof'ev}}, \bibinfo {author} {\bibfnamefont {O.}~\bibnamefont
  {Ruebenacker}}, \ and\ \bibinfo {author} {\bibfnamefont {B.}~\bibnamefont
  {Svistunov}},\ }\href {\doibase 10.1103/PhysRevA.69.053625} {\bibfield
  {journal} {\bibinfo  {journal} {Phys. Rev. A}\ }\textbf {\bibinfo {volume}
  {69}},\ \bibinfo {pages} {053625} (\bibinfo {year} {2004})}\BibitemShut
  {NoStop}%
\bibitem [{\citenamefont {Arnold}\ \emph {et~al.}(2001)\citenamefont {Arnold},
  \citenamefont {Moore},\ and\ \citenamefont
  {Tom{\'{a}}{\v{s}}ik}}]{Arnold2001}%
  \BibitemOpen
  \bibfield  {author} {\bibinfo {author} {\bibfnamefont {P.}~\bibnamefont
  {Arnold}}, \bibinfo {author} {\bibfnamefont {G.}~\bibnamefont {Moore}}, \
  and\ \bibinfo {author} {\bibfnamefont {B.}~\bibnamefont
  {Tom{\'{a}}{\v{s}}ik}},\ }\href {\doibase 10.1103/PhysRevA.65.013606}
  {\bibfield  {journal} {\bibinfo  {journal} {Phys. Rev. A}\ }\textbf {\bibinfo
  {volume} {65}},\ \bibinfo {pages} {013606} (\bibinfo {year}
  {2001})}\BibitemShut {NoStop}%
\bibitem [{Note2()}]{Note2}%
  \BibitemOpen
  \bibinfo {note} {Similarly to the single-component case, the saddle point
  equation has to be evaluated from the unperturbed grand-canonical potential
  with the gapped spectrum, and \protect \textit {not} from Eqs.~\protect
  \textup {\hbox {\mathsurround \z@ \protect \normalfont (\ignorespaces \ref
  {Eq:K_mix}\unskip \@@italiccorr )}} and~\protect \textup {\hbox
  {\mathsurround \z@ \protect \normalfont (\ignorespaces \ref {Eq:E_mix}\unskip
  \@@italiccorr )}} where we have assumed the lowest order identity $\Lambda _i
  = g_{ii}n_i$ to hold. The details of the calculation can be found in
  Appendix~\ref {App:mixt}.}\BibitemShut {Stop}%
\bibitem [{\citenamefont {Pethick}\ and\ \citenamefont
  {Smith}(2008)}]{Pethick}%
  \BibitemOpen
  \bibfield  {author} {\bibinfo {author} {\bibfnamefont {C.~J.}\ \bibnamefont
  {Pethick}}\ and\ \bibinfo {author} {\bibfnamefont {H.}~\bibnamefont
  {Smith}},\ }\href@noop {} {\emph {\bibinfo {title} {{Bose-Einstein
  Condensation in Dilute Gases}}}}\ (\bibinfo  {publisher} {Cambridge
  University Press},\ \bibinfo {address} {Cambridge},\ \bibinfo {year}
  {2008})\BibitemShut {NoStop}%
\bibitem [{\citenamefont {Bienaim{\'{e}}}\ \emph {et~al.}(2016)\citenamefont
  {Bienaim{\'{e}}}, \citenamefont {Fava}, \citenamefont {Colzi}, \citenamefont
  {Mordini}, \citenamefont {Serafini}, \citenamefont {Qu}, \citenamefont
  {Stringari}, \citenamefont {Lamporesi},\ and\ \citenamefont
  {Ferrari}}]{Bienaime2016}%
  \BibitemOpen
  \bibfield  {author} {\bibinfo {author} {\bibfnamefont {T.}~\bibnamefont
  {Bienaim{\'{e}}}}, \bibinfo {author} {\bibfnamefont {E.}~\bibnamefont
  {Fava}}, \bibinfo {author} {\bibfnamefont {G.}~\bibnamefont {Colzi}},
  \bibinfo {author} {\bibfnamefont {C.}~\bibnamefont {Mordini}}, \bibinfo
  {author} {\bibfnamefont {S.}~\bibnamefont {Serafini}}, \bibinfo {author}
  {\bibfnamefont {C.}~\bibnamefont {Qu}}, \bibinfo {author} {\bibfnamefont
  {S.}~\bibnamefont {Stringari}}, \bibinfo {author} {\bibfnamefont
  {G.}~\bibnamefont {Lamporesi}}, \ and\ \bibinfo {author} {\bibfnamefont
  {G.}~\bibnamefont {Ferrari}},\ }\href {\doibase 10.1103/PhysRevA.94.063652}
  {\bibfield  {journal} {\bibinfo  {journal} {Phys. Rev. A}\ }\textbf {\bibinfo
  {volume} {94}},\ \bibinfo {pages} {063652} (\bibinfo {year}
  {2016})}\BibitemShut {NoStop}%
\bibitem [{\citenamefont {Fava}\ \emph {et~al.}(2018)\citenamefont {Fava},
  \citenamefont {Bienaim\'e}, \citenamefont {Mordini}, \citenamefont {Colzi},
  \citenamefont {Qu}, \citenamefont {Stringari}, \citenamefont {Lamporesi},\
  and\ \citenamefont {Ferrari}}]{Fava2018}%
  \BibitemOpen
  \bibfield  {author} {\bibinfo {author} {\bibfnamefont {E.}~\bibnamefont
  {Fava}}, \bibinfo {author} {\bibfnamefont {T.}~\bibnamefont {Bienaim\'e}},
  \bibinfo {author} {\bibfnamefont {C.}~\bibnamefont {Mordini}}, \bibinfo
  {author} {\bibfnamefont {G.}~\bibnamefont {Colzi}}, \bibinfo {author}
  {\bibfnamefont {C.}~\bibnamefont {Qu}}, \bibinfo {author} {\bibfnamefont
  {S.}~\bibnamefont {Stringari}}, \bibinfo {author} {\bibfnamefont
  {G.}~\bibnamefont {Lamporesi}}, \ and\ \bibinfo {author} {\bibfnamefont
  {G.}~\bibnamefont {Ferrari}},\ }\href {\doibase
  10.1103/PhysRevLett.120.170401} {\bibfield  {journal} {\bibinfo  {journal}
  {Phys. Rev. Lett.}\ }\textbf {\bibinfo {volume} {120}},\ \bibinfo {pages}
  {170401} (\bibinfo {year} {2018})}\BibitemShut {NoStop}%
\bibitem [{\citenamefont {Hall}\ \emph {et~al.}(1998)\citenamefont {Hall},
  \citenamefont {Matthews}, \citenamefont {Ensher}, \citenamefont {Wieman},\
  and\ \citenamefont {Cornell}}]{Hall1998}%
  \BibitemOpen
  \bibfield  {author} {\bibinfo {author} {\bibfnamefont {D.~S.}\ \bibnamefont
  {Hall}}, \bibinfo {author} {\bibfnamefont {M.~R.}\ \bibnamefont {Matthews}},
  \bibinfo {author} {\bibfnamefont {J.~R.}\ \bibnamefont {Ensher}}, \bibinfo
  {author} {\bibfnamefont {C.~E.}\ \bibnamefont {Wieman}}, \ and\ \bibinfo
  {author} {\bibfnamefont {E.~A.}\ \bibnamefont {Cornell}},\ }\href {\doibase
  10.1103/PhysRevLett.81.1539} {\bibfield  {journal} {\bibinfo  {journal}
  {Phys. Rev. Lett.}\ }\textbf {\bibinfo {volume} {81}},\ \bibinfo {pages}
  {1539} (\bibinfo {year} {1998})}\BibitemShut {NoStop}%
\bibitem [{\citenamefont {Papp}\ \emph {et~al.}(2008)\citenamefont {Papp},
  \citenamefont {Pino},\ and\ \citenamefont {Wieman}}]{Papp2008}%
  \BibitemOpen
  \bibfield  {author} {\bibinfo {author} {\bibfnamefont {S.~B.}\ \bibnamefont
  {Papp}}, \bibinfo {author} {\bibfnamefont {J.~M.}\ \bibnamefont {Pino}}, \
  and\ \bibinfo {author} {\bibfnamefont {C.~E.}\ \bibnamefont {Wieman}},\
  }\href {\doibase 10.1103/PhysRevLett.101.040402} {\bibfield  {journal}
  {\bibinfo  {journal} {Phys. Rev. Lett.}\ }\textbf {\bibinfo {volume} {101}},\
  \bibinfo {pages} {040402} (\bibinfo {year} {2008})}\BibitemShut {NoStop}%
\bibitem [{\citenamefont {{Van Schaeybroeck}}(2013)}]{Schaeybroeck2013}%
  \BibitemOpen
  \bibfield  {author} {\bibinfo {author} {\bibfnamefont {B.}~\bibnamefont {{Van
  Schaeybroeck}}},\ }\href {\doibase 10.1016/j.physa.2013.04.026} {\bibfield
  {journal} {\bibinfo  {journal} {Physica A}\ }\textbf {\bibinfo {volume}
  {392}},\ \bibinfo {pages} {3806} (\bibinfo {year} {2013})}\BibitemShut
  {NoStop}%
\bibitem [{\citenamefont {Romito}\ \emph {et~al.}(2020)\citenamefont {Romito},
  \citenamefont {Lobo},\ and\ \citenamefont {Recati}}]{Romito2019}%
  \BibitemOpen
  \bibfield  {author} {\bibinfo {author} {\bibfnamefont {D.}~\bibnamefont
  {Romito}}, \bibinfo {author} {\bibfnamefont {C.}~\bibnamefont {Lobo}}, \ and\
  \bibinfo {author} {\bibfnamefont {A.}~\bibnamefont {Recati}},\ }\href
  {http://arxiv.org/abs/2002.03955} {\  (\bibinfo {year} {2020})},\ \Eprint
  {http://arxiv.org/abs/2002.03955} {arXiv:2002.03955} \BibitemShut {NoStop}%
\bibitem [{\citenamefont {Fil}\ and\ \citenamefont
  {Shevchenko}(2005)}]{Fil2005}%
  \BibitemOpen
  \bibfield  {author} {\bibinfo {author} {\bibfnamefont {D.~V.}\ \bibnamefont
  {Fil}}\ and\ \bibinfo {author} {\bibfnamefont {S.~I.}\ \bibnamefont
  {Shevchenko}},\ }\href {\doibase 10.1103/PhysRevA.72.013616} {\bibfield
  {journal} {\bibinfo  {journal} {Phys. Rev. A}\ }\textbf {\bibinfo {volume}
  {72}},\ \bibinfo {pages} {013616} (\bibinfo {year} {2005})}\BibitemShut
  {NoStop}%
\bibitem [{\citenamefont {Parisi}\ and\ \citenamefont
  {Giorgini}(2017)}]{Parisi2017}%
  \BibitemOpen
  \bibfield  {author} {\bibinfo {author} {\bibfnamefont {L.}~\bibnamefont
  {Parisi}}\ and\ \bibinfo {author} {\bibfnamefont {S.}~\bibnamefont
  {Giorgini}},\ }\href {\doibase 10.1103/PhysRevA.95.023619} {\bibfield
  {journal} {\bibinfo  {journal} {Phys. Rev. A}\ }\textbf {\bibinfo {volume}
  {95}},\ \bibinfo {pages} {023619} (\bibinfo {year} {2017})}\BibitemShut
  {NoStop}%
\bibitem [{\citenamefont {Kim}\ \emph {et~al.}(2020)\citenamefont {Kim},
  \citenamefont {Hong},\ and\ \citenamefont {Shin}}]{Kim2020}%
  \BibitemOpen
  \bibfield  {author} {\bibinfo {author} {\bibfnamefont {J.~H.}\ \bibnamefont
  {Kim}}, \bibinfo {author} {\bibfnamefont {D.}~\bibnamefont {Hong}}, \ and\
  \bibinfo {author} {\bibfnamefont {Y.}~\bibnamefont {Shin}},\ }\href {\doibase
  10.1103/PhysRevA.101.061601} {\bibfield  {journal} {\bibinfo  {journal}
  {Phys. Rev. A}\ }\textbf {\bibinfo {volume} {101}},\ \bibinfo {pages}
  {061601(R)} (\bibinfo {year} {2020})}\BibitemShut {NoStop}%
\bibitem [{\citenamefont {Larsen}(1963)}]{Larsen1963}%
  \BibitemOpen
  \bibfield  {author} {\bibinfo {author} {\bibfnamefont {D.~M.}\ \bibnamefont
  {Larsen}},\ }\href@noop {} {\bibfield  {journal} {\bibinfo  {journal} {Ann.
  Phys. (Berlin)}\ }\textbf {\bibinfo {volume} {24}},\ \bibinfo {pages} {89}
  (\bibinfo {year} {1963})}\BibitemShut {NoStop}%
\bibitem [{\citenamefont {Timmermans}(1998)}]{Timmermans1998}%
  \BibitemOpen
  \bibfield  {author} {\bibinfo {author} {\bibfnamefont {E.}~\bibnamefont
  {Timmermans}},\ }\href {\doibase 10.1103/PhysRevLett.81.5718} {\bibfield
  {journal} {\bibinfo  {journal} {Phys. Rev. Lett.}\ }\textbf {\bibinfo
  {volume} {81}},\ \bibinfo {pages} {5718} (\bibinfo {year}
  {1998})}\BibitemShut {NoStop}%
\end{thebibliography}%

\end{document}